\documentclass[12pt]{article}
\usepackage{amssymb,amsmath,bm,bbm}
\usepackage{epsf}
\usepackage{epsfig}
\usepackage{afterpage}
\usepackage{longtable}
\usepackage{cite}
\usepackage{latexsym, mathrsfs}
\usepackage{color}
\usepackage{axodraw}

 \usepackage{theorem}
\usepackage[latin1]{inputenc}
 \usepackage{amsfonts}
\usepackage{amstext}
  \usepackage{graphicx}
 \usepackage{epic}
 \usepackage{psfrag}
 \usepackage{scrpage2}
\usepackage{textcomp}

\newcommand{\rep}[1]{\mathbf{#1}}
\newcommand{\bidoublet}{(\rep{2},\rep{2})}
\newcommand{\singlet}{(\rep{1},\rep{1})}

\newcommand{\Ltriplet}{(\rep{3},\rep{1})}
\newcommand{\Rtriplet}{(\rep{1},\rep{3})}
\DeclareMathOperator{\Tr}{Tr}

\def\simge{\mathrel{\rlap{\raise 0.511ex \hbox{$>$}}{\lower 0.511ex \hbox{$\sim$}}}}
\def\simle{\mathrel{\rlap{\raise 0.511ex \hbox{$<$}}{\lower 0.511ex \hbox{$\sim$}}}}
\def\slash#1{\setbox0=\hbox{$#1$}\dimen0=\wd0
      \setbox1=\hbox{/} \dimen1=\wd1 \ifdim\dimen0>\dimen1
      \rlap{\hbox to \dimen0{\hfil/\hfil}} #1                        \else
      \rlap{\hbox to \dimen1{\hfil$#1$\hfil}}
      /   \fi}

\newcommand{\lsim}{
\mathrel{\hbox{\rlap{\hbox{\lower4pt\hbox{$\sim$}}}\hbox{$<$}}}}

\newcommand{\gsim}{
\mathrel{\hbox{\rlap{\hbox{\lower4pt\hbox{$\sim$}}}\hbox{$>$}}}}

\def\eps{\varepsilon}

\newcommand{\tev}{\, {\rm TeV}}
\newcommand{\gev}{\, {\rm GeV}}

\newcommand{\MKf}{M^{\text{KK}}}

\allowdisplaybreaks[2]

\newcommand{\be}{\begin{equation}}
\newcommand{\ee}{\end{equation}}
\newcommand{\bea}{\begin{eqnarray}}
\newcommand{\eea}{\end{eqnarray}}
\newcommand{\nn}{\nonumber}
\newcommand{\bi}{\begin{itemize}}
\newcommand{\ei}{\end{itemize}}
\newcommand{\ord}{{\cal O}}

\newcommand{\newsection}[1]{\section{#1}\setcounter{equation}{0}}

\begin{document}
\begin{titlepage}
\vspace*{-0.5truecm}

%{\Large \today}
\begin{flushright}
TUM-HEP-711/09\\
MPP-2009-17
\end{flushright}

\vfill

\begin{center}
\boldmath
{\Large\textbf{Electroweak and Flavour Structure of a\vspace{0.2truecm} \\
Warped Extra Dimension  
with Custodial Protection}}
\unboldmath
\end{center}

\vspace{0.4truecm}

\begin{center}
{\bf Michaela { E.~}Albrecht$^a$, Monika Blanke$^{a,b}$, Andrzej J.~Buras$^{a,c}$,\\ Bj\"orn Duling$^a$ and Katrin Gemmler$^{a}$}
\vspace{0.4truecm}

 $^a${\sl Physik Department, Technische Universit\"at M\"unchen,
D-85748 Garching, Germany}\vspace{0.1truecm}

 {\sl $^b$Max-Planck-Institut f{\"u}r Physik (Werner-Heisenberg-Institut), \\
D-80805 M{\"u}nchen, Germany}\vspace{0.1truecm}

 $^c${\sl TUM Institute for Advanced Study, Technische Universit\"at
   M\"unchen,
 \\ D-80333 M\"unchen, Germany}
\end{center}
\vspace{0.6cm}
\begin{abstract}
\vspace{0.2cm}
\noindent 
We present the electroweak and flavour structure of a model with a warped
extra dimension and the bulk gauge group 
$SU(3)_c\times SU(2)_L\times SU(2)_R\times P_{LR}\times U(1)_X$.
The presence of $SU(2)_R$ implies {an unbroken} custodial symmetry in the Higgs system
allowing to eliminate large contributions to the $T$ parameter, whereas the
$P_{LR}$ symmetry and the enlarged fermion representations provide a custodial 
symmetry for flavour diagonal and flavour changing couplings of the SM $Z$
boson
to left-handed {down-type} quarks. We {diagonalise} analytically  the mass matrices of
charged and neutral gauge bosons including the first KK modes. We present the
mass matrices for quarks including heavy KK modes and discuss the neutral and
charged currents involving light and heavy fields. We give the corresponding 
complete set of Feynman rules {in the unitary gauge}.

\end{abstract}

\vfill\vfill\vfill\vfill
\end{titlepage}

\setcounter{page}{1}
\pagenumbering{arabic}

\newsection{Introduction}\label{sec:intro}

Models with a  warped extra dimension, {also called Randall-Sundrum (RS) models} {\cite{Randall:1999ee,Chang:1999nh,Csaki:2005vy,Gherghetta:2006ha}}, in which all Standard Model (SM) fields are allowed to propagate in the bulk, offer natural solutions to many outstanding puzzles of contemporary particle physics. In addition to providing a geometrical solution to the hierarchy problem related to the vast difference between the Planck scale and the electroweak (EW) scale, they also allow to naturally generate hierarchies in fermion masses and weak mixing angles \cite{Grossman:1999ra,Gherghetta:2000qt}, suppress flavour changing neutral current (FCNC) interactions \cite{Huber:2003tu,Agashe:2004cp,Moreau:2006np}, construct realistic models of EW symmetry breaking (EWSB) \cite{Agashe:2003zs,Csaki:2003zu,Agashe:2004rs,Cacciapaglia:2006gp,Contino:2006qr,Carena:2006bn} and achieve gauge coupling unification\cite{Agashe:2002pr,Agashe:2005vg}.

The question then arises whether some imprints of this new physics scenario could be in the reach of the LHC, while satisfying all existing experimental constraints, coming in particular from EW precision tests, from the data on FCNC processes in both quark and lepton sectors and also the data on the very highly suppressed electric dipole moments (for recent reviews, see \cite{Amsler:2008zz,Buchalla:2008jp,Raidal:2008jk}).

A necessary, though not always sufficient, condition for direct signals of RS
{models} at the LHC is the existence of Kaluza-Klein (KK) modes with $\ord(1\tev)$
masses. Early studies of EW precision observables (EWPO)
\cite{Agashe:2003zs,Agashe:2005dk} have shown that with the SM gauge group in
the bulk such low masses of KK particles are inconsistent  in particular with
the bounds on the oblique parameter $T$ and the well-measured $Zb_L\bar b_L$
coupling.

In a number of very interesting papers \cite{Agashe:2003zs,Csaki:2003zu,Agashe:2004rs,Agashe:2006at,Contino:2006qr,Carena:2006bn,Cacciapaglia:2006gp} these two obstacles have been basically overcome by enlarging the bulk {symmetry} to
\be\label{eq:Gbulk}
G_\text{bulk}=SU(3)_c\times SU(2)_L\times SU(2)_R\times {  U(1)_X \times P_{LR}}
\ee
and enlarging the fermion representations, so that the discrete left-right symmetry $P_{LR}$, exchanging $SU(2)_L$ and $SU(2)_R$, is preserved. The presence of the additional gauge group $SU(2)_R$ implies the existence of an unbroken custodial $SU(2)$ symmetry in the Higgs sector, so that tree level contributions to the $T$ parameter can be safely neglected. The $P_{LR}$ symmetry and the related enlarged fermion representations eliminate the problematic contributions to the  $Zb_L\bar b_L$ coupling.

Interestingly, the presence of new light KK modes necessary to solve the ``$Zb_L\bar b_L$ problem'' implies significant contributions to the $T$ parameter at the one loop level \cite{Carena:2007ua}. However with an appropriate choice of quark bulk mass parameters, an agreement with the EW precision data in the presence of light KK modes can be obtained {\cite{Djouadi:2006rk,Bouchart:2008vp}}. In fact, while the masses of the KK gauge bosons are forced to be at least $(2-3)\tev$ to be consistent with the data on the oblique parameter $S$, fermionic KK modes with masses even below $1\tev$ can be made consistent with the measured EWPO.

The suppression of FCNC transitions to an acceptable level in the presence of
light KK modes turns out to be much more challenging if the hierarchy of
fermion masses and weak mixings is supposed to come solely from geometry so
that the fundamental 5D Yukawa couplings are anarchic. In fact, recent studies
demonstrate that in this case the data on the CP-violating parameter $\eps_K$
imply a lower bound on the lightest gauge KK modes in the ballpark of {$20\tev$}
\cite{Csaki:2008zd,Blanke:2008zb}, the corresponding bound from $\mu\to e\gamma$ is
above 
$10\tev$ \cite{Agashe:2006iy,Davidson:2007si}\footnote{We would like to mention that this bound can be avoided by new choices of lepton representations~\cite{Agashe:2009tu}.} and even stronger bounds come
from electric dipole moments \cite{Agashe:2004cp,Iltan:2007sc}. {Moreover
it has been pointed out in \cite{Agashe:2008uz} that the flavour problem
in these models becomes even more serious when $\eps_K$ and $B\to X_s\gamma$
decays are considered simultaneously.} {Note however that the bound in question can be somewhat relaxed by appropriately chosen brane kinetic terms \cite{Csaki:2008zd} and/or by allowing the Higgs boson to propagate in the bulk \cite{Agashe:2008uz}.}

In view of this situation a number of proposals has been made in order to overcome these ``FCNC problems'' of RS models that are directly related to the breakdown of the universality of gauge boson--fermion couplings implied by the geometric explanation of the hierarchical structure of fermion masses and mixings. This breakdown implies the violation of the GIM mechanism \cite{Glashow:1970gm} and consequently tree level FCNC transitions that are inconsistent with the data for light KK scales, provided that anarchic 5D Yukawa couplings are chosen and the relevant couplings are $\ord(1)$.

In \cite{Cacciapaglia:2007fw} a class of RS models has been considered that
makes use of bulk and brane flavour symmetries in order to prevent the theory
from large FCNCs. It has been shown that if flavour mixing is introduced via
UV brane kinetic terms, the GIM mechanism is realized
and a minimal flavour violating (MFV) model
\cite{Buras:2000dm,Buras:2003jf,D'Ambrosio:2002ex,Hall:1990ac,Chivukula:1987py}
can be obtained. However, the natural explanation of fermionic hierarchies
had to be abandoned in that setup. A different strategy has been followed in
\cite{Fitzpatrick:2007sa}, where the field theoretical concept of MFV has been
promoted to the 5D theory, i.\,e. the bulk mass matrices are expressed in
terms of the 5D Yukawa couplings. Low energy flavour violation can be further
suppressed by a single parameter that dials the amount of violation in the up
or down sector. If this parameter is ensured to be small, no flavour or CP problem arises even with KK masses as
low as $2\tev$. A more thorough analysis, including the presentation of a
possible dynamical origin of such a model, has been given in
\cite{CGPWS}. 
Another economical model based on {a $U(3)_d$} bulk flavour symmetry has been proposed in \cite{Santiago:2008vq}. Here the right-handed down quark bulk masses are enforced to be degenerate, so that the contributions of the $\mathcal{Q}_{LR}$ operator to $\eps_K$ are generated only by suppressed mass insertions on the IR brane.
A recent approach \cite{Csaki:2008eh} presents a simple model where the key ingredient are two horizontal $U(1)$ symmetries. The SM fields are embedded  into the 5D fields motivated by protecting $Zb_L\bar b_L$. The horizontal $U(1)$ symmetries force an alignment of bulk masses and down Yukawas which strongly suppresses FCNCs in the down sector. FCNCs in the up sector, however, can be close 
to the experimental limits.

In two recent papers \cite{Blanke:2008zb,Blanke:2008yr} we took a different strategy and
investigated to which extent 
a hierarchy in the 5D Yukawa couplings has to be reintroduced in order to
achieve consistency with the existing data on FCNC processes in the presence
of KK modes in the reach of the LHC.
In particular in \cite{Blanke:2008zb} we have demonstrated that
there exist regions in parameter space with only modest fine-tuning
in the 5D Yukawa couplings involved which allow to obtain a satisfactory
description of the quark masses and weak mixing angles and to satisfy all
existing $\Delta F=2$ and
EW precision constraints for scales $M_\text{KK}\simeq 3\tev$
in {the} reach of the LHC.
 As the dominant part of the observed hierarchy in masses and mixings is still
 explained
 through the AdS$_5$ geometry, the resulting hierarchies are significantly
 milder  than in the SM and other usual 4D approaches.

 Subsequently, confining the numerical analysis to the
regions of parameter space allowed by $\Delta F=2$ observables and with 
only modest
fine-tuning,
we have presented in \cite{Blanke:2008yr}
a complete study of rare $K$ and $B$ meson decays 
including
  $K^+\to \pi^+\nu\bar\nu$, $K_L\to\pi^0\nu\bar\nu$, $K_L\to\pi^0 \ell^+\ell^-$,
$K_L\to \mu^+\mu^-$, $B_{s,d}\to \mu^+\mu^-$, $B\to K\nu\bar\nu$, 
$B\to K^*\nu\bar\nu$  and
$B_{s,d}\to X_{s,d}\nu\bar\nu$.

In this context it should be emphasised that the presence of FCNC transitions
already at the tree level in the model in question,
as opposed to the MSSM and Little Higgs models, 
necessarily implies other patterns in CP-violating observables and rare decay
branching
 ratios. In particular in RS models not only non-MFV interactions are present, 
like for instance in the Little Higgs models
with T-Parity, but also new operators 
become important that are strongly suppressed in the {latter}. As
found in \cite{Blanke:2008zb,Blanke:2008yr}
such new contributions lead to interesting deviations from the SM and in
particular from models with Constrained MFV
\cite{Buras:2000dm,Buras:2003jf,Blanke:2006ig} in observables that are still
poorly measured and which allow for large new physics contributions.

{The main results of \cite{Blanke:2008zb} can be briefly summarised as follows:
\begin{itemize}
\item 
 The EW tree level contributions to $\Delta F=2$ observables mediated by the new
  weak gauge boson {$Z_H$}, while subleading in the case of
 $\varepsilon_{\rm K}$ and $\Delta M_{\rm K}$, turn out to be of  
roughly the 
same size
 as the KK gluon contributions in the case of $B_{d,s}$ physics observables.
\item
 The contributions of KK {gauge boson} tree level exchanges involving new
 flavour and CP-violating interactions allow not only to satisfy all
 existing $\Delta F=2$ constraints but also to remove a number of
 tensions between  the SM and the data, claimed in particular in
 $\varepsilon_{K}$, $S_{\psi {K_S}}$ and $S_{\psi \phi}$
 \cite{Buras:2008nn,Lunghi:2008aa,Lenz:2006hd,Bona:2008jn}.
\item Interestingly the model allows naturally for
 $S_{\psi \phi}$ as high as 0.4 that is hinted at by the most recent CDF and
  D{\O} data \cite{Aaltonen:2007he,:2008fj,Brooijmans:2008nt} and which is
 by an
 order of magnitude larger than the SM expectation: $(S_{\psi \phi})_\text{SM}\simeq
 0.04$.
\item
The $P_{LR}$ symmetry implies automatically the protection of
flavour violating
 $Zd_L^i\bar d_L^j$ couplings so that tree level $Z$ contributions to 
 all processes in which flavour changes appear in the down quark sector
are  dominantly represented by $Zd_R^i\bar d_R^j$ couplings.
\item 
However, the tree level $Z$ contributions to $\Delta F=2$  processes are of
 higher order in $v/M_{\rm KK}$ and can be neglected.
\end{itemize}
On the other hand the main messages from \cite{Blanke:2008yr} are as follows:
\begin{itemize}
\item
New physics contributions to rare $K$ and $B$ decays,
as opposed to $\Delta F=2$ transitions, are governed
by
 tree level contributions from $Z$ boson exchanges 
(dominated by $Z d_R^i \bar d_R^j$ couplings) with the new heavy  
  EW gauge bosons playing a subdominant role. 
\item
Imposing
all existing constraints from $\Delta F=2$ transitions 
we find that a number of branching ratios for rare $K$ decays 
can differ significantly from
the SM predictions, while the corresponding effects in rare $B$ decays 
are modest. In particular the branching ratios for $K_L\to\pi^0\nu\bar\nu$
and $K^+\to\pi^+\nu\bar\nu$ can be by a factor of three and two larger than the
SM predictions, respectively.
The latter enhancement could be welcomed one day if the central
  experimental value \cite{Artamonov:2008qb} will remain in the ballpark 
of $15\cdot 10^{-11}$ and its
error will decrease. 
\item
However, it is very unlikely to get simultaneously large 
NP effects in rare $K$ decays and $S_{\psi\phi}$, which constitutes a
good test of the model.
\item
Sizable departures from the MFV relations 
between $\Delta M_{s,d}$ and
    $Br(B_{s,d} \to \mu^+ \mu^-)$ and between $S_{\psi K_S}$ and the $K \to
    \pi \nu \bar \nu$ decay rates are possible. 
\item
The pattern of
deviations from the SM differs from the deviations found in {the LHT} model \cite{Blanke:2006eb}.
\end{itemize}

 It is interesting that in spite of many new flavour parameters present
 in this model a clear pattern of new flavour violating effects has been 
 identified in \cite{Blanke:2008zb,Blanke:2008yr}{:
 large} effects in $\Delta F=2$ transitions, large effects
 in $\Delta F=1$ rare $K$ decays, small effects in $\Delta F=1$ rare $B$
 decays and the absence of simultaneous large effects in the $K$ and $B$
 system. This pattern implies that an observation of a large $S_{\psi\phi}$
 asymmetry would in the context of this model preclude sizable NP effects
 in rare $K$ decays. On the other hand, finding $S_{\psi\phi}$ to 
 be SM-like will open the road to large NP effects in rare $K$ decays, even
 if such large effects are only a possibility and are not guaranteed.
 On the other hand, an observation of large NP effects in rare $B$ decays would
 put this model in serious difficulties. 
}

 In \cite{Blanke:2008zb,Blanke:2008yr} only a brief description of the RS model
{in question} has been presented as only gauge boson exchanges were relevant at the
tree level. {In particular details on the fermion sector have not been presented there.} For the subsequent phenomenological
studies {like the $b\to s\gamma$ and $\mu\to e \gamma$ transitions} it is of interest to have a more detailed presentation {which}
is the main goal of our paper.
 We  formulate a particular RS model based on 
the bulk gauge group $G_\text{bulk}$ in \eqref{eq:Gbulk} and having
appropriate quark representations in order to avoid  tensions with EWPO. 
We work out the general structure of the gauge and fermion sectors, discuss
the new sources of flavour violation, and we give a collection of Feynman
rules\footnote{Some of these Feynman rules have already been presented in \cite{Agashe:2007ki,Agashe:2008jb}.} that can be used to calculate all  observables of interest. 
In fact a subset of the Feynman rules presented here has already been
used in \cite{Blanke:2008zb,Blanke:2008yr}.

Throughout our analysis we follow the perturbative approach, i.\,e. we first solve the 5D equations of motion and perform the KK decomposition in the absence of EWSB, as also done e.\,g. in \cite{Agashe:2007ki,Muck:2001yv} and then treat the Higgs vacuum expectation value (VEV) as a small perturbation that induces mixing among the various modes. The complementary approach, solving the equations of motion already in the presence of EWSB, has been followed e.\,g. in
\cite{Huber:2000fh,Csaki:2002gy,Burdman:2002gr,Csaki:2003dt}. Recently, a very detailed theoretical discussion of the latter approach has been presented in \cite{Casagrande:2008hr}. In Appendix \ref{app:approaches} we show that both approaches are indeed equivalent; for an independent discussion see also \cite{Goertz:2008vr}.

The present paper is organised as follows. In Section \ref{sec:gauge} we
present in detail the gauge sector of the model and in particular the effects 
of EWSB. The final formulae for gauge boson masses and mixings in
the charged and neutral sectors are collected in Appendix \ref{app:EWSB}. Next
in Section \ref{sec:quarks} we set up the quark representations under the bulk
gauge group. 
In Section \ref{sec:flavour},
one of the main sections of our paper, we work out the flavour structure of the
quark sector. After a detailed discussion of quark mass matrices and Yukawa
couplings in the flavour eigenbasis we {outline} the diagonalisation of these
matrices and study the structure of weak neutral and charged
currents. Subsequently the couplings of KK gluons and photons are considered.
This section forms the basis of the Feynman rules in the quark sector that are collected in
Appendix \ref{app:FR}. We end this section by listing the sources of flavour
violation in this model{, with the pattern of flavour violation, in particular in $\Delta F=1$ processes, governed by the custodial protection present in the model}.
In Section \ref{sec:parameters} we list the parameters of the model
and present a useful parameterisation for the 5D Yukawa couplings {in terms of parameters accessible at low energies}.
In Section \ref{sec:leptons} we discuss one possible realisation 
of the lepton sector and present a dictionary that allows in a straightforward
manner to obtain the Feynman rules for the leptons from those of quarks.
We close the paper with a brief summary in Section \ref{sec:concl}.

\newsection{Gauge Sector}\label{sec:gauge}

\subsection{Preliminaries}

We consider an $SU(3)_c\times SU(2)_L\times SU(2)_R\times U(1)_X\times P_{LR}$ gauge theory on a slice of AdS$_5$ with the metric \cite{Randall:1999ee}
\be\label{eq:RS}
ds^2=e^{-2ky}\eta_{\mu\nu}dx^\mu dx^\nu - dy^2\,,
\ee
with the fifth coordinate being restricted to the interval $0\le y\le L$, and
$k\sim\ord(M_\text{Pl})$. {In order to simplify the phenomenological
  discussion in {\cite{Blanke:2008zb,Blanke:2008yr}} we chose to work with the $(+----)$ sign convention for the metric, i.\,e. $\eta_{\mu\nu}=\text{diag}(1,-1,-1,-1)$.}
The gauge bosons and fermions are allowed to propagate in the 5D bulk, while the Higgs field will be localised on or near the  IR brane ($y=L$).

In the EW sector, we consider the gauge symmetry \cite{Agashe:2003zs,Csaki:2003zu,Agashe:2006at}
\be\label{eq:bulk-group}
O(4)\times U(1)_X\sim SU(2)_L\times SU(2)_R\times P_{LR}\times U(1)_X\,,
\ee
where $P_{LR}$ is the discrete symmetry interchanging the two $SU(2)$ groups. This means for instance that $g_{L}=g_{R}\equiv g$. The gauge group \eqref{eq:bulk-group} is broken by boundary conditions (BCs) on the UV brane ($y=0$) to
 the Standard Model (SM) gauge group, i.\,e.
\be
SU(2)_L\times SU(2)_R\times P_{LR}\times U(1)_X \xrightarrow{\text{ UV brane }} SU(2)_L\times U(1)_Y\,.
\ee
This breakdown is achieved by the following assignment of BCs\footnote{These BCs can be naturally achieved by adding a scalar $SU(2)_R$ doublet with $Q_X=1/2$ charge on the UV brane, that develops a VEV $v_\text{UV}\to\infty$ (see \cite{Csaki:2003dt,Csaki:2005vy} for details).}
\bea
W_{L\mu}^a (++)\,,\qquad && B_\mu (++)\,, \\
W_{R\mu}^b (-+)\,,\qquad && Z_{X\mu} (-+)\,,
\eea
{where the first (second) sign denotes the BC on the UV (IR) brane: $+$ stands for a Neumann BC while $-$  stands for a Dirichlet BC. Furthermore $a=1,2,3$ and $b=1,2$. The fields $B_\mu$ and $Z_{X\mu}$ are given in terms of the original fields $W_{R\mu}^3$ and $X_\mu$ as follows:}
\bea
Z_{X\mu} &=& \cos\phi\, W_{R\mu}^{3} - \sin\phi\, X_\mu\,,\\
B_{\mu} &=& \sin\phi\, W_{R\mu}^{3} + \cos\phi\, X_\mu\,,
\eea
where
\be\label{eq:phi}
\cos\phi = \frac{g}{\sqrt{g^2+g_X^2}}\,,
\qquad
\sin\phi = \frac{g_{X}}{\sqrt{g^2+g_X^2}}\,.
\ee
Here, $g$ and $g_X$ are the 5D gauge couplings of $O(4)$ and $U(1)_X$, respectively. Note that the BCs for a gauge field $V_\mu$ imply automatically opposite BCs for its 5th component $V_5$. In what follows we choose to work in the gauge $V_5=0$ and $\partial_\mu V^\mu=0$.

The fields with $(++)$ BCs have, in addition to the massive KK modes, zero modes which are massless at this stage and are identified with the SM gauge bosons $W_{L\mu}^a$ and $B_\mu$ {of $SU(2)_L\times U(1)_Y$}.
The fields with $(-+)$ BCs contain only massive KK modes.

Before EWSB the profiles of gauge boson zero modes along the extra dimension are flat. The profiles of KK gauge bosons are given by \cite{Gherghetta:2000qt} { (see also Appendix \ref{app:eom} for details)}
\be\label{eq:gaugeprofile}
f^{(n)}_\text{gauge}(y)= \frac{e^{ky}}{N_n}\left[J_1\left(\frac{m_n}{k}e^{ky}\right)+b_1(m_n)Y_1\left(\frac{m_n}{k}e^{ky}\right)\right]\,,
\ee
where $J_1(x)$ and  $Y_1(x)$ are the Bessel functions of first and second kind, and explicit expressions for $b_1(m_n)$ and $N_n$ can be found in Appendix \ref{app:eom}. The bulk masses are approximately given by \cite{Gherghetta:2000qt}
\be
m^\text{gauge}_n \simeq \left(n - \frac{1}{4}\right)\pi k e^{-kL}\qquad (n=1,2,\dots)
\ee
for the modes with a $+$ BC on the IR brane that we are presently interested in. {The accuracy of this approximate formula improves significantly with increasing $n$, hence for the first KK modes it is safer to work with the exact KK masses. These can be found numerically to be
\be
m^\text{gauge}_1(++) \simeq 2.45  f \equiv M_{++}
\ee
for gauge bosons with $(++)$ BCs, and 
\be
m^\text{gauge}_1(-+) \simeq 2.40  f \equiv M_{-+}
\ee
for gauge bosons with $(-+)$ BCs. Here we have introduced the effective new physics scale $f=ke^{-kL}$ and set $e^{-kL}\simeq 10^{-16}$ in order to solve the hierarchy problem.} The $\sim2\%$ suppression in the latter case is a direct consequence of the different BC on the UV brane~\cite{Agashe:2007ki}. 
Note that the KK masses for the gauge bosons depend neither on the gauge group nor on the size of the gauge coupling, but are universal for all gauge bosons with the same BCs. Only after EWSB, the weak KK gauge boson masses will receive small additional corrections. 
As can easily be seen from \eqref{eq:gaugeprofile}, the gauge KK modes are localised near the IR brane.

To further proceed it will be useful to follow \cite{Agashe:2007ki} and define the fields
\be
W_{L\mu}^{\pm}=\frac{W_{L\mu}^{1}\mp iW_{L\mu}^{2} }{\sqrt{2}}\,,\qquad 
W_{R\mu}^{\pm}=\frac{W_{R\mu}^{1}\mp iW_{R\mu}^{2} }{\sqrt{2}}\,,
\ee
and
\bea
Z_{\mu} &=& \cos\psi\, W_{L\mu}^{3} - \sin\psi\, B_\mu\,,\\
A_{\mu} &=& \sin\psi\, W_{L\mu}^{3} + \cos\psi\, B_\mu\,,
\eea
where again $\sin\psi$ is given in terms of gauge couplings
 (see \eqref{eq:phi} for the definition of $\phi$)
\be\label{eq:psi}
\cos\psi=\frac{1}{\sqrt{1+\sin^2\phi}}\,,\qquad \sin\psi=\frac{\sin\phi}{\sqrt{1+\sin^2\phi}}\,.
\ee
Because of {the} mixing between the various gauge boson zero and KK modes $\sin\psi
\ne \sin\theta_W${, but corrections appear first at  order $\ord(v^2/f^2)$}. {{Their} impact on EW precision studies is beyond the scope of this paper and will be studied elsewhere.}

{We note that the above relations can be modified by the presence of additional gauge kinetic terms on the UV and IR branes, that are allowed by the symmetries of the model. In order not to complicate our analysis, we will neglect such terms and work exclusively with the action given in Appendix \ref{app:action}. A generalisation of our results to include also the effects of possible brane terms is straightforward.}
{In Section \ref{sec:brane} we comment on the effects of such terms on flavour phenomenology.}

\subsection{Electroweak Symmetry Breaking}\label{sec:EWSB}

{As discussed in the previous section, the bulk gauge symmetry $G_\text{bulk}$ in \eqref{eq:Gbulk} is broken to the SM gauge group
\be
G_\text{UV} = SU(3)_c \times SU(2)_L \times U(1)_Y \equiv G_\text{SM}
\ee
by means of the BCs of the EW gauge bosons on the UV brane. In order to achieve the standard EWSB, $SU(2)_L\times U(1)_Y\to U(1)_Q$, 
a Higgs boson is introduced that is localised either on or near the IR brane,  transforming as a {self-dual} bidoublet of $SU(2)_L\times SU(2)_R$
\be\label{eq:H}
H=\begin{pmatrix} \pi^+/\sqrt{2} && -(h^0-i\pi^0)/{2} \\
(h^0+i\pi^0)/{2} && \pi^-/\sqrt{2} \end{pmatrix}\,,
\ee
and being a singlet under $U(1)_X$, $Q_X(H)=0$.
 In the {case of a 5D Higgs field living in the bulk}, the whole bidoublet has to obey $(++)$ BCs in order to yield a light zero mode.

When its neutral component $h^0$ develops a 4D effective VEV, on or near the IR brane the symmetry breaking
\be
SU(2)_L \times SU(2)_R \times P_{LR} \to SU(2)_V \times P_{LR}
\ee
takes place. We see explicitly that in the Higgs sector of the theory an unbroken custodial symmetry $SU(2)_V$ remains, being responsible for the protection of the $T$ parameter. Similarly the $P_{LR}$ symmetry, protecting the $Z d_L^i\bar d_L^j$ coupling, remains unbroken.

Combining then the symmetry breakings by BCs on the UV brane and by the Higgs VEV in the IR, we see that the low energy effective theory is described by the spontaneous breaking 
\be
SU(2)_L\times U(1)_Y\to U(1)_Q\,,
\ee
as required by phenomenology. The symmetry breaking structure of the model 
 is displayed in Fig.~\ref{fig:SBpattern}.

\begin{figure}
\center{\includegraphics[width = .5\textwidth,angle=270]{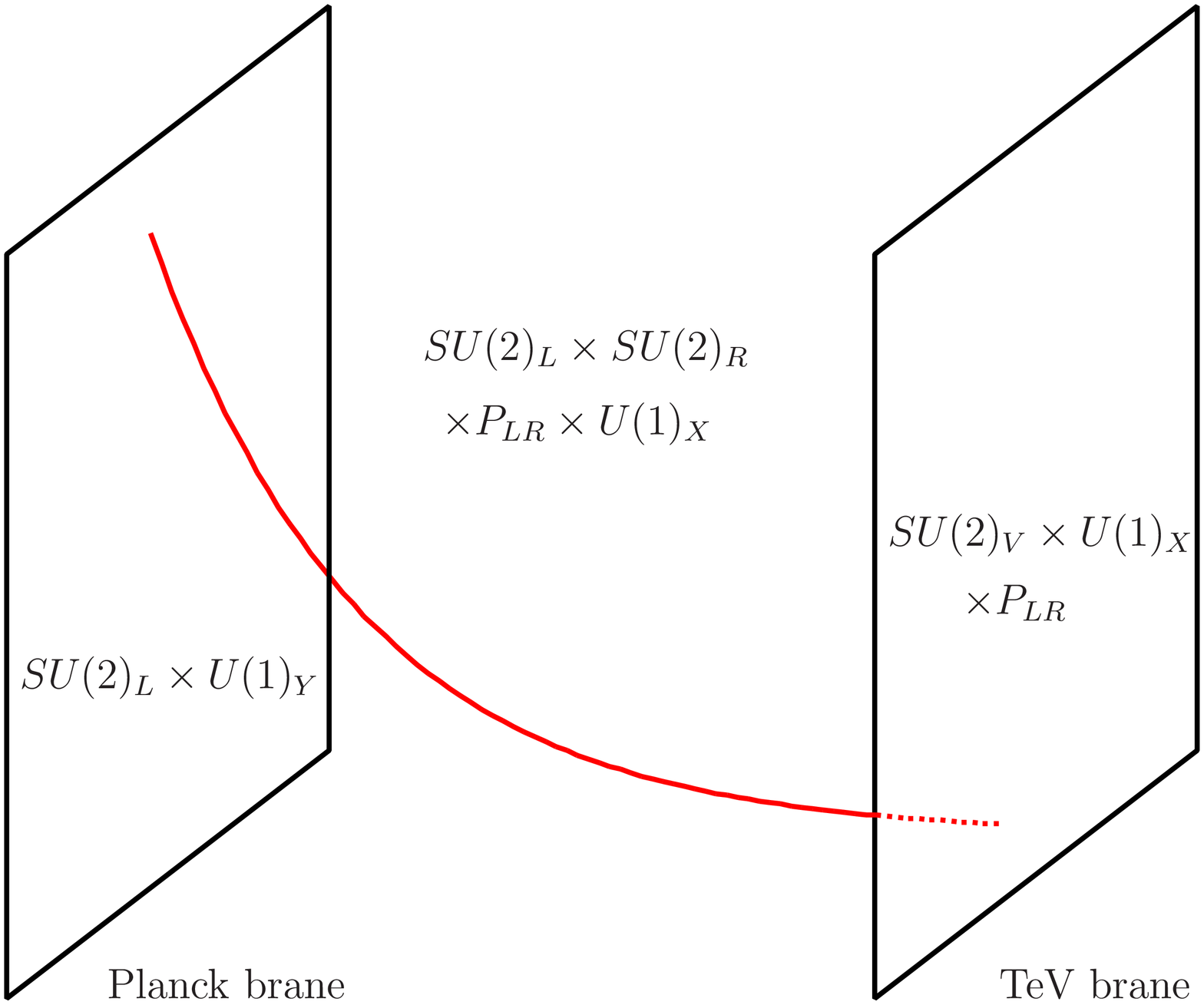}}
\caption{\label{fig:SBpattern}\it
EW symmetry breaking pattern of the RS model with custodial protection.}
\end{figure}

}

Now due to the unbroken gauge invariance of QED and QCD, the gluon and photon fields including their KK modes do not couple to the Higgs boson at leading order in perturbation theory and hence do not mix with each other or with $Z_\mu^{(0)}$, $Z_\mu^{(1)}$ and $Z_{X\mu}^{(1)}$ and the higher KK modes {of $Z$ and $Z_X$}. Therefore, even after EWSB
\bea
M_{A^{(0)}}= 0\,,\qquad && M_{A^{(1)}}= M_{++}\,,\\
M_{G^{(0)}}= 0\,,\qquad && M_{G^{(1)}}= M_{++}\,,
\eea
and the corresponding states remain mass eigenstates. On the other hand the kinetic term for the Higgs field (see Appendix \ref{app:action})
\be\label{eq:Higgskin}
S_\text{Higgs}= \int d^4x \int_0^L dy\,\sqrt{ G} \Tr \left[(D_M H(x^\mu,y))^\dagger (D^M H(x^\mu,y))\right]
\ee
leads to $v^2$-corrections to the masses of $W_{L\mu}^{(0)\pm}$, $W_{L\mu}^{(1)\pm}$ and $W_{R\mu}^{(1)\pm}$ as well as of $Z_\mu^{(0)}$, $Z_\mu^{(1)}$ and $Z_{X\mu}^{(1)}$, and mixing between states of the same electric charge is induced. 
Here
\be
H(x^\mu,y)=\frac{1}{\sqrt{L}}H(x^\mu)h(y) + \text{heavy KK modes}\,,
\ee
where $h(y)$ is the Higgs shape function along the extra dimension. We assume $h(y)$ to be of the form
\be\label{eq:h(y)}
h(y)= \sqrt{2(\beta-1) kL}\,e^{kL}\, e^{\beta k (y-L)}\qquad (\beta\gg 1)
\ee
where in the limit $\beta\to\infty$ the case of an IR brane localised Higgs is {recovered.} {The case of a bulk Higgs has first been considered in \cite{Davoudiasl:2005uu,Cacciapaglia:2006mz}. Furthermore 
\be
\langle H(x^\mu) \rangle = \begin{pmatrix} 0  && -v/{2} \\
v/{2} && 0 \end{pmatrix}\,,
\ee 
and $v=246\gev$ denotes the {effective 4D VEV of the  zero mode of $h^0$ in \eqref{eq:H}}.

Restricting the discussion to $n=0,1$ for simplicity, the gauge boson interactions with the Higgs  resulting from \eqref{eq:Higgskin} lead to two mass matrices $\mathcal{M}_\text{charged}^2$ and $\mathcal{M}_\text{neutral}^2$ \cite{Agashe:2007ki}
\begin{gather}
\begin{pmatrix} W_L^{(0)+} &&  W_L^{(1)+} &&W_R^{(1)+} \end{pmatrix}\mathcal{M}_\text{charged}^2 \begin{pmatrix} W_L^{(0)-} \\  W_L^{(1)-} \\ W_R^{(1)-} \end{pmatrix}\,,\label{eq:Mcharged}\\
\frac{1}{2}\begin{pmatrix} Z^{(0)} &&  Z^{(1)} && Z_X^{(1)} \end{pmatrix}\mathcal{M}_\text{neutral}^2 \begin{pmatrix} Z^{(0)} \\  Z^{(1)} \\ Z_X^{(1)} \end{pmatrix}\,,\label{eq:Mneutral}
\end{gather}
with $\mathcal{M}_\text{charged}^2$ and $\mathcal{M}_\text{neutral}^2$ given explicitly in Appendix \ref{app:EWSB}.

In order to determine the physical mass eigenstates and the corresponding masses, $\mathcal{M}_\text{charged}^2$ and $\mathcal{M}_\text{neutral}^2$ have to be diagonalised {by means of orthogonal transformations}:
\bea
\mathcal{G}_W\, \mathcal{M}_\text{charged}^2 \,\mathcal{G}_W^T  &=& 
\text{diag}(M_{W}^2, M_{W_H}^2, M_{W'}^2)\,,\\
\mathcal{G}_Z\, \mathcal{M}_\text{neutral}^2 \,\mathcal{G}_Z^T &=&
\text{diag}(M_Z^2, M_{Z_H}^2,M_{Z'}^2)\,.
\eea
The mass eigenstates $(W^\pm,W_H^\pm,W^{\prime\pm})$ and $(Z,Z_H,Z')$ are then related to the gauge eigenstates of the KK modes via
\be\label{eq:GWGZ}
\begin{pmatrix} W^\pm \\W_H^\pm\\W^{\prime\pm}\end{pmatrix}
 = \mathcal{G}_W  \begin{pmatrix} W_L^{(0)\pm} \\W_L^{(1)\pm}\\W_R^{(1)\pm}\end{pmatrix}\,,\qquad
\begin{pmatrix} Z \\Z_H\\Z'\end{pmatrix}= \mathcal{G}_Z \begin{pmatrix} Z^{(0)} \\Z^{(1)}\\Z_X^{(1)}\end{pmatrix} 
\,.
\ee
The explicit form of {the orthogonal matrices} $\mathcal{G}_W$ and $\mathcal{G}_Z$ 
can be found in Appendix \ref{app:EWSB}.

\newsection{Fermion Sector -- Quarks}\label{sec:quarks}

\subsection{Preliminaries}

In order to preserve the $O(4)\sim SU(2)_L\times SU(2)_R\times P_{LR}$
symmetry, that is  necessary for the suppression of dangerous contributions to EW
precision observables
\cite{Agashe:2003zs,Csaki:2003zu,Agashe:2006at,Contino:2006qr,Carena:2006bn,Carena:2007ua},
we will choose a particular simple set of representations of the $O(4)$
group. Although in order to satisfy EW precision measurements only the third
quark generation needs to preserve the $P_{LR}$ symmetry, the incorporation of
CKM mixing requires the same choice of $O(4)$ representations also for the
first two quark generations. 
{This is crucial {for having} a custodial protection for the 
flavour violating couplings $Z d_L^i\bar d_L^j$ \cite{Blanke:2008yr} as well.}

{In this section we restrict our attention to the quark sector of the model.}
The lepton sector will be discussed separately in Section \ref{sec:leptons}.

The particular fermion assignment given below has been motivated by the analyses of \cite{Cacciapaglia:2006gp,Carena:2006bn,Contino:2006qr,Carena:2007ua,Medina:2007hz}. In particular the representations given below can easily be embedded into complete $SO(5)$ multiplets used in  \cite{Contino:2006qr,Carena:2007ua,Medina:2007hz} in the context of models with gauge-Higgs unification.

We introduce  three $O(4)$ multiplets per generation $(i=1,2,3)$:
\bea\label{eq:xi1L}
\xi^i_{1L}&=&\begin{pmatrix}\chi^{u_i}_{L}(-+)_{5/3} && q_L^{u_i}(++)_{2/3} \\
\chi^{d_i}_{L}(-+)_{2/3} && q_L^{d_i}(++)_{-1/3}\end{pmatrix}_{2/3}\,,\\
\xi^i_{2R} &=& u^i_R (++)_{2/3}\,,\label{eq:xi2R}\\
\xi^i_{3R} &=& T^i_{3R} \oplus T^i_{4R} = \begin{pmatrix}
\psi^{\prime i}_R(-+)_{5/3} \\ U^{\prime i}_R (-+)_{2/3} \\ D^{\prime i}_R (-+)_{-1/3} \end{pmatrix}_{2/3} \oplus 
\begin{pmatrix}
\psi^{\prime\prime i}_R(-+)_{5/3} \\ U^{\prime\prime i}_R (-+)_{2/3} \\ D^{ i}_R (++)_{-1/3} \end{pmatrix}_{2/3}\,.\label{eq:xi3R}
\eea
The corresponding states of opposite chirality are  given by
\bea
\xi^i_{1R}&=&\begin{pmatrix}\chi^{u_i}_{R}(+-)_{5/3} && q_R^{u_i}(--)_{2/3} \\
\chi^{d_i}_{R}(+-)_{2/3} && q_R^{d_i}(--)_{-1/3}\end{pmatrix}_{2/3}\,,\\
\xi^i_{2L} &=& u^i_L (--)_{2/3}\,,\\
\xi^i_{3L} &=& T^i_{3L} \oplus T^i_{4L} = \begin{pmatrix}
\psi^{\prime i}_L(+-)_{5/3} \\ U^{\prime i}_L (+-)_{2/3} \\ D^{\prime i}_L (+-)_{-1/3} \end{pmatrix}_{2/3} \oplus 
\begin{pmatrix}
\psi^{\prime\prime i}_L(+-)_{5/3} \\ U^{\prime\prime i}_L (+-)_{2/3} \\ D^{ i}_L (--)_{-1/3} \end{pmatrix}_{2/3}\,.\label{eq:xi3L}
\eea

The following comments are in order:
\bi
\item
All fields in \eqref{eq:xi1L}--\eqref{eq:xi3L} are triplets under $SU(3)_c$, i.\,e. they carry QCD colour.
\item
$\xi^i_{1L}$ and $\xi^i_{1R}$ are bidoublets of $SU(2)_L\times SU(2)_R$, with $SU(2)_L$ acting vertically and $SU(2)_R$ horizontally.
\item
$u^i_L$ and $u^i_R$ are singlets of $O(4)$.
\item 
{$T^i_{3L,R}\oplus T^i_{4L,R}$ transform as $\Ltriplet\oplus\Rtriplet$} under $SU(2)_L\times SU(2)_R$. The embedding of the right-handed down-type quarks into triplet representations $\Rtriplet$ is necessary in order to allow for a $U(1)_X$ invariant Yukawa coupling.
\item 
The charges {$Q_X$} assigned to the various multiplets are $U(1)_X$ charges.
\item
{The charges $Q$ assigned to separate fields are electric charges, given as
\be
Q= T^3_L + T^3_R +Q_X\,,
\ee
where $T^3_L$ and $T^3_R$ denote the third component of the $SU(2)_L$ and $SU(2)_R$ isospins, respectively.}
\item
Only the fields obeying $(++)$ BCs have massless zero modes. {Up to small
  mixing effects with other massive modes due to the transformation to mass
  eigenstates discussed in Section \ref{sec:flavour}, these zero modes can be
  identified 
 with the usual SM quarks.}
\item
The remaining fields are KK modes {with approximately vectorlike couplings}. We thus have in this model additional heavy fermionic states. These are
\bea
Q=5/3&:&\qquad \chi^{u_i(n)},\psi^{\prime i(n)},\psi^{\prime\prime i(n)}\,, \\
Q=2/3&:&\qquad  q^{u_i(n)},  u^{i(n)}, U^{\prime i(n)}, U^{\prime\prime i(n)}, \chi^{d_i(n)}\,,\\
Q=-1/3&:&\qquad q^{d_i(n)}, D^{i(n)},D^{\prime i(n)}\,,
\eea
where $n=1,2,\dots$.
\item Left- and right-handed fermion fields are defined via $\psi_{L,R}=\mp\gamma^5\psi_{L,R}$.
\ei

\subsection{KK Decomposition and Bulk Profiles}

\subsubsection{Zero Modes}

The profiles of left-handed fermionic zero modes with respect to the flat metric are given by \cite{Grossman:1999ra,Gherghetta:2000qt}
\be\label{eq:fermionprofile}
\hat f^{(0)}_{L}(y,c)=\sqrt{\frac{(1-2c)kL}{e^{(1-2c)kL}-1}}\,e^{(\frac{1}{2}- c)ky}\,,
\ee
where $ck$ is the bulk mass of the 5D fermion field. In the case of right-handed zero modes, $c$ has to be replaced by $-c$ in the above formula and in the discussion following below. We note that
\bi
\item
For $c>1/2$ the normalisation factor in \eqref{eq:fermionprofile} is $\ord(1)$ and $f_L^{(0)}(y,c)$ is peaked around $y=0$, i.\,e. fermions with bulk mass parameter $c>1/2$ are placed close to the UV brane.
\item
For $c<1/2$, as is the case of the top quark, the second term in the denominator of \eqref{eq:fermionprofile} can be neglected and we obtain
\be
\hat f_L^{(0)}(y,c)\simeq\sqrt{{(1-2c)kL}}\,e^{(\frac{1}{2}-c)k(y-L)}\,.
\ee
Thus the shape function is strongly peaked towards $y=L$, i.\,e. the IR brane.
\item
One should stress that generally the $c_{L}$ and $c_{R}$ of left- and right-handed SM fermions can differ from each other, as these fermions are zero modes of different 5D representations. As both $c_{L}$ and $c_{R}$ enter the formula for the Yukawa couplings and fermion masses, this freedom can help to satisfy certain features in EW precision studies \cite{Cacciapaglia:2006gp,Carena:2006bn,Contino:2006qr,Carena:2007ua}  and in flavour physics { \cite{Agashe:2004cp,Casagrande:2008hr}} while keeping the fermion masses of their natural size. This is in particular relevant for the third {quark generation}.
\ei

\subsubsection{KK Modes}

The shape functions for fermionic KK modes with respect to the warped metric
 are given by \cite{Gherghetta:2000qt} {(see also Appendix  \ref{app:eom} for details)}
\be\label{eq:fermionKK} 
f_{L,R}^{(n)}(y,c,\text{BC})=\frac{e^{ky/2}}{N_n}\left[J_\alpha\left(\frac{m_n}{k}e^{ky}\right)+b_\alpha(m_n) Y_\alpha\left(\frac{m_n}{k}e^{ky}\right)\right]\,,
\ee
where $\alpha=|c\pm1/2|$ for left-(right-)handed modes, and expressions for $N_n$ and $b_\alpha(m_n)$ can be found in Appendix \ref{app:eom}. The KK masses are approximately given by 
\be\label{eq:mnfermion}
m_n^\text{fermion} \simeq \left(n+\frac{1}{2}\left(\left|c+\frac{1}{2}\right|-1\right)\mp\frac{1}{4}\right)\pi f \,,
\ee
where the $\mp$ sign corresponds to a $\pm$ BC for the left-handed fermion on the IR brane. Again, as in the case of gauge KK modes, the accuracy of \eqref{eq:mnfermion} improves with increasing $n$.

The following comments are in order \cite{Gherghetta:2000qt,Gherghetta:2006ha}:
\bi
\item
The bulk mass parameter $c$ is universal for the full tower of KK modes and equal to the $c$ describing the localisation of the zero mode if such mode exists.
\item
In spite of the same $c$ even for $c>1/2$ the form of \eqref{eq:fermionKK} implies that {\it all} KK modes are localised near the IR brane. There is no freedom to delocalise the massive KK modes away from the IR brane, as was the case for the zero mode.
\ei

\subsection{Yukawa Couplings and Fermion Masses}

The SM fermions acquire masses via their Yukawa interactions with the Higgs
in the process of  EWSB. The effective 4D Yukawa matrices $Y_{ij}$ are then given by
\be\label{eq:Yij}
Y_{ij} \propto \int_0^L \frac{dy}{L^{3/2}}\lambda_{ij} h(y) f_L^{(0)}(y,c^i) f_R^{(0)}(y,c^j )\,,
\ee
where $\lambda_{ij}$ is the fundamental 5D Yukawa coupling and $h(y)$ is the Higgs shape function along the extra dimension, as given in \eqref{eq:h(y)}. 

We stress that together with $h(y)$ from \eqref{eq:h(y)} the fermionic zero mode functions with respect to the warped metric, as given in \eqref{eq:ferm-zero-mode}, have to be used in order to determine the effective 4D Yukawa couplings $Y_{ij}$ in \eqref{eq:Yij}.

In the special case of an IR brane localised Higgs we obtain 
\be
Y_{ij} \propto \frac{\lambda_{ij}}{N_{i,L}N_{j,R}} \,e^{(1-c^i_{L}+c^j_{R})kL}\,
\ee
{where $N_{i,L},N_{j,R}$ are normalisation factors of the fermion shape functions on the IR brane}.
We note that in this case, up to $\lambda_{ij}$, $Y_{ij}$ has a factorised form, as emphasised in  \cite{Huber:2003tu}. 
In the more general case of the Higgs field propagating in the bulk this factorisation is broken, but only weakly for a large range of $c^i_L,c^j_R$  \cite{Moreau:2005kz}.

On the other hand we stress that in the presence of non-diagonal entries of $\lambda_{ij}$, the factorisation must always be broken as not all entries of $\lambda_{ij}$ can be equal. Indeed this special choice of 5D Yukawa couplings would lead to two zero eigenvalues of the corresponding mass matrix. In addition some non-degeneracy between the entries of $\lambda_{ij}$ is needed to cure the ``$|V_{ub}|$ problem'' identified in \cite{Huber:2003tu}. Indeed this is completely analogous to the case of the Froggatt-Nielsen flavour symmetry \cite{Froggatt:1978nt}, where also a slight structure in the Yukawa coupling matrices is needed to obtain the correct size of $|V_{ub}|$. {A detailed discussion of the close analogy between the Froggatt-Nielsen mechanism and bulk fermions in RS models has been presented in \cite{Blanke:2008zb}.}

\newsection{Flavour Structure}\label{sec:flavour}

\subsection{Preliminaries}

The flavour structure of this class of models is rather complicated, although as emphasised by Agashe {\it et~al.} \cite{Agashe:2004cp} in certain approximations it is quite simple. For the time being we will however not make any approximations.

The procedure to find all interactions including flavour violating ones is basically the following:

\subsubsection*{Step 1}

We begin with the interaction terms {in $\mathcal{L}_\text{fermion}$ and $\mathcal{L}_\text{Yuk}$ in \eqref{eq:S}} in terms of flavour eigenstates for fermions and in the gauge eigenbasis for the gauge bosons. 

To this end we start with the fundamental 5D interactions and then perform the KK decomposition as described in Appendix~\ref{app:eom}. We thus obtain effective 4D couplings that are non-local quantities along the extra dimension, resulting from the overlap of the gauge boson and fermion shape functions. Schematically the interactions between different KK levels $(k,m,n=0,1,\dots)$ are given by
\be\label{eq:gijn}
g_{kmn}= \frac{g}{L^{3/2}} \int_0^L dy\, e^{ky} {  f^{(k)}(y,c) f^{(m)}(y,c)}f^{(n)}_\text{gauge}(y)\,.
\ee
{Note that only fermions within the same gauge multiplet are coupled to each other in this way, so that their bulk mass parameters are necessarily equal.}
{In the Feynman rules collected in Appendix \ref{app:FR} these overlap integrals appear as $\mathcal{R}$, $\mathcal{P}$ and $\mathcal{S}$.}

As the gauge boson zero mode has a flat shape function, the coupling {for equal fermion KK levels $k=m$ to the gauge boson zero mode $n=0$} reduces to the 4D gauge coupling $g/\sqrt{L}$, {while the integrals with $k\not=m$ vanish
 due to the normalisation} of the fermion shape function. {In the same way, couplings of different fermionic KK levels to the gauge boson zero mode $(n=0)$ vanish due to the orthogonality of the fermion shape functions.} However, due to the effects of EWSB, the {EW} gauge boson zero mode mixes with its KK modes, so that eventually flavour non-universalities in the couplings of the SM weak gauge bosons {and non-zero couplings between various KK levels} will arise. 

A different treatment of the effects of EWSB on the gauge boson zero modes has
been discussed in \cite{Huber:2000fh,Csaki:2002gy,Burdman:2002gr}. In their
description the Higgs VEV is inserted already at the level of the 5D equations
of motion, which leads to a distortion of the gauge boson zero mode in the
vicinity of the IR brane. On the other hand no mixing between the various KK
levels appears. In Appendix \ref{app:EWSB} we review both approaches in
detail, study their advantages and shortcomings and show that both
interpretations are indeed physically equivalent. {A similar 
independent discussion has been presented in \cite{Goertz:2008vr}.}

\subsubsection*{Step 2}

Next we have to consider the mass matrices of gauge bosons and fermions. Here the new aspects relative to the SM are
\bi 
\item in the gauge sector:
\bi
\item The extended gauge group $SU(2)_L\times SU(2)_R\times U(1)_X$ leads to the presence of additional (heavy) {gauge bosons}.
\item In addition, also the heavy KK modes of the SM gauge bosons, including also gluons and the photon, are present.
\item EWSB induces mixing of the SM zero modes with the additional heavy KK modes of the same electric charge.
\ei
\item in the fermion sector:
\bi
\item
The enlarged fermionic representations imply the presence of new {heavy  fermions} that could be much lighter than the gauge KK modes \cite{Contino:2006qr,Carena:2006bn}.
\item Also the heavy KK modes of the SM fermions have to be considered.
\item
Again mixing of the SM zero modes with the heavy KK states is induced.
\ei
\ei

\subsubsection*{Step 3}

Finally all interactions have to be rewritten in terms of mass eigenstates for gauge bosons and fermions. Therefore the corresponding mass matrices need to be diagonalised. Note that mixing takes place not only between different flavours, but also between different KK levels\footnote{This should be contrasted with models with flat extra dimensions where the presence of KK parity has eliminated such mixing. For a recent attempt to introduce KK parity in the RS framework, see \cite{Agashe:2007jb}.}. As the new physics scale $f$ is experimentally constrained to be $f\simge 1\tev$, within a good approximation it is sufficient to consider only the contributions of $n=0,1$ modes and neglect all higher KK levels. Therefore in what follows we restrict our discussion to this simplified case. The generalisation of our formulae to include also higher KK modes is then straightforward.

\subsection{Quark Mass Matrices}\label{sec:qmm}

The transformation to mass eigenstates in the gauge sector {is} performed in Section \ref{sec:gauge} and Appendix \ref{app:EWSB}. The goal of the present section is  to construct and diagonalise the mass matrices for the quark fields given in \eqref{eq:xi1L}--\eqref{eq:xi3L}. To this end we will only consider zero modes and the lowest ($n=1$) KK modes. As there are only few fields among the ones in \eqref{eq:xi1L}--\eqref{eq:xi3L} that have zero modes, we will assign to  them {the} superscript ${(0)}$. For the excited KK modes we will just use the notation of \eqref{eq:xi1L}--\eqref{eq:xi3L}, making the $n=1$ index implicit.

We will have to deal with three mass matrices corresponding to the electric charges $+5/3$, $+2/3$ and $-1/3$. To this end we group the fermion modes into the following vectors:

For the $+5/3$ charge mass matrix we have
\bea\label{eq:psiL53}
\Psi_L(5/3) &=& \left( \chi^{u_i}_L(-+), \psi^{\prime i}_L(+-), \psi^{\prime\prime i}_L(+-)\right)^T\,,\\
\Psi_R(5/3) &=& \left( \chi^{u_i}_R(+-), \psi^{\prime i}_R(-+), \psi^{\prime\prime i}_R(-+)\right)^T\,,
\eea
where the flavour index $i=1,2,3$ runs over the three quark generations. Thus we deal with 9-dimensional vectors. Note that in this sector only massive excited KK states are present.

For the charge $+2/3$ mass matrix the corresponding vectors read
\bea\label{eq:psiL23}
\Psi_L(2/3) &=& \left( q_L^{u_i(0)}(++), q_L^{u_i}(++),U_L^{\prime i}(+-),U_L^{\prime\prime i}(+-), \chi^{d_i}_L(-+), u_L^i(--)  \right)^T\,,\qquad\\
\Psi_R(2/3) &=& \left( u_R^{i(0)}(++), q_R^{u_i}(--), 
U_R^{\prime i}(-+),U_R^{\prime\prime i}(-+),
\chi^{d_i}_R(+-), u_R^i(++)  \right)^T\,.
\eea
Here the first components are zero modes, and $i=1,2,3$ so that we really deal with 18-dimensional vectors. 

The $-1/3$ charge vectors read
\bea\label{eq:psiL-13}
\Psi_L(-1/3) &=& \left( q_L^{d_i(0)}(++), q_L^{d_i}(++), D_L^{\prime i}(+-),D_L^{ i}(--) \right)^T\,,\label{eq:psiL13}\\
\Psi_R(-1/3) &=& \left( D_R^{i(0)}(++), q_R^{d_i}(--), D_R^{\prime i}(-+),D_R^{ i}(++) \right)^T\,.\label{eq:psiR13}
\eea
Again the first entries are zero modes, the remaining ones massive KK modes, and $i=1,2,3$, so that in this case a 12-dimensional vector is obtained.

In order to construct the mass matrices let us briefly recall certain properties, known already from numerous studies in the literature:
\begin{enumerate}
\item
We have three bulk mass matrices $c_1,c_2,c_3$ corresponding to the $O(4)$ representations $\xi^i_1,\xi^i_2,\xi^i_3$ ($i=1,2,3$ {is the flavour index}), respectively. Note that for a given $O(4)$ multiplet with fixed flavour index all bulk mass parameters for different components of the multiplet are equal to each other. 
\item
In general, $c_k$ are arbitrary hermitian $3\times3$ matrices{, where $k=1,2,3$ corresponds to the $O(4)$ multiplet $\xi_k$.} In the following we choose to work in the basis where they are real and diagonal, i.\,e. each of them is described by three real parameters $c^i_k$, where $i$ is the flavour index. This can always be achieved by appropriate field redefinitions of the $\xi^i$ multiplets. Explicitly we then have:
\be\label{eq:c1}
c_1\equiv \text{diag}(c^1_1,c^2_1,c^3_1)\,,
\ee
and similarly for $c_2$ and $c_3$.
\item
The allowed Yukawa couplings{, giving mass to the fermion zero modes after EWSB,} have to preserve the full $O(4)\sim  SU(2)_L\times SU(2)_R\times P_{LR}$ gauge symmetry. The possible gauge invariant terms in the full 5D theory can be found in Appendix \ref{app:action}.
\item
The effective 4D Yukawa matrices will involve the fermion and Higgs shape functions. We will denote {the fermionic ones} by $f^{Q}_{L,k}(y)$ and  $f^{Q}_{R,l}(y)$, corresponding to the $k$-th and $l$-th component of $\Psi_L(Q)$ and $\Psi_R(Q)$ in \eqref{eq:psiL53}--\eqref{eq:psiR13}, respectively, and $h(y)$ is the Higgs shape function as given in \eqref{eq:h(y)}.

Having at hand this information and restricting ourselves to $n=0,1$ for simplicity, we obtain the following {effective} 4D Yukawa couplings
\bea
\left[Y_{ij}^{(5/3)}\right]_{kl} &=& 
\frac{1}{\sqrt{2}L^{3/2}}\int_0^L dy\, \lambda^d_{ij} f^{5/3}_{L,k}(y) f^{5/3}_{R,l}(y)h(y)\,,\\
\left[Y_{ij}^{(2/3)}\right]_{kl} &=& \frac{1}{2L^{3/2}}\int_0^L dy\, \lambda^d_{ij} f^{2/3}_{L,k}(y) f^{2/3}_{R,l}(y)h(y)\,,\\
{}\left[\tilde Y_{ij}^{(2/3)}\right]_{kl} &=& \frac{1}{\sqrt{2}L^{3/2}}\int_0^L dy\, \lambda^u_{ij} f^{2/3}_{L,k}(y) f^{2/3}_{R,l}(y)h(y)\,,\\
{}\left[ Y_{ij}^{(-1/3)}\right]_{kl} &=& \frac{1}{\sqrt{2}L^{3/2}} \int_0^L dy\, \lambda^d_{ij}
f^{-1/3}_{L,k}(y) f^{-1/3}_{R,l}(y)h(y)\,.\qquad
\eea
{Interestingly, the {Yukawa coupling} proportional to $\lambda^d_{ij}$, connecting $\xi^i_1$ with $\xi^j_3$ and being thus responsible for the SM down quark Yukawa coupling, leads to mass terms not only for the charge $-1/3$ quarks, but simultaneously also to mass terms for the $+5/3$ and $+2/3$ quarks. This is a direct consequence of $T^j_3$ and $T^j_4$ being placed in the adjoint representations of $SU(2)_L$ and $SU(2)_R$, respectively, as seen in \eqref{eq:xi3R}, \eqref{eq:xi3L}.

On the other hand,   the term proportional to $\lambda^u_{ij}$, connecting $\xi^i_1$ with $\xi^j_2$ and being thus responsible for the SM up quark Yukawa coupling, contributes only to the mass matrix for the charge $+2/3$ quarks.}

\item
Finally the  fermionic KK masses, which can be obtained from solving the bulk equations of motion, have to be included {in the mass matrices}. Note that both the fermion shape function and the KK mass depend on the bulk mass parameter $c$ {\it and} on the BCs.

In what follows we will use the $3\times3$ KK fermion mass matrices $\MKf_k(\text{BC-L})$, where $k=1,2,3$ labels the representations in \eqref{eq:xi1L}--\eqref{eq:xi3L}, and (BC-L) are the BCs for the left-handed mode.
\end{enumerate}

In terms of the mode vectors \eqref{eq:psiL53}--\eqref{eq:psiR13} we can write
\bea
\mathcal{L}_\text{mass} &=& -\bar\Psi_L(5/3) \,\mathcal{M}(5/3)\,\Psi_R(5/3) +h.c.\nn\\
&&-\bar\Psi_L(2/3) \,\mathcal{M}(2/3)\, \Psi_R(2/3) +h.c.\nn\\
&&-\bar\Psi_L(-1/3) \,\mathcal{M}(-1/3)\, \Psi_R(-1/3) +h.c.\,.
\eea
{In order to distinguish zero modes from the KK fermions we will label
the zero mode components of the vectors \eqref{eq:psiL53}--\eqref{eq:psiR13}
by the index $0$.}
Then the quark mass matrices read
\addtolength{\arraycolsep}{2pt}
\renewcommand{\arraystretch}{1.5}
\be\label{eq:M53}
\mathcal{M}(5/3) = \left(\begin{array}{ccc} \MKf_1(-+) & v\left[Y_{ij}^{(5/3)}\right]_{12} & -v\left[Y_{ij}^{(5/3)}\right]_{13} \\
v\left[Y_{ij}^{(5/3)}\right]_{21}^\dagger & \MKf_3(+-) & 0 \\
-v\left[Y_{ij}^{(5/3)}\right]_{31}^\dagger &0& \MKf_3(+-)\end{array}\right)\,,
\ee
\bea\label{eq:M23}
&&\mathcal{M}(2/3) =\\
\nn&& \hspace{-.65cm}\left(\begin{array}{cccccc} 
v\left[\tilde Y_{ij}^{(2/3)}\right]_{00} & 0 & -v\left[Y_{ij}^{(2/3)}\right]_{02} & v\left[Y_{ij}^{(2/3)}\right]_{03} &0& v\left[\tilde Y_{ij}^{(2/3)}\right]_{05} \\
v\left[\tilde Y_{ij}^{(2/3)}\right]_{10} & \MKf_1(++) & -v\left[Y_{ij}^{(2/3)}\right]_{12} & v\left[Y_{ij}^{(2/3)}\right]_{13} & 0 & v\left[\tilde Y_{ij}^{(2/3)}\right]_{15} \\
0& -v\left[Y_{ij}^{(2/3)}\right]_{21}^\dagger &\MKf_3(+-)& 0 & -v\left[Y_{ij}^{(2/3)}\right]_{24}^\dagger & 0 \\
0 & v\left[Y_{ij}^{(2/3)}\right]_{31}^\dagger & 0 & \MKf_3(+-) &  v\left[Y_{ij}^{(2/3)}\right]_{34}^\dagger & 0 \\
-v\left[\tilde Y_{ij}^{(2/3)}\right]_{40} & 0 & -v\left[Y_{ij}^{(2/3)}\right]_{42} & v\left[Y_{ij}^{(2/3)}\right]_{43} & \MKf_1(-+) & -v\left[\tilde Y_{ij}^{(2/3)}\right]_{45} \\
0 & v\left[\tilde Y_{ij}^{(2/3)}\right]_{51}^\dagger & 0 & 0 & -v\left[\tilde Y_{ij}^{(2/3)}\right]_{54}^\dagger & \MKf_2(--)
 \end{array}\right)
\eea
\be\label{eq:M13}
\mathcal{M}(-1/3) = \left(\begin{array}{cccc}
v\left[ Y_{ij}^{(-1/3)}\right]_{00} & 0 & -v\left[ Y_{ij}^{(-1/3)}\right]_{02} & v\left[ Y_{ij}^{(-1/3)}\right]_{03} \\
v\left[ Y_{ij}^{(-1/3)}\right]_{10} & \MKf_1(++) & -v\left[ Y_{ij}^{(-1/3)}\right]_{12} & v\left[ Y_{ij}^{(-1/3)}\right]_{13} \\
0 & -v\left[ Y_{ij}^{(-1/3)}\right]_{21}^\dagger & \MKf_3(+-) & 0 \\
0 & v\left[ Y_{ij}^{(-1/3)}\right]_{31}^\dagger & 0 & \MKf_3(--)
\end{array}\right)\,.
\ee
\addtolength{\arraycolsep}{-2pt}
\renewcommand{\arraystretch}{1.0}

These three matrices have to be diagonalised via a bi-unitary transformation
to find the quark mass eigenstates. Due to the large size of the mass
matrices this diagonalisation has to be done numerically. 

Let us make a few remarks:
\bi
\item
The mass eigenstates of $5/3$ charge are all heavy.
\item
In $\mathcal{M}(2/3)$ and $\mathcal{M}(-1/3)$ the off-diagonal entries in the
first column and row lead to mixing between light zero modes and heavy KK
modes. This mixing will be {suppressed by $\ord(v^2/{f^2})$}.
\item
{In the case of the Higgs field being confined exactly to the IR brane, only
Yukawa couplings to those fermion modes are non-vanishing that obey a $+$ BC on the IR brane. In that case some of the entries in the above mass matrices in  \eqref{eq:M53}--\eqref{eq:M13} vanish:
\begin{gather}
\mathcal{M}(5/3)_{21} =  \mathcal{M}(5/3)_{31} = 0\,,\\
\mathcal{M}(2/3)_{21} =\mathcal{M}(2/3)_{31} = \mathcal{M}(2/3)_{51} =\mathcal{M}(2/3)_{24} =\mathcal{M}(2/3)_{34} = \mathcal{M}(2/3)_{54} =0\,,\\
\mathcal{M}(-1/3)_{21} =  \mathcal{M}(-1/3)_{31} = 0\,.
\end{gather}
As pointed out in \cite{Blanke:2008zb} and discussed in detail in \cite{BDG}, this difference has profound implications on the size of flavour violating Higgs couplings. 
}
\ei

We can then diagonalise the $+5/3$, $+2/3$ and $-1/3$ charge matrices by 
\bea
M_\text{diag}(5/3) &=& \mathcal{X}_L^\dagger \,\mathcal{M}(5/3) \,\mathcal{X}_R\,,\\
M_\text{diag}(2/3) &=& \mathcal{U}_L^\dagger \,\mathcal{M}(2/3) \,\mathcal{U}_R\,,\\
M_\text{diag}(-1/3) &=& \mathcal{D}_L^\dagger \,\mathcal{M}(-1/3) \,\mathcal{D}_R\,.
\eea
The corresponding rotations of the $\Psi_{L,R}$ {vectors of fermion modes} are
\bea
\label{eq:XL}
\Psi_{L,R}(5/3)_\text{mass} &=& \mathcal{X}_{L,R}^\dagger \,\Psi_{L,R}(5/3)\,,\\
\label{eq:UL}
\Psi_{L,R}(2/3)_\text{mass} &=& \mathcal{U}_{L,R}^\dagger \,\Psi_{L,R}(2/3)\,,\\
\label{eq:DL}
\Psi_{L,R}(-1/3)_\text{mass} &=& \mathcal{D}_{L,R}^\dagger \,\Psi_{L,R}(-1/3)\,.
\eea
Note that $\mathcal{X}_{L,R}$, $\mathcal{U}_{L,R}$ and $\mathcal{D}_{L,R}$ are unitary $9\times9$, $18\times18$ and $12\times12$  matrices, respectively.

\subsection{Weak Currents}\label{sec:currents}

\subsubsection{Neutral Currents }

Here we have to consider the currents involving the three gauge bosons $Z^{(0)}$, $Z^{(1)}$ and $Z_X^{(1)}$ with the corresponding mass eigenstates $Z$, $Z_H$ and $Z'$ as defined in \eqref{eq:GWGZ}.

In order to simplify the presentation we first perform the rotation to mass eigenstates in the gauge boson sector, as described in \eqref{eq:GWGZ} and Appendix \ref{app:EWSB}. The currents involving the neutral gauge boson mass eigenstates $Z,Z_H,Z'$ and the quarks given still in the flavour eigenstate basis are then given as follows:
\bea
J_\mu(Z) &=& \bar\Psi_L(5/3)\,\gamma_\mu A^{5/3}_L(Z)\,\Psi_L(5/3) + \bar\Psi_R(5/3)\,\gamma_\mu A^{5/3}_R(Z)\,\Psi_R(5/3)\nn\\
&&{}+ \bar\Psi_L(2/3)\,\gamma_\mu A^{2/3}_L(Z)\,\Psi_L(2/3)  + \bar\Psi_R(2/3)\,\gamma_\mu A^{2/3}_R(Z)\,\Psi_R(2/3)\\
&&{}+ \bar\Psi_L(-1/3)\,\gamma_\mu A^{-1/3}_L(Z)\,\Psi_L(-1/3)
+  \bar\Psi_R(-1/3)\,\gamma_\mu A^{-1/3}_R(Z)\,\Psi_R(-1/3)\nn
\eea
and similarly for $J_\mu(Z_H)$, $J_\mu(Z')$. 

 The $A^{Q}_{L,R}$ $(Q=2/3,-1/3,5/3)$ matrices are flavour-diagonal  matrices 
and have dimensions $18\times18$, $12\times12$ and $9\times9$, respectively.
{With flavour-diagonal we mean that all $3\times 3$ sub-matrices are
 diagonal. On the other hand some non-vanishing elements mixing different 
fermions with the same flavour exist. The $3\times 3$ sub-matrices} are not proportional to the unit matrix due to the universality breakdown in the gauge couplings.
Indeed their entries have the generic structure as in \eqref{eq:gijn}. However not all entries will differ from each other as not all {fermionic} shape functions are different {from each other}. 
{Recall that we work in the basis where the bulk mass matrices $c_k$ are diagonal in flavour space. Explicit
 expressions for the $A^Q_{L,R}$ matrices can easily be obtained from the Feynman rules given in Appendix \ref{app:FR}, that involve as usual gauge boson 
mass eigenstates but still quark flavour eigenstates.}

As $c^i_k$ are flavour non-universal, non-universalities in the gauge couplings are generated already at this stage and will remain after the rotation to the mass eigenbasis.

Let us then write the currents in the mass eigenbasis for fermions:
\bea
J_\mu(Z) &=& \bar\Psi_L(5/3)_\text{mass}\,\gamma_\mu B^{5/3}_L(Z)\,\Psi_L(5/3)_\text{mass} \nn\\ &&{}
+\bar\Psi_R(5/3)_\text{mass}\,\gamma_\mu B^{5/3}_R(Z)\,\Psi_R(5/3)_\text{mass}\nn\\
&&{}+\bar\Psi_L(2/3)_\text{mass}\,\gamma_\mu B^{2/3}_L(Z)\,\Psi_L(2/3)_\text{mass} \nn\\ &&{}
+\bar\Psi_R(2/3)_\text{mass}\,\gamma_\mu B^{2/3}_R(Z)\,\Psi_R(2/3)_\text{mass}\nn\\
&&{}+ \bar\Psi_L(-1/3)_\text{mass}\,\gamma_\mu B^{-1/3}_L(Z)\,\Psi_L(-1/3)_\text{mass} \nn\\ &&{}
+\bar\Psi_R(-1/3)_\text{mass}\,\gamma_\mu B^{-1/3}_R(Z)\,\Psi_R(-1/3)_\text{mass}\,,\label{eq:JmuZ}
\eea
with analogous expressions for $J_\mu(Z_H)$ and $J_\mu(Z')$. Then
\bea
B^{5/3}_{L,R}(Z)&=&\mathcal{X}_{L,R}^\dagger\,A^{5/3}_{L,R}(Z)\,\mathcal{X}_{L,R}\,,\\
\label{eq:B23}
B^{2/3}_{L,R}(Z)&=&\mathcal{U}_{L,R}^\dagger\,A^{2/3}_{L,R}(Z)\,\mathcal{U}_{L,R}\,,\\\label{eq:B-13}
B^{-1/3}_{L,R}(Z)&=&\mathcal{D}_{L,R}^\dagger\,A^{-1/3}_{L,R}(Z)\,\mathcal{D}_{L,R}\,,
\eea
and similarly for $Z_H$ and $Z'$. Note that now all $B$ matrices are non-diagonal also in flavour space so that the neutral gauge bosons in question mediate FCNC transitions both between different KK levels and between different flavours.
Formula \eqref{eq:JmuZ} and similar expressions for $J_\mu(Z_H)$ and $J_\mu(Z')$ summarise tree level weak FCNC transitions in the model under consideration. Tree level FCNC transitions mediated by KK photons and gluons are discussed in Section \ref{se:KKphot-glu}.

\subsubsection{Charged Currents}

Similarly to the case of neutral currents we first rotate from the gauge eigenstates $W_L^{(0)},W_L^{(1)},W_R^{(1)}$ to the mass eigenstates $W,W_H,W'$ by means of \eqref{eq:GWGZ} and the explicit expressions in Appendix \ref{app:EWSB}. Then
\bea
J_\mu(W^+) &=& \bar\Psi_L(2/3)\,\gamma_\mu G_L(W^+)\,\Psi_L(-1/3)
+ \bar\Psi_R(2/3)\,\gamma_\mu G_R(W^+)\,\Psi_R(-1/3)\nn\\
&& {}+\bar\Psi_L(5/3)\,\gamma_\mu \tilde G_L(W^+)\,\Psi_L(2/3)
+ \bar\Psi_R(5/3)\,\gamma_\mu \tilde G_R(W^+)\,\Psi_R(2/3)+h.c.\,,\quad\qquad
\eea
and similarly for $G_{L,R}(W_H^+),G_{L,R}(W^{\prime+}), \tilde
G_{L,R}(W_H^+),\tilde G_{L,R}(W^{\prime+})$. Evidently the matrices
$G_{L,R},\tilde G_{L,R}$ are not square matrices because the number of $Q=
+5/3$, $+2/3$ and $-1/3$ quarks differ. In the model under consideration they
are $18\times12$ and $9\times 18$ matrices, respectively. In addition the
$3\times3$ diagonal sub-matrices are not proportional to the unit matrix due
to the non-universality of gauge couplings.
{Explicit
 expressions for the $G_{L,R}$ and $\tilde G_{L,R}$ 
 matrices can be obtained from the Feynman rules given  in Appendix \ref{app:FR}, that involve as usual gauge boson 
mass eigenstates but still quark flavour eigenstates.}

Moreover due to the {mixing of the SM quarks with the additional heavy} $+2/3$ and $-1/3$ fields, effects of non-unitarity will appear in the {$3\times 3$} CKM matrix.  Note that the way we define the fields $\Psi_L(2/3)$ and $\Psi_L(-1/3)$ in \eqref{eq:psiL23} and \eqref{eq:psiL-13}, the standard CKM matrix will eventually be the $3\times3$ sub-matrix placed in the upper left corner of the final mixing matrix.

Our next step then is to go to the mass eigenbasis for the fermions. In this basis the $G_{L,R}(W,W_H,W'), \tilde G_{L,R}(W,W_H,W')$  are replaced by 
\bea
H_{L,R}(W^+) &=& \mathcal{U}_{L,R}^\dagger\,G_{L,R}(W^+)\,\mathcal{D}_{L,R}\,,\\
\tilde H_{L,R}(W^+) &=& \mathcal{X}_{L,R}^\dagger\,\tilde G_{L,R}(W^+)\,\mathcal{U}_{L,R}\,,
\eea
and similarly for $W_H, W'$.

Therefore the final expression for the charged currents in the mass eigenbasis is
\bea
J_\mu(W^\pm) &=&  \bar\Psi_L(2/3)_\text{mass}\,\gamma_\mu H_L(W^+)\,\Psi_L(-1/3)_\text{mass} \nn\\
&&{} + \bar\Psi_R(2/3)_\text{mass}\,\gamma_\mu H_R(W^+)\,\Psi_R(-1/3)_\text{mass}\nn\\
&&{}+ 
\bar\Psi_L(5/3)_\text{mass}\,\gamma_\mu \tilde H_L(W^+)\,\Psi_L(2/3)_\text{mass} \nn\\
&&{} + \bar\Psi_R(5/3)_\text{mass}\,\gamma_\mu \tilde H_R(W^+)\,\Psi_R(2/3)_\text{mass}
+h.c.\,,\qquad\qquad\label{eq:JmuW}
\eea
and similarly for $J_\mu(W_H^\pm)$ and $J_\mu(W^{\prime\pm})$. The CKM matrix is then given by 
{\be
V_\text{CKM}= \left(\frac{g}{\sqrt{2L}}\right)^{-1} H_L(W^\pm)_{11}\,,
\ee
where $H_L(W^\pm)_{11}$ denotes the upper left $3\times 3$ sub-matrix of $H_L(W^\pm)$.}

\subsection{Photonic and Gluonic Currents }\label{se:KKphot-glu}

The photonic and gluonic currents mediating FCNCs are similar to the neutral EW currents discussed above, but because of the absence of spontaneous symmetry breaking in that case, the various KK modes do not mix with each other. Consequently only the massive KK modes contribute to FCNC processes.

The massive photonic current reads
\be
J_\mu(A^{(1)})=  \bar\Psi_{L,R}(Q)\,\gamma_\mu A_{L,R}^Q(A^{(1)}) \,\Psi_{L,R}(Q)\,,
\ee
where $Q$ denotes the electric charge. The massive gluonic current, on the other hand, reads
\be
J^{A}_\mu(G^{(1)})=  \bar\Psi_{L,R}(Q)\,\gamma_\mu t^{A}A^Q_{L,R}(G^{(1)})\,\Psi_{L,R}(Q)\,,
\ee
where $A=1,\dots,8$ and  we have made the QCD colour indices of the quark fields implicit. Interestingly, in spite of the universality of the gauge boson shape functions, the matrices $A^Q_{L,R}(Z^{(1)})$ and $A^Q_{L,R}(Z_X^{(1)})$ are not proportional to $A^Q_{L,R}(A^{(1)})$, $A^Q_{L,R}(G^{(1)})$ due to the different fermionic representations of $SU(2)_L\times SU(2)_R$. However,
\be
\frac{A^Q_{L,R}(G^{(1)})}{g_s} = \frac{A^Q_{L,R}(A^{(1)})}{Qe}\,.
\ee 
Although the $SU(3)_c$ and $U(1)_Q$ gauge symmetries remain unbroken, $A^Q_{L,R}(G^{(1)})$ and $A^Q_{L,R}(A^{(1)})$ do depend both on $Q$ and on the fermion chirality due to the different fermionic shape functions involved. 
After rotation to the fermion mass eigenbasis, the currents are given by
\be\label{eq:Jphoton}
J_\mu(A^{(1)})=  \bar\Psi_{L,R}(Q)_\text{mass}\,\gamma_\mu B_{L,R}^Q(A^{(1)})\, \Psi_{L,R}(Q)_\text{mass}\,,
\ee
and
\be\label{eq:Jgluon}
J^{A}_\mu(G^{(1)})= {\bar\Psi_{L,R}(Q)}_\text{mass}\,\gamma_\mu t^{A}B_{L,R}^Q(G^{(1)})\,{\Psi_{L,R}(Q)}_\text{mass}\,.
\ee
Again, the matrices $B^Q_{L,R}(A^{(1)})$ and $B^Q_{L,R}(G^{(1)})$   are proportional to each other. Their relation to $A^Q_{L,R}(A^{(1)})$, $A^Q_{L,R}(G^{(1)})$ is given in \eqref{eq:B23} and \eqref{eq:B-13}.
 {Explicit
 expressions for the $A^Q_{L,R}(G^{(1)})$ and $A^Q_{L,R}(A^{(1)})$ matrices can be obtained from the Feynman rules given in Appendix \ref{app:FR}.}

As due to the absence of spontaneous symmetry breaking in this sector the gauge bosons are already  in their mass eigenstates, the expressions \eqref{eq:Jphoton} and \eqref{eq:Jgluon} are already the final expressions for the (massive) photonic and gluonic currents.

Because the matrices $B_{L,R}^Q$ are non-diagonal in flavour space, $G^{(1)}$ and $A^{(1)}$ mediate tree level FCNC processes.

\subsection{Sources of Flavour Violation}

\subsubsection{Preliminaries}

Due to non-universalities of the couplings of
quarks to KK gauge bosons, implied by the manner the hierarchies of masses and
mixings are explained in this NP scenario, FCNC transitions mediated by KK
gauge bosons appear already at the tree level. The mixing of {the} $Z$ boson with the 
KK gauge bosons in the process of EWSB implies also
tree level $Z$ contributions. Fortunately the model has a custodial protection 
symmetry not only for the $Zb_L\bar b_L$ coupling but as pointed out in
\cite{Blanke:2008zb} also for the $Zd_L^i\bar d_L^j$ couplings. Consequently, 
the tree level
$Z$ exchanges in processes with external down-type quarks while implying interesting
effects
in rare $K$ and $B$ decays \cite{Blanke:2008yr}  are not problematic. In particular
 no fine-tuning is necessary
to satisfy present constraints on the branching ratios of these decays.

The pattern of flavour violation in the present model goes far beyond the one of the SM and also the one  characteristic for models with MFV. There are new flavour violating parameters in addition to the SM Yukawa couplings and the resulting CKM matrix, and in particular new CP-violating phases. The counting of all these parameters is given {in the next section}. Moreover, new operators contribute that are either absent or strongly suppressed within the SM. 

There are basically two main origins of these non-MFV effects:
\begin{enumerate}
\item
The explanation of hierarchies of fermion masses through the differences of fermionic bulk masses and shape functions leads to the non-universality of fermion-gauge interactions and consequently FCNC transitions at tree level.
\item
The requirement of consistency with the well-measured EWPO, including the {$Z\to b_L\bar b_L$} transition, brings in not only new heavy gauge bosons, but also new heavy fermions. The presence of the latter and their mixing with the standard quarks and leptons implies a small non-unitarity of the CKM matrix and consequently still another source of tree level FCNC transitions. Moreover these new particles can contribute at loop level to FCNC processes. 
\end{enumerate}
Let us briefly elaborate on all these effects.

\subsubsection{Tree-level Exchange of KK Gluons and KK Photons}

As the zero and KK  modes of gluons and photons do not feel EWSB, no mixing between the various KK modes appears. Consequently the couplings of the zero modes remain flavour conserving. 

{On the other hand the shape functions of the gluonic and photonic massive KK modes are peaked towards the IR brane. Consequently the different shape functions of light fermions  then imply
flavour violating couplings of the fermion zero modes to the KK gluons and photons, given by the overlaps of the respective shape functions.} These are summarised in \eqref{eq:Jphoton} and  \eqref{eq:Jgluon}. As we have seen, the explicit appearance of the new flavour mixing matrices $\mathcal{U}_{L,R}$ and $\mathcal{D}_{L,R}$, that are unobservable in the SM and all other MFV models, introduces new flavour violating parameters.

 The exchange of massive KK gluons leads  to tree level contributions to
 $K^0-\bar K^0$ and $B^0_{d,s}-\bar B^0_{d,s}$ mixings and non-leptonic decays
 {discussed originally in \cite{Burdman:2003nt} and in more details recently in
\cite{Csaki:2008zd,Blanke:2008zb,Bauer:2008xb}.} The massive KK modes of the
 photon contribute in addition also to semi-leptonic decays such as $B\to
 X_s\ell^+\ell^-$ and $K_L\to\pi^0 \ell^+\ell^-$ \cite{Blanke:2008yr}.

For the first two quark generations the universality of the couplings in question is only slightly broken and tree level FCNC transitions are a priori suppressed. Moreover the small overlap of the shape functions for these quarks, that are peaked towards the UV brane, with the shape functions of the KK gauge bosons, peaked towards the IR brane, suppresses the relevant gauge couplings. This so-called RS-GIM mechanism \cite{Agashe:2004cp} helps to suppress FCNCs in the $K$ meson system, but still does not eliminate severe constraints on the model from $\eps_K$ { \cite{Csaki:2008zd,Blanke:2008zb}}, in particular when tree level contributions from KK gluons are involved. These effects are also present in processes involving the third quark generation, where the universality breakdown is stronger. On the other hand, also the experimental constraints are weaker in that case.

{ 

\boldmath
\subsubsection{Tree Level Exchanges of $Z$, $Z_H$ and $Z'$}
\unboldmath
Flavour violation is more involved in this case because of the spontaneous
breaking of EW symmetry that introduces
\bi
\item
mixing between $Z^{(0)}$  and the KK modes $Z^{(1)}$ and $Z_X^{(1)}$,
\item
mixing between SM fermions and KK fermions.
\ei
Concerning the first effect, even if the $Z^{(0)}$ gauge boson does not
mediate any FCNCs before EWSB, such transitions are mediated by the 
$Z^{(1)}$ and $Z_X^{(1)}$ due to their non-universal couplings to light 
fermions.
 The mixing of $Z^{(0)}$ with $Z^{(1)}$ and $Z_X^{(1)}$ in the process of EWSB
 then implies that the light mass eigenstate $Z$ does indeed mediate tree
 level FCNCs, and, together with $Z_H$ and $Z'$, can in principle have a  
significant impact on rare FCNC processes.\footnote{Equivalently, as discussed in Appendix \ref{app:EWSB}, the tree level
transitions mediated by $Z$ can be traced back to the distortion of its shape
function in the vicinity of the IR brane after EWSB has taken place
\cite{Burdman:2002gr}.}

Now, our detailed study in \cite{Blanke:2008yr,Blanke:2008zb} shows that 
in the model in question
the flavour violating couplings of $Z$ and $Z'$ to left-handed down quarks 
are protected by the custodial symmetry $P_{LR}$ of the model so that tree level 
contributions of $Z$ to all flavour violating processes (dominantly 
represented by the $Zd_R^i\bar d_R^j$ couplings) can be kept under control,
while $Z'$ contributions are fully negligible.
It turns out then that while new contributions to $\varepsilon_K$ and
$\Delta M_K$ are dominated by KK gluon exchanges, corresponding 
contributions to the $\Delta B=2$ observables are governed by KK gluon
and $Z_H$ gauge boson exchanges, while the tree level $Z$ contributions being of
higher order in $v^2/M_\text{KK}^2$ are negligible. On the other hand new physics 
contributions to rare
$K$ and $B$ decays are governed by the right-handed couplings of $Z$.

Similarly the mixing between SM fermions and KK fermions generates additional
contributions to flavour violating $Z$ couplings, but numerical studies \cite{Blanke:2008zb,Blanke:2008yr} and a dedicated analysis 
in \cite{BDG} show that these effects are subleading with respect to
the ones originating from the mixing in the gauge sector.

Flavour violation in the neutral EW sector is given in a compact way by 
\eqref{eq:JmuZ}, with similar expressions holding for $Z_H$ and $Z'$.

\subsubsection{Impact on Charged Current Interactions}

The two types of effects, mixing in the gauge sector and mixing in the fermion
sector between zero modes and heavy KK modes also have an impact on charged
current interactions of ordinary quarks, although these effects are not
as important as in the neutral sector because flavour violation in
charged current interactions appears in the SM already at the tree level. Still
a number of novel effects can be identified:
\bi
\item
The presence of new heavy charged gauge bosons $W_H$ and $W'$ introduces new
flavour violating interactions. 
In particular charged weak interactions between right-handed ordinary quarks
are present,  leading to new effective operators that were absent in the SM. 
\item
Moreover,  due to the imposed $P_{LR}$ symmetry and the corresponding
fermionic 
representations of the EW gauge group, also ordinary $W$ bosons mediate
right-handed weak interactions, as seen explicitly in \eqref{eq:JmuW}. 
\item
However, charged current interactions, both of $W^\pm$ and of the new heavy $W_H^\pm, W^{\prime\pm}$ gauge bosons, involving right-handed zero modes appear 
only due to  the mixing of the fermion zero modes with their heavy KK modes, 
which is generally found to be a subleading effect
\cite{Moreau:2005kz,Huber:2003tu,Blanke:2008zb,Blanke:2008yr,BDG}.
\item
Of some interest is also the violation of the unitarity of the CKM matrix that
originates both from the mixing in the charged gauge boson sector and in
particular in the mixing of the fermionic KK modes with the ordinary quarks 
as seen in
\eqref{eq:UL}--\eqref{eq:DL}. For masses of fermionic KK modes
significantly below $1\tev$, as identified in
\cite{Contino:2006qr,Carena:2006bn}, such effects have to be taken into
account, although they are generally smaller than the ones related to the
breakdown of universality 
{\cite{Moreau:2005kz,Huber:2003tu,Blanke:2008zb,Blanke:2008yr}.} 
{A different conclusion has however been reached in \cite{Bauer:2008xb,Casagrande:2008hr}.}
\ei

{A dedicated study of the impact of KK fermions on the low energy couplings of the SM fermions to SM gauge bosons and the Higgs boson is presented in \cite{BDG}, where further references to related literature can be found.}

}

{

\subsubsection{Tree Level Higgs Exchanges}

An order of magnitude estimate for the absolute size of the tree level flavour changing Higgs couplings can be obtained in the mass insertion approximation (MIA) as done in Appendix~C of \cite{Blanke:2008zb}. In the MIA, tree level flavour changing Higgs couplings arise from the flavour changing interactions of the Higgs boson with a light fermion and a heavy KK fermion {or with two different heavy fermions}. 
Adding one, or 
two, {such interactions}
on the heavy fermion line then gives rise to flavour changing couplings of the Higgs boson to two light fermions. Naively, one would expect the coupling in the second case 
to be comparable in size to the first case since there the coupling is additionally reduced by a chiral suppression factor. 
However, evaluating the Dirac structure of the corresponding diagrams reveals that also in the second case a strong chiral suppression is active which is even stronger than in the first case. This is a result of the assignment of BCs to the fermion representations, see \eqref{eq:xi1L}--\eqref{eq:xi3L}, which suppresses (or forbids in the case of a brane localized Higgs field) certain {transitions on the IR brane}. Hence the largest contribution to tree level flavor changing Higgs couplings in the MIA is expected from diagrams with only one {flavour changing} transition, and one finds the overall suppression factor to be proportional to the IR brane overlaps of the involved fermions, to $v/M_\text{KK}$, and finally to $m/M_\text{KK}$, where $m$ denotes the mass of the involved light fermions.
From these considerations one can conclude that the Higgs contributions to FCNC processes are negligible in the model in question, even for $M_\text{KK}$ as low as $2.45 \tev$, as we have also verified numerically. These findings are
supported by the analysis in the effective Lagrangian approach in \cite{BDG}.

An alternative derivation of the flavour changing Higgs coupling has been presented in \cite{Casagrande:2008hr}.
}

\subsubsection{One Loop Effects}

Until now our discussion concentrated on tree level FCNC
contributions. However, important new effects can also arise at the one loop
level, in particular when new contributions are absent or strongly suppressed
at tree level. This is the case of dipole operators that are relevant for
radiative decays such as $b \to s\gamma$ { \cite{Agashe:2004cp,Agashe:2008uz}} and $\mu\to
e\gamma$ \cite{Agashe:2006iy}. 
{These operators receive new contributions from the heavy KK gauge bosons and KK fermions running in the loop, as well as from the modifications in the SM couplings that appear due to the mixing of zero and KK modes both in the gauge and in the fermion sector.}

{\subsubsection{Effects of Brane-Localised Terms}\label{sec:brane}

While throughout the present analysis we have omitted brane-kinetic terms, we would like to stress that they are generally expected to be present, not being forbidden by any symmetry.
The impact of brane-kinetic terms for the gauge fields on FCNC observables has been discussed in detail in \cite{Csaki:2008zd,Blanke:2008zb}. Here we just mention for the sake of completeness that the strength of the KK gauge couplings can be strengthened or weakened by as much as a factor of two, depending on the value of the brane-localised coupling constant. Accordingly the generic bounds arising from FCNC observables on the KK scale can be worsened or ameliorated by up to a factor of two \cite{Csaki:2008zd,Agashe:2008uz}.

In principle also the inclusion of brane-localised mass terms can affect flavour phenomenology. This happens e.\,g.\ in the gauge-Higgs unification scenario, where the structure of effective Yukawa couplings is obtained with the help of quark mass terms on the IR brane \cite{Contino:2003ve,Agashe:2004rs}. In this case it turns out \cite{Csaki:2008zd} that while the flavour structure remains unchanged at the qualitative level, an enhancement of flavour violating effects appears, leading to somewhat more stringent phenomenological constraints.

Thus we conclude that while the inclusion of brane-localised terms can have an $\ord(1)$ impact on flavour violating observables, the qualitative picture of RS flavour physics remains unaffected. 
In fact this is straightforward to understand from the effective 4D two-site approach \cite{Contino:2006nn}: IR brane-localised terms correspond to couplings within the strongly coupled sector of the model and may thus modify certain predictions at the $\ord(1)$ level. The qualitative flavour structure however is determined by the mixing of the elementary fermions with the composite degrees of freedom. This mixing is characterised by the bulk mass parameters and can therefore only be affected indirectly by the IR brane-localised physics via radiative corrections.
}

\newsection{Parameter Counting}\label{sec:parameters}

In this section we list all parameters of the model\footnote{Clearly the number of free parameters would be larger if we were to consider the more general case including all possible brane Lagrangians.}, paying particular attention to those parameters relevant for flavour physics. Subsequently we develop a useful parameterisation for the 5D Yukawa coupling matrices $\lambda^{u,d}$ {in terms of parameters that can in principle be determined from low energy experiments}.

\subsection{Gauge Sector}

In the gauge sector, we have the three gauge couplings
\be
g_s\,,\qquad g\,,\qquad g_X\,,
\ee
for $SU(3)_c$, $SU(2)_L\times SU(2)_R$ and $U(1)_X$, respectively. The $P_{LR}$ symmetry ensures the equality of $SU(2)_L$ and $SU(2)_R$ couplings. 

{We would like to stress that throughout this paper $g_s$, $g$ and $g_X$ denote the 5D gauge couplings that are not dimensionless. Usually the impact of brane kinetic terms is neglected and the simple tree level matching condition
\be
g_s^{4D} = \frac{g_s}{\sqrt{L}}
\ee
is imposed, with similar equations holding also for $g^{4D}$ and $g_X^{4D}$. For a discussion of possible brane kinetic terms modifying this matching, see e.\,g. \cite{Csaki:2008zd,Blanke:2008zb}.}

\subsection{Higgs Sector}

The number of parameters present in the Higgs sector depends on the realization of the Higgs mechanism in the RS bulk. For instance in gauge-Higgs unification models \cite{Contino:2003ve} the Higgs sector is completely determined by the gauge couplings of the theory, so that no new parameters enter. On the other hand, if a Higgs potential is introduced at tree level, the number of parameters depends on its exact realization (bulk and/or boundary potential etc.). As we do not specify the mechanism of EWSB in our analysis but simply assume the presence of a Higgs field $H(x^\mu,y)$ with 4D VEV $\langle h^0(x)\rangle = v$, see Section \ref{sec:EWSB}, and bulk shape function
$h(y) \propto e^{\beta k (y-L)}$, we effectively introduce two new parameters
\be
v\,,\qquad \beta\,.
\ee
{In our phenomenological analyses \cite{Blanke:2008zb,Blanke:2008yr} we have restricted our attention to the case of a brane Higgs field, i.\,e. $\beta\to\infty$.}

If we were to study the Higgs sector of our model in more detail, we would also have to introduce the Higgs mass $m_H$ as an additional free parameter.

\subsection{Geometry}

Here we have the two parameters
\be
k\,,\qquad L\,,
\ee
which are correlated through $e^{kL}\sim\ord(10^{16})$ necessary to explain the hierarchy between the Planck and the EWSB scale. In order to simplify our {phenomenological analysis \cite{Blanke:2008zb,Blanke:2008yr}}, we fix $e^{kL}=10^{16}$ and treat
\be
f=ke^{-kL}
\ee
as the only free parameter coming from space-time geometry. This approximation is justified as physical observables depend only weakly on the exact value of ${kL}$.  Recently, however, it has been observed \cite{Davoudiasl:2008hx} that abandoning the aim to solve the gauge hierarchy problem and allowing $e^{kL}\sim\ord(10^{3})$ can solve some of the generic problems of RS models and allow for a smaller gauge KK scale in accordance with EWPO. On the other hand, the authors of \cite{Bauer:2008xb} claim that the ``$\eps_K$ problem'' \cite{Csaki:2008zd,Blanke:2008zb,Agashe:2008uz} can not be solved in this Little RS scenario.

\subsection{Quark Flavour Parameters}

Our counting of flavour parameters in the quark sector follows the one presented in \cite{Agashe:2004cp}. We recall it here for completeness.

First the $3\times 3$ complex 5D Yukawa coupling matrices
\be
\lambda^u_{ij}\,,\qquad \lambda^d_{ij}
\ee
contain each 9 real parameters and 9 complex phases. This is precisely the case of the SM. {We note that $\lambda^{u,d}_{ij}$ are not dimensionless.}

New flavour parameters enter through the three hermitian $3\times 3$ bulk mass matrices 
\be
c_1\,,\qquad c_2\,,\qquad c_3\,,
\ee
which bring in additional 18 real parameters and 9 complex phases.

In total thus we have at this stage 36 real parameters and 27 complex phases. Not all of these however are physical and some of them can be eliminated by the {quark} flavour symmetry $U(3)^3$ of the 5D theory which exists in the limit of vanishing $\lambda^{u,d}_{ij}$ and $c_i$. Note that this flavour symmetry is identical to the one present in the SM, and as in the SM 9 real parameters and 17 phases can be eliminated by making use of this symmetry. Note that one phase cannot be removed as it corresponds to the unbroken $U(1)_B$ baryon number symmetry.

We are then left with 27 real parameters and 10 complex phases to be compared with 9 real parameters and one complex phase in the SM. Evidently the new 18 real parameters and 9 phases come from the three bulk mass matrices $c_1$, $c_2$ and $c_3$.

\subsection{Flavour Parameters at Low Energies}

As we have seen, the flavour sector of the model comes along with a quite large number of parameters. One possibility, adopted in \cite{Blanke:2008zb,Blanke:2008yr}, is to work in the special basis where the bulk mass matrices are real and diagonal, and to parameterise the fundamental Yukawa couplings $\lambda^{u,d}$ in terms of physical parameters only. Details on such a parameterisation can be found in \cite{Blanke:2008zb}. The first necessary step in a phenomenological analysis is then to fit the SM quark masses and CKM mixing parameters, that have been determined experimentally. While such an approach shows clearest how the fundamental parameters enter low energy flavour observables, it is in practise complicated and numerically time-consuming. 

Therefore it is desirable to have at hand a parameterisation in which the quark masses and CKM parameters enter explicitly and do not have to be fitted. The remaining $18+9$ parameters can then be scanned over, having to fulfill only the (stringent) $\Delta F=2$ constraints.\footnote{We would like to thank Yuval Grossman for stressing the importance of a description in terms of parameters accessible at low energies.} In some analogy to the Casas-Ibarra parameterisation \cite{Casas:2001sr} in the lepton sector, we therefore aim to derive a parameterisation of the RS flavour sector in terms of the SM quark masses, the CKM parameters, and the parameters of the new flavour mixing matrices $\mathcal{D}_L$, $\mathcal{U}_R$ and $\mathcal{D}_R$.

To this end we start by working in the ``special basis'' in which the bulk mass matrices are real and diagonal. In what follows we will work with
\bea
F_Q&=& \text{diag} \left(f^{(0)}_L(y=L,c_1^1), f^{(0)}_L(y=L,c_1^2), f^{(0)}_L(y=L,c_1^3) \right)\, \frac{e^{kL/2}}{\sqrt{L}}\,  ,\\
F_u&=& \text{diag} \left(f^{(0)}_R(y=L,c_2^1), f^{(0)}_R(y=L,c_2^2), f^{(0)}_R(y=L,c_2^3) \right)\, \frac{e^{kL/2}}{\sqrt{L}}\,   ,\\
F_d&=& \text{diag} \left(f^{(0)}_R(y=L,c_3^1), f^{(0)}_R(y=L,c_3^2), f^{(0)}_R(y=L,c_3^3) \right)\, \frac{e^{kL/2}}{\sqrt{L}}\, ,
\eea
i.\,e. $F_{Q,u,d}$ are diagonal $3\times 3$ matrices whose entries are the fermion zero mode shape functions on the IR brane.

Neglecting now the mixing with fermionic KK modes and approximating the Higgs field to be exactly localised on the IR brane, we can write
\bea
\text{diag} (m_u,m_c,m_t) &=& \frac{v}{\sqrt{2}} \tilde{\mathcal{U}}_L^\dagger F_Q \lambda^u F_u \tilde{ \mathcal{U}}_R\,,\\
\text{diag} (m_d,m_s,m_b) &=& \frac{v}{\sqrt{2}} \tilde{\mathcal{D}}_L^\dagger F_Q \lambda^d F_d \tilde{ \mathcal{D}}_R\,,
\eea
where $\tilde{ \mathcal{U}}_{L,R},\tilde{ \mathcal{D}}_{L,R}$ are the upper left $3\times3$ blocks of the corresponding matrices in \eqref{eq:UL}, \eqref{eq:DL}.
The CKM matrix is given by 
\be
\tilde{ \mathcal{U}}_L^\dagger\tilde{\mathcal{D}}_L=V_\text{CKM}\,.
\ee

The 5D Yukawa couplings can then be written as
\bea
\lambda^u &=&  \frac{\sqrt{2}}{v} F_Q^{-1} \tilde{ \mathcal{U}}_L \text{diag} (m_u,m_c,m_t) \tilde{ \mathcal{U}}_R^\dagger F_u^{-1}\nn\\
&=& \frac{\sqrt{2}}{v} F_Q^{-1} \tilde{ \mathcal{D}}_L V_\text{CKM}^\dagger \text{diag} (m_u,m_c,m_t) \tilde{ \mathcal{U}}_R^\dagger F_u^{-1}\,,\label{eq:lambdau}\\
\lambda^d &=&  \frac{\sqrt{2}}{v} F_Q^{-1} \tilde{\mathcal{D}}_L \text{diag} (m_d,m_s,m_b) \tilde{ \mathcal{D}}_R^\dagger F_d^{-1}\label{eq:lambdad}\,,
\eea
where we have expressed $\tilde{ \mathcal{U}}_L$ through $\tilde{ \mathcal{D}}_L$ and $V_\text{CKM}$. The SM parameters are encoded in the quark masses and $V_\text{CKM}$. 9 new real flavour parameters are present in $F_{Q,u,d}$. The remaining 9 real parameters and 9 complex phases are distributed among $\tilde{\mathcal{D}}_L$, $\tilde{ \mathcal{U}}_R$ and $\tilde{ \mathcal{D}}_R$. 

{In order to obtain a parameterisation of $\tilde{\mathcal{D}}_L$, $\tilde{ \mathcal{U}}_R$ and $\tilde{ \mathcal{D}}_R$ in terms of these $9+9$ physical parameters only, we start by writing\footnote{Note that every unitary $3\times 3$ matrix can be parameterised in such a way.}
\bea
\tilde{\mathcal{D}}_L &=& \renewcommand{\arraystretch}{1.3}\addtolength{\arraycolsep}{1.5pt}
\left(\begin{array}{ccc}
1 & 0 & 0\\
0 & c_{23}^{\mathcal{D}_L} & s_{23}^{\mathcal{D}_L} \exp(- i\delta^{\mathcal{D}_L}_{23})\\
0 & -s_{23}^{\mathcal{D}_L} \exp(i\delta^{\mathcal{D}_L}_{23}) & c_{23}^{\mathcal{D}_L}\\
\end{array}\right)
\hspace{-1mm}\cdot\hspace{-1mm}
 \left(\begin{array}{ccc}
c_{13}^{\mathcal{D}_L} & 0 & s_{13}^{\mathcal{D}_L} \exp( - i\delta^{\mathcal{D}_L}_{13})\\
0 & 1 & 0\\
-s_{13}^{\mathcal{D}_L} \exp( i\delta^{\mathcal{D}_L}_{13}) & 0 & c_{13}^{\mathcal{D}_L}\\
\end{array}\right) \nn\\
&& \cdot\hspace{-1mm}\renewcommand{\arraystretch}{1.3}\addtolength{\arraycolsep}{1.5pt}
 \left(\begin{array}{ccc}
c_{12}^{\mathcal{D}_L} & s_{12}^{\mathcal{D}_L} \exp({- i\delta^{\mathcal{D}_L}_{12}}) & 0\\
-s_{12}^{\mathcal{D}_L} \exp({i\delta^{\mathcal{D}_L}_{12}}) & c_{12}^{\mathcal{D}_L} & 0\\
0 & 0 & 1\\
\end{array}\right)
\hspace{-1mm}\cdot \hspace{-1mm}
\left(\begin{array}{ccc}
\exp({i\varphi_1^{\mathcal{D}_L}}) & 0 & 0 \\
0 & \exp({i\varphi_2^{\mathcal{D}_L}}) & 0 \\
0 & 0 & \exp({i\varphi_3^{\mathcal{D}_L}}) 
\end{array}\right),\nn\\
\eea
i.\,e. as a product of three rotation matrices with a complex phase $\delta_{ij}^{\mathcal{D}_L}$ ($i,j=1,2,3$) in each of them \cite{Blanke:2006xr}, times a diagonal matrix containing three additional phases $\varphi_i^{\mathcal{D}_L}$ ($i=1,2,3$). Further
\be
c_{ij}^{\mathcal{D}_L} = \cos\theta_{ij}^{\mathcal{D}_L}\,,\qquad
s_{ij}^{\mathcal{D}_L} = \sin\theta_{ij}^{\mathcal{D}_L}\qquad (i,j=1,2,3)\,.
\ee
$\tilde{ \mathcal{U}}_R$ and $\tilde{ \mathcal{D}}_R$ are written in a completely analogous way. It is then easy to see that the diagonal phases $\varphi_i^{\mathcal{D}_L}$, $\varphi_i^{\mathcal{U}_R}$ and $\varphi_i^{\mathcal{D}_R}$ ($i=1,2,3$) can be rotated away by appropriate phase redefinitions of $q_L^{i(0)}$, $u_R^{i(0)}$ and  $D_R^{i(0)}$, respectively. 

We are thus left with a parameterisation of $\tilde{\mathcal{D}}_L$, $\tilde{ \mathcal{U}}_R$ and $\tilde{ \mathcal{D}}_R$ in terms of three mixing angles $\theta_{ij}$ and three complex phases $\delta_{ij}$ in each of them, which reads \cite{Blanke:2006xr}
\bea
\tilde{\mathcal{D}}_L &=& \\
&&\hspace*{-1.6cm} \begin{pmatrix}
c_{12}^{\mathcal{D}_L} c_{13}^{\mathcal{D}_L} & s_{12}^{\mathcal{D}_L} c_{13}^{\mathcal{D}_L} e^{-i\delta^{\mathcal{D}_L}_{12}}& s_{13}^{\mathcal{D}_L} e^{-i\delta^{\mathcal{D}_L}_{13}}\\
-s_{12}^{\mathcal{D}_L} c_{23}^{\mathcal{D}_L} e^{i\delta^{\mathcal{D}_L}_{12}}-c_{12}^{\mathcal{D}_L} s_{23}^{\mathcal{D}_L}s_{13}^{\mathcal{D}_L} e^{i(\delta^{\mathcal{D}_L}_{13}-\delta^{\mathcal{D}_L}_{23})} &
c_{12}^{\mathcal{D}_L} c_{23}^{\mathcal{D}_L}-s_{12}^{\mathcal{D}_L} s_{23}^{\mathcal{D}_L}s_{13}^{\mathcal{D}_L} e^{i(\delta^{\mathcal{D}_L}_{13}-\delta^{\mathcal{D}_L}_{12}-\delta^{\mathcal{D}_L}_{23})} &
s_{23}^{\mathcal{D}_L}c_{13}^{\mathcal{D}_L} e^{-i\delta^{\mathcal{D}_L}_{23}}\\
s_{12}^{\mathcal{D}_L} s_{23}^{\mathcal{D}_L} e^{i(\delta^{\mathcal{D}_L}_{12}+\delta^{\mathcal{D}_L}_{23})}-c_{12}^{\mathcal{D}_L} c_{23}^{\mathcal{D}_L}s_{13}^{\mathcal{D}_L} e^{i\delta^{\mathcal{D}_L}_{13}} &
-c_{12}^{\mathcal{D}_L} s_{23}^{\mathcal{D}_L} e^{i\delta^{\mathcal{D}_L}_{23}}-s_{12}^{\mathcal{D}_L} c_{23}^{\mathcal{D}_L}s_{13}^{\mathcal{D}_L} e^{i(\delta^{\mathcal{D}_L}_{13}-\delta^{\mathcal{D}_L}_{12})} &
c_{23}^{\mathcal{D}_L}c_{13}^{\mathcal{D}_L}\\
\end{pmatrix}\nn
\eea

In order to naturally obtain anarchic 5D Yukawa matrices $\lambda^{u,d}$, it {can} in practice be useful to adapt the above parameterisation.} 
While the phases
\be
\delta_{ij}^{\mathcal{D}_L}\,,\qquad
\delta_{ij}^{\mathcal{U}_R}\,,\qquad
\delta_{ij}^{\mathcal{D}_R}
\ee
are all chosen to lie in their natural range $0\le\delta<2\pi$, the case of the mixing angles 
\be
\theta_{ij}^{\mathcal{D}_L}\,,\qquad
\theta_{ij}^{\mathcal{U}_R}\,,\qquad
\theta_{ij}^{\mathcal{D}_R}
\ee
is somewhat different. Here one finds \cite{Agashe:2004cp} that anarchic 5D Yukawa couplings imply the hierarchies
\be
\theta_{ij}^{\mathcal{D}_L}\sim \frac{(F_Q)_{ii}}{(F_Q)_{jj}}\,,\qquad
\theta_{ij}^{\mathcal{U}_R}\sim \frac{(F_u)_{ii}}{(F_u)_{jj}}
\,,\qquad
\theta_{ij}^{\mathcal{D}_R}\sim \frac{(F_d)_{ii}}{(F_d)_{jj}}\,.
\ee
We can now use this knowledge to find a parameterisation that automatically leads to a natural structure for $\lambda^{u,d}$. Therefore we define
\be
\theta_{ij}^{\mathcal{D}_L}= \kappa_{ij}^{\mathcal{D}_L} \frac{(F_Q)_{ii}}{(F_Q)_{jj}}\,,\qquad
\theta_{ij}^{\mathcal{U}_R} =  \kappa_{ij}^{\mathcal{U}_R}\frac{(F_u)_{ii}}{(F_u)_{jj}}
\,,\qquad
\theta_{ij}^{\mathcal{D}_R}=  \kappa_{ij}^{\mathcal{D}_R} \frac{(F_d)_{ii}}{(F_d)_{jj}}\,,
\ee
where $\kappa_{ij}^{\mathcal{D}_L}$, $\kappa_{ij}^{\mathcal{U}_R}$, $\kappa_{ij}^{\mathcal{D}_R}$ are $\ord(1)$ parameters. 

Note that although the parameterisation of the 5D Yukawa matrices $\lambda^{u,d}$ in \eqref{eq:lambdau}, \eqref{eq:lambdad} in terms of $F_{Q,u,d}$ and $\tilde{ \mathcal{D}}_L,\tilde{ \mathcal{U}}_R, \tilde{ \mathcal{D}}_R$ is clearly an approximation in case of a bulk Higgs field, the hierarchies in  $\tilde{ \mathcal{D}}_L,\tilde{ \mathcal{U}}_R, \tilde{ \mathcal{D}}_R$ are still the same, so that the above parameterisation of these matrices can still be used without loss of generality. Small deviations from the exact results will only appear where formulae \eqref{eq:lambdau}, \eqref{eq:lambdad} are used explicitly.

\newsection{The Lepton Sector}\label{sec:leptons}

The embedding of the lepton sector into multiplets of the symmetry $O(4)\sim SU(2)_L\times SU(2)_R\times P_{LR}$ is analogous to the quark sector as given in (\ref{eq:xi1L})-(\ref{eq:xi3R}). Merely the $U(1)_X$ charges have to be modified in order to accommodate the electric charges of the charged leptons and neutrinos. As in the quark sector, there are three $O(4)$ multiplets per generation ($i=1,2,3$):
\bea\label{eq:xi1L_leptons}
\xi^{i,\ell}_{1L}&=&\begin{pmatrix}\chi^{\nu_i}_{L}(-+)_{1} && \ell_L^{\nu_i}(++)_{0} \\
\chi^{e_i}_{L}(-+)_{0} && \ell_L^{e_i}(++)_{-1}\end{pmatrix}_{0}\,,\\
\xi^{i,\ell}_{2R} &=& \nu^i_R (++)_{0}\,,\\
\xi^{i,\ell}_{3R} &=& T^{i,\ell}_{3R} \oplus T^{i,\ell}_{4R} = \begin{pmatrix}
\lambda^{\prime i}_R(-+)_{1} \\ N^{\prime i}_R (-+)_{0} \\ L^{\prime i}_R (-+)_{-1} \end{pmatrix}_{0} \oplus 
\begin{pmatrix}
\lambda^{\prime\prime i}_R(-+)_{1} \\ N^{\prime\prime i}_R (-+)_{0} \\ L^{ i}_R (++)_{-1} \end{pmatrix}_{0}\,.\label{eq:xi3R_leptons}
\eea
The corresponding  states of opposite chirality can be deduced from (\ref{eq:xi1L_leptons})-(\ref{eq:xi3R_leptons}) by exchanging $L\leftrightarrow R$ and flipping the BCs.

The following comments are in order:
\begin{itemize}
 \item As in the quark case the field obeying $(++)$ BCs have massless zero
   modes. Up to mixing with other massive modes, these zero modes are the
   usual SM leptons.
 \item The remaining fields are KK leptons that are vectorial Dirac particles. We thus have in this model additional heavy lepton states. These are
\bea
Q=1&:&\qquad \chi^{\nu_i(n)},\lambda^{\prime i(n)},\lambda^{\prime\prime i(n)}\,, \\
Q=0&:&\qquad  \ell^{\nu_i(n)},  \nu^{i(n)}, N^{\prime i(n)}, N^{\prime\prime i(n)}, \chi^{e_i(n)}\,,\\
Q=-1&:&\qquad \ell^{e_i(n)}, L^{i(n)}, L^{\prime i(n)}\,,
\eea
where $n=1,2,...$.
\end{itemize}

The vectors necessary to construct the mass matrices are
\bea\label{eq:psiL53_leptons}
\Psi_L^\ell(1) &=& \left( \chi^{\nu_i}_L(-+), \lambda^{\prime i}_L(+-), \lambda^{\prime\prime i}_L(+-)\right)^T\,,\\
\Psi_R^\ell(1) &=& \left( \chi^{\nu_i}_R(+-), \lambda^{\prime i}_R(-+), \lambda^{\prime\prime i}_R(-+)\right)^T\,,\\
\Psi_L^\ell(0) &=& \left( \ell_L^{\nu_i(0)}(++), \ell_L^{\nu_i}(++),N_L^{\prime i}(+-),N_L^{\prime\prime i}(+-), \chi^{e_i}_L(-+), \nu_L^i(--)  \right)^T\,,\qquad\\
\Psi_R^\ell(0) &=& \left( \nu_R^{i(0)}(++), \ell_R^{\nu_i}(--), 
N_R^{\prime i}(-+),N_R^{\prime\prime i}(-+),
\chi^{e_i}_R(+-), \nu_R^i(++)  \right)^T\,,\\
\Psi_L^\ell(-1) &=& \left( \ell_L^{e_i(0)}(++), \ell_L^{e_i}(++), L_L^{\prime i}(+-),L_L^{ i}(--) \right)^T\,,\label{eq:psiL13_leptons}\\
\Psi_R^\ell(-1) &=& \left( L_R^{i(0)}(++), \ell_R^{e_i}(--), L_R^{\prime i}(-+),L_R^{ i}(++) \right)^T\,.\label{eq:psiR13_leptons}
\eea 
Then the structure of the mass matrices is as given for quarks in (\ref{eq:M53})-(\ref{eq:M13}). The corresponding Yukawa couplings and shape functions are of course those for leptons.

{The weak currents have the same structure as in the case of quarks} except that the couplings in (\ref{eq:g53})-(\ref{eq:kappa5}) are modified as now in all these formulae $Q_X=0$. 

For the coupling to the $Z$ boson,
\be
g_Z(\psi_i)=\frac{g}{\sqrt{L}\cos\psi}\left(T_L^3-Q\sin^2\psi\right)
\ee
also applies to leptons so that all the couplings, both left-handed or right-handed, to the $Z$ boson can easily be found.

For the coupling to the $Z_X$ boson {we find}
\bea
\kappa_1^\ell&=&g_{Z_X}(\ell^{\nu_i})=g_{Z_X}(\ell^{e_i})=-\frac{1}{2\sqrt{L}}g\cos\phi\,,\\
\kappa_2^\ell&=&g_{Z_X}(\chi^{\nu_i})= g_{Z_X}(\chi^{e_i})= \frac{1}{2\sqrt{L}}g\cos\phi\,,\\
\kappa_3^\ell&=&g_{Z_X}(\nu^i)=g_{Z_X}(\lambda^{\prime i})=g_{Z_X}(N^{\prime i})=g_{Z_X}(N^{\prime\prime i})=g_{Z_X}(L^{\prime i})=0\,,\\
\kappa_4^\ell&=&g_{Z_X}(\lambda^{\prime\prime i})=\frac{1}{\sqrt{L}}g\cos\phi\,,\\
\kappa_5^\ell&=&g_{Z_X}(L^i)=-\frac{1}{\sqrt{L}}g\cos\phi\,.
\eea
The expressions for $\kappa_i^\ell$ ($i=1,...,5$) can be combined into the following formula
\be
\kappa^\ell=T^3_R\, \frac{g}{\sqrt{L}} \cos\phi\,.
\ee

The Feynman rules for leptons can then be obtained directly from the Feynman rules for quarks by simply mapping the quark fields in (\ref{eq:psiL53})-(\ref{eq:psiR13}) onto the leptonic fields in (\ref{eq:psiL53_leptons})-(\ref{eq:psiR13_leptons}) and replacing the quark couplings $\kappa_i$ and $g_Z(\psi_i)$ by the corresponding leptonic couplings. {In addition the leptons are $SU(3)_c$ singlets, so their couplings to the gluons and KK gluons vanish.} Similarly the quark shape functions should be replaced {by} the leptonic ones with the gauge sector and Higgs sector being unchanged.

\newsection{Summary}\label{sec:concl}
In the present paper we have worked out explicitly 
the electroweak and flavour structure
of a particular warped extra dimension model with a custodial protection
not only for the flavour diagonal coupling $Zb_L\bar b_L$ as introduced in 
\cite{Agashe:2006at} but in particular
for flavour non-diagonal couplings $Zd^i_L\bar d^j_L$ pointed out in
\cite{Blanke:2008zb,Blanke:2008yr}. The most important result of the present
paper are the Feynman rules collected in Appendix \ref{app:FR}. These rules
allowed already to perform two detailed phenomenological analyses of 
$\Delta F=2$ FCNC processes in the quark sector \cite{Blanke:2008zb} and 
those $\Delta F=1$ rare $K$ and $B$ decays in which new physics contributions
enter already at the tree level \cite{Blanke:2008yr}. Short reviews of these
results can be found in {\cite{Buras:2009dy,Duling:2009sf,Gori:2009tr}}. The analyses of
processes with dipole operators like $B\to X_s\gamma$, $\mu\to e\gamma$,
and electric dipole moments, where new physics enters first at the one-loop
{level will} be presented elsewhere.

\subsection*{Acknowledgements}

We would like to thank Stefania Gori, Tillmann Heidsieck and Andreas Weiler for useful discussions.
This research was
partially supported by the Graduiertenkolleg GRK 1054, the Deutsche Forschungsgemeinschaft (DFG) under contract BU 706/2-1, {the DFG Cluster of Excellence `Origin
and  Structure of the Universe' and by} the German Bundesministerium f{\"u}r 
Bildung und Forschung under contract 05HT6WOA.

\begin{appendix}

\newsection{Fundamental 5D Action}\label{app:action}

The fundamental 5D action of the $SU(3)_c\times SU(2)_L\times SU(2)_R\times U(1)_X\times P_{LR}$ model under consideration can be decomposed as
\be\label{eq:S}
S = \int d^4x \int_0^L dy\,\left(\mathcal{L}_\text{gauge} +  \mathcal{L}_\text{fermion} + \mathcal{L}_\text{Higgs} +  \mathcal{L}_\text{Yuk}\right)\,,
\ee
with the various contributions being discussed in what follows. 

{We note that it is possible to extend the theory by additional contributions to the action that are confined to the UV or IR brane. Indeed such terms, if consistent with the symmetries of the model, will be generated through loop corrections anyway. In order to keep the presentation as clear as possible, we do however not consider this most general case, but restrict our attention to the bulk action given in \eqref{eq:S}. 
}

\subsection{Gauge Sector}

The kinetic terms for the gauge fields are given by
\be
\mathcal{L}_\text{gauge} = \sqrt{G}\left[-\frac{1}{4}G_{MN}^A G^{MN,A} -
\frac{1}{4}L_{MN}^a L^{MN,a}-
\frac{1}{4}R_{MN}^\alpha R^{MN,\alpha}-
\frac{1}{4}X_{MN}X^{MN}\right]\,,
\ee
where 
\be
G_{MN}^A=\partial_MG_N^A- \partial_NG_M^A-g_s f^{ABC}G_M^BG_N^C \qquad(A=1,\dots,8)
\ee
 corresponds to $SU(3)_c$ and $g_s$ is the 5D strong coupling constant. 
\bea
L_{MN}^a&=&\partial_MW_{L,N}^a- \partial_NW_{L,M}^a-g \varepsilon^{abc}W_{L,M}^bW_{L,N}^c \qquad(a=1,2,3)\\
R_{MN}^\alpha&=&\partial_MW_{R,N}^\alpha- \partial_NW_{R,M}^\alpha-g \varepsilon^{\alpha\beta\gamma}W_{R,M}^\beta W_{R,N}^\gamma
\qquad(\alpha=1,2,3)
\eea
correspond to $SU(2)_L$ and $SU(2)_R$, respectively, with equal gauge coupling $g$, and 
\be
X_{MN}=\partial_MX_{N}- \partial_NX_{M}
\ee
is the field strength tensor of $U(1)_X$, whose coupling constant is given by $g_X$.
Here and in the following $G=\det G_{MN}=e^{-8ky}$ has to be included in order to obtain an invariant integration measure.

{We denote} $SU(2)_L$ indices by small Latin letters $a,b,\dots$ and  $SU(2)_R$ indices by small Greek letters $\alpha,\beta,\dots$. $SU(3)_c$ indices are denoted by capital Latin letters $A,B,\dots$, but are usually made implicit in order to simplify the notation.

\subsection{Fermion Sector}

\subsubsection{Quarks}

The quark sector contains fields with the following transformation properties under $SU(2)_L\times SU(2)_R\times U(1)_X$
\bea
(\xi^i_1)_{a\alpha} &\sim& \bidoublet_{2/3}\,,\\
\xi^i_2 &\sim& \singlet_{2/3}\,,\\
\xi^i_3 =(T^i_3)_a\oplus (T^i_4)_\alpha &\sim& \Ltriplet_{2/3} \oplus \Rtriplet_{2/3}\,,
\eea
where again $SU(2)_L$ indices are denoted by Latin letters while $SU(2)_R$ indices are denoted by Greek letters.  
All these multiplets transform as triplets under $SU(3)_c$. The fermionic Lagrangian is then given by
\bea
\mathcal{L}_\text{fermion} &=&  \frac{1}{2}\sqrt{G}\sum_{i=1}^3 \Big[ 
(\bar \xi^i_1)_{a\alpha} i \Gamma^M ({D}^1_M)_{ab,\alpha\beta} (\xi^i_1)_{b\beta} + (\bar\xi^i_1)_{a\alpha} (i\Gamma^M \omega_M - c^i_1 k) (\xi^i_1)_{a\alpha} \nn\\
&&\quad {} + \bar \xi^i_2 ( i \Gamma^M {D}^2_M+ i \Gamma^M \omega_M - c^i_2 k) \xi^i_2\nn\\
&&\quad {}+ (\bar T^i_3)_a i \Gamma^M ({D}^3_M)_{ab}(T^i_3)_b  + (\bar T^i_3)_a(i\Gamma^M\omega_M- c^i_3 k) (T^i_3)_a \nn\\
&&\quad {} +
(\bar T^i_4)_\alpha i \Gamma^M ({D}^4_M)_{\alpha\beta}(T^i_4)_\beta  + (\bar T^i_4)_\alpha(i \Gamma^M\omega_M- c^i_3 k) (T^i_4)_\alpha \Big] + h.c. \,,\label{eq:Lf}
\eea
where summation over repeated indices is understood. Writing out the ``$+h.c.$'' term explicitly, one finds that the two terms including the spin connection $\omega_M$ cancel each other \cite{Csaki:2003sh}. 
Here, $\Gamma^M=E_A^M\gamma^A$ with $\gamma^A=\{\gamma^\mu,-i \gamma^5\}$,\footnote{Here, $\gamma^5 = i\gamma^0\gamma^1\gamma^2\gamma^3$ is defined in the usual 4D way.} and $E^M_A$ is the {inverse} vielbein defined through
\be
G^{MN}=E_A^M E_B^N \eta^{AB}\,,
\ee
i.\,e. it connects the warped space to the flat tangent space. For the case of the RS metric \eqref{eq:RS}, we have
\be
E^M_A=
\begin{cases}
1 & \text{for }A=M=5 \,,\\
e^{ky} & \text{for }A=M=\mu \,,\\
0&\text{else}\,,
\end{cases}
\ee
and {the vielbein $e^A_M$} is given by
\be
e^A_M=
\begin{cases}
1 & \text{for }A=M=5 \,,\\
e^{-ky} & \text{for }A=M=\mu \,,\\
0&\text{else}\,.
\end{cases}
\ee

$\omega_M$ is the spin connection defined through
\be
\omega_M=e^A_N (\partial_M E^N_B + \Gamma^N_{MK} E^K_B)\frac{{\sigma_A}^{B}}{2}\,,
\ee
{with $\sigma_{AB}=\frac{1}{4}[\gamma_A,\gamma_B]$ and 
$\Gamma^N_{MK}=\frac{1}{2}G^{NR}(\partial_K G_{MR} +\partial_M G_{KR} - \partial_R G_{MK})$,}
which yields in case of the RS metric \eqref{eq:RS}
\be
\omega_M = 
\begin{cases}
\frac{i}{2}k e^{-ky} \gamma_\mu \gamma^5 & \text{for } M=\mu \,,\\
0&  \text{for } M=5 \,.
\end{cases}
\ee

The covariant derivatives $D_M^i$ are given by
\bea
(D_M^1)_{ab,\alpha\beta} &=& 
 (\partial_M  +  i g_s t^A G_M^A + i g_X Q_X X_M)\delta_{ab}\delta_{\alpha\beta} \nn\\
&&{} + i g (\tau^c)_{ab} W_{L,M}^c \delta_{\alpha\beta} +  i g (\tau^\gamma)_{\alpha\beta} W_{R,M}^\gamma \delta_{ab}\,,\label{eq:DM1}\\
D_M^2 &=&
 \partial_M +i g_s t^A G_M^A + i g_X Q_X X_M\,,\\
(D_M^3)_{ab} &=& (\partial_M +i g_s t^A G_M^A + i g_X Q_X X_M)\delta_{ab}  + g \varepsilon^{abc}  W_{L,M}^c\,,\\
(D_M^4)_{\alpha\beta} &=& (\partial_M 
+i g_s t^A G_M^A + i g_X Q_X X_M)\delta_{\alpha\beta} + g \varepsilon^{\alpha\beta\gamma}  W_{R,M}^\gamma\,.\label{eq:DM4}
\eea
$t^A=\lambda^A/2$ $(A=1,\dots,8)$ are the generators of the fundamental representation of $SU(3)_c$, where $\lambda^A$ are the known Gell-Mann matrices. $\tau^{a}=\sigma^{a}/2$ ($\tau^{\alpha}=\sigma^{\alpha}/2$) are the generators of the fundamental $SU(2)_{L}$ ($SU(2)_{R}$) representations, respectively, where $\sigma^{a},\sigma^\alpha$ are the Pauli matrices, and  $-i\varepsilon^{abc}$ and $-i\varepsilon^{\alpha\beta\gamma}$ are the generators of the adjoint triplet representations of $SU(2)_L$ and $SU(2)_R$, respectively. Recall that despite  having the same matrix structure, the $SU(2)_L$ and $SU(2)_R$ generators act on different internal spaces.

In addition, the components of the $T^i_{3,4}$ triplets, as given in \eqref{eq:xi3R}, \eqref{eq:xi3L}, are not those components associated to $a,\alpha=1,2,3$. Instead
\be
(T^i_{3})_a = \begin{pmatrix}\frac{1}{\sqrt{2}}
(\psi^{\prime i}+ D^{\prime i})   \\ \frac{i}{\sqrt{2}}(\psi^{\prime i}- D^{\prime i})\\ U^{\prime i} \end{pmatrix}\,,\qquad
(T^i_{4})_\alpha = \begin{pmatrix}\frac{1}{\sqrt{2}}
(\psi^{\prime\prime i}+ D^{ i})  \\ \frac{i}{\sqrt{2}}(\psi^{\prime\prime i}- D^{ i}) \\ U^{\prime\prime i} \end{pmatrix}\,.
\ee
Recall that the same structure appears also in the gauge sector, where $W^{1,2}_{L,R}$ are related to $W^\pm_{L,R}$ via
\be
W^\pm_{L,R} = \frac{W^1_{L,R}\mp iW^2_{L,R}}{\sqrt{2}}\,.
\ee

\subsubsection{Leptons}
In order to preserve the minimality of the model, we take the lepton sector in complete analogy to the quark sector. The only necessary modifications are:
\bi
\item
Leptons transform as singlets under $SU(3)_c$, i.\,e. the coupling to gluons, $+i g_s t^A G_M^A$ in \eqref{eq:DM1}--\eqref{eq:DM4} has to be removed.
\item
In order to obtain correct electric charges for the leptons, $Q_X=0$ has to be imposed, so that leptons do not couple to the $X_M$ gauge boson of $U(1)_X$. Effectively thus also the $+i g_X Q_X X_M$ term  in \eqref{eq:DM1}--\eqref{eq:DM4} is absent in the case of leptons.
\ei
\subsection{Higgs Sector}

The Lagrangian describing the Higgs bidoublet $H$, given in \eqref{eq:H}, reads
\be\label{eq:LHiggs}
\mathcal{L}_\text{Higgs} =\sqrt{G} \left[ (D_M H)^\dagger_{a\alpha} (D^M H)_{a\alpha} - V(H)\right]\,,
\ee
with 
\be
(D_M H)_{a\alpha}=\partial_M H_{a\alpha}  +  ig (\tau^c)_{ab} W^c_{L,M}H_{b\alpha} + ig (\tau^\gamma)_{\alpha\beta} W^\gamma_{R,M}H_{a\beta}
\ee
and $V(H)$ being the potential that eventually leads to EWSB. 

Note that in case of a bulk Higgs field, $H$ contains in addition to the zero mode also massive KK modes. The potential $V(H)$ then has to be constructed in such a way that only the zero mode obtains a VEV, as otherwise the consistency with EW precision tests would be spoiled. 
{However, their couplings are, due to the similar profile, roughly the same as the Higgs zero mode couplings. In addition, the scalar KK modes are even heavier than the gauge and fermionic KK modes, so that in most phenomenological applications the Higgs KK modes can be safely neglected.}
Therefore, we will not give an explicit expression for $V(H)$, but merely assume that it leads to a VEV for the zero mode and the particular shape function $h(y)$, as given in \eqref{eq:h(y)}. 

The kinetic term in $\mathcal{L}_\text{Higgs}$ is responsible for the effects of EWSB in the gauge sector. Those will be discussed in detail in Appendix~\ref{app:EWSB}.

\subsection{Yukawa Sector}

Finally, we need to construct the Higgs couplings to  fermion fields, which
will yield the masses of the SM fermions after EWSB. For simplicity, we
restrict ourselves to the quark sector, the Yukawa couplings for the lepton
sector can then be obtained in a completely analogous way. {A dictionary 
that allows to obtain Feynman rules for leptons from the rules
for quarks is given in {Section \ref{sec:leptons}.}}
 
The most general Yukawa coupling including the Higgs bidoublet $H$ and the quark fields $\xi^i_{1,2,3}$ is given by
\bea
\mathcal{L}_\text{Yuk} &=&  -\sqrt{2}\sqrt{G}\sum_{i,j=1}^3 \Big[ -\lambda^u_{ij} (\bar\xi^i_1)_{a\alpha} H_{a\alpha} \xi^j_2\nn\\
&&\hspace{.7cm}{}
 + \sqrt{2}\lambda^d_{ij} \left[(\bar\xi^i_1)_{a\alpha} (\tau^c)_{ab} (T^{j}_3)^c H_{b\alpha} + (\bar \xi^i_1)_{a\alpha} (\tau^\gamma)_{\alpha\beta} (T^{j}_4)^\gamma H_{a\beta}\right] + h.c.\Big]\,,\qquad
\eea
where again summation over repeated indices is understood, and the normalisation factor $\sqrt{2}$ enters the second term in order to canonically normalise the fermion triplets $T^j_{3,4}$. {The overall signs of the two contributions are chosen such that the {00 components of the $\mathcal{M}(2/3)$ and $\mathcal{M}(-1/3)$ in \eqref{eq:M23}, \eqref{eq:M13} carry an overall plus sign.}\footnote{Recall that the fermionic mass term possesses an overall minus sign.}}

Interestingly, while the first coupling, proportional to $\lambda^u_{ij}$, contributes, after EWSB, only to the mass matrix of $+2/3$ charge quarks, the second term, proportional to $\lambda^d_{ij}$, contributes to all $+5/3$, $+2/3$ and $-1/3$ mass matrices.

\newsection{Bulk Profiles of Wave Functions and KK Masses}\label{app:eom}

In this appendix we briefly review the determination of the bulk shape functions and the KK masses and collect all necessary formulae. We follow closely the presentation in \cite{Gherghetta:2000qt}, using the conventions of these authors, except for the opposite sign of the metric and the fact that we work on the interval $0\le y\le L$ rather than on the orbifold $0\le y< 2\pi R$. We confirmed the results of \cite{Gherghetta:2000qt} except for some approximate formulae given in that paper which turn out to be too rough when compared with the exact numerical results.

\subsection{Bulk Equations of Motion}

By setting all interaction terms in \eqref{eq:S} to zero, the 5D bulk equations of motion (EOM) can straightforwardly be obtained from the variational principle $\delta S=0$, which yields generally
\be\label{eq:gen-eom} 
\left[ - e^{2ky}\eta^{\mu\nu}\partial_\mu\partial_\nu + e^{sky}\partial_5(e^{-sky}\partial_5) - M_\Phi^2\right] \Phi(x^\mu,y) =0\,.
\ee
In the case of gauge fields, $\Phi\equiv V_\mu$, $s=2$ and $M_\Phi^2=0$, while in the case of fermions, $\psi_{L,R}$ has to be rescaled by $\Phi\equiv  e^{-2ky}\psi_{L,R}$ with $s=1$ and $M_\Phi^2=c(c\pm1)k^2$ for left-/right-handed modes. The only physical scalar field is the Higgs, $\Phi\equiv H$, for which $s=4$ and $M_\Phi^2$ depends on the exact form of $V(H)$. 

Note that the minus sign in front of the first term in \eqref{eq:gen-eom}, which does not appear in equation (11) of \cite{Gherghetta:2000qt}, is due to our sign convention for the metric tensor.

\subsection{Gauge Fields}

With the KK decomposition
\be
V_\mu(x^\mu,y) = \frac{1}{\sqrt{L}}\sum_{n=0}^\infty V_\mu^{(n)}(x^\mu)f^{(n)}_\text{gauge}(y)\,,
\ee
we  obtain for the gauge KK modes \cite{Gherghetta:2000qt}
\bea
f^{(0)}_\text{gauge}(y) &=& 1\,,\\
f^{(n)}_\text{gauge}(y) &=& \frac{e^{ky}}{N_n}\left[J_1\left(\frac{m_n}{k}e^{ky}\right)+b_1(m_n)Y_1\left(\frac{m_n}{k}e^{ky}\right)\right]\qquad (n=1,2,\dots)\,,
\eea
where $f^{(0)}_\text{gauge}(y)$ exists only for $(++)$ BCs. The $f^{(n)}_\text{gauge}(y)$ satisfy the orthonormality condition
\be\label{eq:norm-gauge}
\frac{1}{L}\int_0^L dy\, f^{(n)}_\text{gauge}(y) f^{(m)}_\text{gauge}(y) = \delta_{nm}\,.
\ee
$b_1(m_n)$ and $m_n$ are determined through the boundary conditions on the branes. For $(++)$ fields, which means
\be
\partial_y f^{(n)}_\text{gauge}(y)\Big|_{y=0,L}=0\,,
\ee
one obtains \cite{Gherghetta:2000qt}
\be
b_1(m_n) = -\frac{J_1(m_n/k)+ m_n/k \,J'_1(m_n/k)}{Y_1(m_n/k)+ m_n/k\, Y'_1(m_n/k)} = b_1(m_ne^{kL})\,,
\ee
which can only be solved numerically for $m_n$ and $b_1(m_n)$. For large values of $n$, the result can be well approximated by \cite{Gherghetta:2000qt}
\be
b_1(m_n) = 0\,,\qquad m^\text{gauge}_n \simeq \left(n - \frac{1}{4}\right)\pi k e^{-kL}\qquad (n=1,2,\dots)\,,
\ee
however, for small values of $n$ it is safer to use the exact numerical result.
For $(-+)$ fields, meaning
\be
f^{(n)}_\text{gauge}(y)\Big|_{y=0} = \partial_y f^{(n)}_\text{gauge}(y)\Big|_{y=L}=0\,,
\ee
one finds instead
\be
b_1(m_n) = -\frac{J_1(m_n/k)}{Y_1(m_n/k)} =  -\frac{J_1(m_ne^{kL}/k)+ m_ne^{kL}/k \,J'_1(m_ne^{kL}/k)}{Y_1(m_ne^{kL}/k)+ m_ne^{kL}/k\, Y'_1(m_ne^{kL}/k)} \,.
\ee
The numerical solution yields a $\sim 2\%$ suppression of $m^\text{gauge}_1$ in that case, with respect to the $(++)$ one.
We do not consider gauge fields with a Dirichlet BC on the IR brane here, as they do not appear in our model.

Finally, $N_n$ has to be determined from the normalisation condition \eqref{eq:norm-gauge}. For fields (also fermions and scalars) with a Neumann BC on the IR brane, $N_n$ is approximately given by \cite{Gherghetta:2000qt}
\be
N_n\simeq \frac{e^{kL/2}}{\sqrt{\pi L m_n}}\,.
\ee
Note that this approximation is however {\it not} valid in case of a Dirichlet BC on the IR brane.

\subsection{Fermion Fields}

In this case the KK decomposition reads
\be
\psi_{L,R}(x^\mu,y) = \frac{e^{2ky}}{\sqrt{L}}\sum_{n=0}^\infty \psi_{L,R}^{(n)}(x^\mu)f^{(n)}_{L,R}(y)\,,
\ee
and the fermionic KK modes are \cite{Gherghetta:2000qt}
\bea\label{eq:ferm-zero-mode}
f^{(0)}_L(y) &=& \sqrt{\frac{(1-2c)kL}{e^{(1-2c)kL}-1}} e^{-cky}\,,\\
f^{(n)}_L(y) &=& \frac{e^{ky/2}}{N_n}\left[J_{\alpha}\left(\frac{m_n}{k}e^{ky}\right)+b_\alpha(m_n)Y_\alpha\left(\frac{m_n}{k}e^{ky}\right)\right]\qquad (n=1,2,\dots)\,,
\eea
where $\alpha = |c+1/2|$ and again $f^{(0)}_L(y)$ exists only for $(++)$ BCs for the left-handed mode. The right-handed mode obeys automatically opposite BCs and $f^{(n)}_R(y)$ can be obtained by replacing $c$ by $-c$ in the above formulae.
The $f^{(n)}_{L,R}(y)$ satisfy the orthonormality condition
\be\label{eq:norm-ferm}
\frac{1}{L}\int_0^L dy\, e^{ky} f^{(n)}_{L,R}(y) f^{(m)}_{L,R}(y) = \delta_{nm}\,.
\ee
Note that the fermionic zero mode profile in \eqref{eq:fermionprofile} has been given with respect to the flat tangent space metric, i.\,e. the factor $e^{ky}$ in \eqref{eq:norm-ferm} has been absorbed into the shape functions, in order to make the localisation of the zero mode more explicit.

Again $b_\alpha(m_n)$ and $m_n$ are determined through the BCs on the branes. 
In the case of left-handed fermions, a $-$ BC means
\be
f^{(n)}_L(y)\Big|_\text{brane}=0\,,
\ee
while the $+$ BC is modified with respect to the gauge fields and reads
\be
(\partial_y + ck) f^{(n)}_L(y)\Big|_\text{brane}=0\,.
\ee
For right-handed fields, the replacement $c\to -c$ has to be made.
$b_\alpha(m_n)$ and $m_n$ are derived completely analogously to the gauge case. Also here the resulting equations can only be solved numerically.

\subsection{Higgs Field}

A bulk Higgs field also needs to be KK expanded:
\be
H(x^\mu,y) = \frac{1}{\sqrt{L}}\sum_{n=0}^\infty H^{(n)}(x^\mu)f^{(n)}_H(y)\,.
\ee
As we do not specify the Higgs potential, we can not solve the bulk equations of motion explicitly for that case. Instead we merely assume the zero mode profile
\be
f^{(0)}_H(y)\equiv h(y)= \sqrt{2(\beta-1) kL}\,e^{kL}\, e^{\beta k (y-L)}\qquad (\beta\gg 1)\,,
\ee
which fulfils the normalisation condition
\be
\frac{1}{L}\int_0^L dy\,e^{-2ky} [h(y)]^2 =1\,.
\ee
As the scalar KK modes turn out to be much heavier than the gauge and fermionic resonances \cite{Gherghetta:2000qt},  they can usually be neglected in phenomenological analyses.

\newsection{Gauge Sector Modifications from EWSB}\label{app:EWSB}

In this Appendix we study explicitly the effects of EWSB on the weak gauge boson sector. We derive expressions for the mass matrices $\mathcal{M}^2_\text{charged}$ and $\mathcal{M}^2_\text{neutral}$ by treating the effects of EWSB as a small perturbation, and work out the rotation matrices $\mathcal{G}_W$ and $\mathcal{G}_Z$, neglecting the $n>1$ KK levels.

Subsequently, we compare our results with the ones obtained from the alternative approach followed in \cite{Huber:2000fh,Csaki:2002gy,Burdman:2002gr}. In that case, the effects of EWSB are already included in the derivation of the 5D equations of motion, so that in principle exact results may be obtained, as has briefly been noted in \cite{Csaki:2003dt}.

\subsection{Gauge Boson Mass Matrices and Mixings}

\subsubsection{Charged Electroweak Gauge Bosons}

The mass matrix $\mathcal{M}_\text{charged}^2$, describing the charged EW gauge bosons $W_L^{(0)\pm}$, $W_L^{(1)\pm}$ and $W_R^{(1)\pm}$ as defined in \eqref{eq:Mcharged}, can be determined from the Higgs kinetic term \eqref{eq:LHiggs}. One finds
\be\renewcommand{\arraystretch}{1.4}\addtolength{\arraycolsep}{3pt}
\mathcal{M}_\text{charged}^2=\left(\begin{array}{ccc}
\frac{g^2v^2}{4L} &  \frac{g^2v^2}{4L}\mathcal{I}^+_1 & -\frac{g^2v^2}{4L}
\mathcal{I}^-_1 \\
\frac{g^2v^2}{4L}\mathcal{I}^+_1 & M_{++}^2 + \frac{g^2v^2}{4L}\mathcal{I}^{++}_2 & -\frac{g^2v^2}{4L}\mathcal{I}^{-+}_2 \\
-\frac{g^2v^2}{4L}\mathcal{I}^-_1 & -\frac{g^2v^2}{4L}\mathcal{I}^{-+}_2 &
M_{--}^2 + \frac{g^2v^2}{4L}\mathcal{I}^{--}_2
\end{array}\right)\,,
\renewcommand{\arraystretch}{1.0}\addtolength{\arraycolsep}{-3pt}
\ee
where the overlap integrals $\mathcal{I}^\pm_1$ and $\mathcal{I}^{ij}_2$ are
given by 
\begin{gather}
\mathcal{I}_1^+ = \frac{1}{L} \int_0^L dy \,e^{-2ky} g(y) h(y)^2\,,\qquad
\mathcal{I}_1^- = \frac{1}{L} \int_0^L dy \,e^{-2ky} \tilde g(y) h(y)^2\,,\\
\mathcal{I}_2^{++} = \frac{1}{L} \int_0^L dy \,e^{-2ky} g(y)^2 h(y)^2\,,\qquad
\mathcal{I}_2^{--} = \frac{1}{L} \int_0^L dy \,e^{-2ky} \tilde g(y)^2 h(y)^2\,,\\
\mathcal{I}_2^{-+} = \frac{1}{L} \int_0^L dy \,e^{-2ky} g(y)\tilde g(y) h(y)^2\,.
\end{gather}
Here we introduced the short-hand notation
\be
g(y) = f^{(1)}_\text{gauge}(y,(++))
\ee
for the bulk shape function of $Z^{(1)}$ and $W_L^{(1)}$, as well as for the KK gluons $G^{(1)A}$ and photon $A^{(1)}$, and
\be
\tilde g(y) = f^{(1)}_\text{gauge}(y,(-+))
\ee
for the bulk shape function of $Z_X^{(1)}$ and $W_R^{(1)}$.

{In order to obtain transparent expressions for mass eigenvalues and mass
  eigenstates we introduce first the following parameterisation}
\be\label{eq:Ma}
M^2_{++}=M^2+av^2, \qquad M^2_{--}=M^2-av^2,
\ee
\be
\mathcal{I}^{--}_2=\mathcal{I}_2, \quad 
\mathcal{I}^{-+}_2=\mathcal{I}_2 \left(1+\delta^{-+}\frac{v^2}{f^2}\right), \quad
\mathcal{I}^{++}_2=\mathcal{I}_2 \left(1+\delta^{++}\frac{v^2}{f^2}\right),
\ee
{where numerically the parameter $a=\ord(1)$ for $f=\ord(1\tev)$ and the
  coefficients $\delta^{ij}$ turn out to be smaller than unity.

Now, our goal is to calculate $\ord(v^2/f^2)$ corrections to the couplings of
$W^\pm$ and $Z$ but only $\ord(1)$ couplings involving heavy gauge boson mass
eigenstates as their contributions in Feynman diagrams will be suppressed by
their large masses in the propagators. It turns out then that to this order in
$v^2/f^2$ the coefficients $\delta^{-+}$ and $\delta^{++}$ 
can be set to zero so that only a
universal $\mathcal{I}_2$ will enter the expressions below.

Next we introduce the function
\be\label{eq:Bpsi}
B(\zeta)=\sqrt{16 a^2 L^2   \cos^2\zeta+
8aLg^2\mathcal{I}_2 \sin^2\zeta +g^4\mathcal{I}^2_2\cos^2\zeta},
\ee
that will also be useful in the case of the diagonalisation of the neutral
EW gauge boson mass matrix. In the case of charged gauge bosons $B(\zeta=0)$ enters the expressions for masses and mixings, while in the case of neutral gauge bosons $B(\zeta=\psi)$ is relevant, where $\psi$ has been defined in \eqref{eq:psi}.}

 Diagonalising\footnote{We would
    like to thank Stefania Gori for help in finding an efficient method
    for analytic diagonalisation of the matrices involved and Zhi-zhong Xing for
    bringing the paper \cite{Kopp:2006wp} to our attention.} then $\mathcal{M}_\text{charged}^2$ leads to
\be\renewcommand{\arraystretch}{1.4}\addtolength{\arraycolsep}{3pt}
\mathcal{G}_W= \left(\begin{array}{ccc}
1 & -\frac{g^2v^2}{4LM^2} \mathcal{I}^+_1 & \frac{g^2v^2}{4LM^2}\mathcal{I}^-_1
 \\
\frac{g^2v^2}{4LM^2}
\left(\mathcal{I}^+_1\cos\chi-\mathcal{I}^-_1\sin\chi\right) & \cos\chi & \sin\chi \\
-\frac{g^2v^2}{4LM^2}
 \left(\mathcal{I}^+_1\sin\chi+\mathcal{I}^-_1\cos\chi\right) &
 -\sin\chi & \cos\chi \end{array}\right)\renewcommand{\arraystretch}{1.0}\addtolength{\arraycolsep}{-3pt}\,,
\ee
where
\be
\cos\chi=\sqrt{\frac{1}{2}-\frac{2aL}{B(0)}}, \qquad
\sin\chi=\sqrt{\frac{1}{2}+\frac{2aL}{B(0)}}\,.
\ee
The corresponding masses are given by
\bea
M^2_W &=& \frac{g^2v^2}{4L}-\frac{g^4v^4}{16L^2M^2}
\left((\mathcal{I}^+_1)^2 +(\mathcal{I}^-_1)^2\right)
\,,\\
M^2_{W_H} &=& M^2 + \frac{v^2}{4L}\left(g^2\mathcal{I}_2-B(0)\right)\,,\\
M_{W'}^2 &=&  M^2 + \frac{v^2}{4L}\left(g^2\mathcal{I}_2+B(0)\right)                 \,.
\eea

\subsubsection{Neutral Electroweak Gauge Bosons}

Also the mass matrix $\mathcal{M}_\text{neutral}^2$, describing the neutral EW gauge bosons $Z^{(0)}$, $Z^{(1)}$ and $Z_X^{(1)}$ as defined in \eqref{eq:Mneutral}, can be determined from the Higgs kinetic term \eqref{eq:LHiggs}. Here we obtain
\be\renewcommand{\arraystretch}{1.4}\addtolength{\arraycolsep}{3pt}
\mathcal{M}_\text{neutral}^2=\left(\begin{array}{ccc}
\frac{g^2v^2}{4L\cos^2\psi} & \frac{g^2v^2\mathcal{I}^+_1}{4L\cos^2\psi} & 
-\frac{g^2v^2\cos\phi\mathcal{I}^-_1}{4L\cos\psi} \\
\frac{g^2v^2\mathcal{I}^+_1}{4L\cos^2\psi} &
M_{++}^2+\frac{g^2v^2 \mathcal{I}^{++}_2}{4L\cos^2\psi} &
-\frac{g^2v^2\cos\phi\mathcal{I}^{-+}_2}{4L\cos\psi} \\
-\frac{g^2v^2\cos\phi\mathcal{I}^+_1}{4L\cos\psi} & 
-\frac{g^2v^2\cos\phi\mathcal{I}^{-+}_2}{4L\cos\psi} &
M_{--}^2+ \frac{g^2v^2\cos^2\phi\mathcal{I}^{--}_2}{4L}
\end{array}\right)\renewcommand{\arraystretch}{1.0}\addtolength{\arraycolsep}{-3pt}\,,
\ee
with the angles $\phi$ and $\psi$ given in \eqref{eq:phi}, \eqref{eq:psi}.
Diagonalisation of $\mathcal{M}_\text{neutral}^2$ gives then
\be\renewcommand{\arraystretch}{1.4}\addtolength{\arraycolsep}{3pt}
\mathcal{G}_Z= \left(\begin{array}{ccc}
1 & - \frac{g^2v^2\mathcal{I}^+_1}{4L M^2\cos^2\psi } & \frac{g^2v^2 \mathcal{I}^-_1 \cos\phi}{4L M^2\cos{\psi}} \\
\frac{g^2v^2}{4LM^2\cos^2\psi}
\left(\mathcal{I}^+_1\cos\xi-\cos\phi\cos\psi\mathcal{I}^-_1\sin\xi\right) &
\cos\xi & \sin\xi \\
-
\frac{g^2v^2}{4LM^2\cos^2\psi}
\left(\mathcal{I}^+_1\sin\xi+\cos\phi\cos\psi\mathcal{I}^-_1\cos\xi\right)  & -\sin\xi  & \cos\xi
\end{array}\right)\renewcommand{\arraystretch}{1.0}\addtolength{\arraycolsep}{-3pt}\,,
\ee
where
\be
\cos\xi=\sqrt{\frac{B(\psi)\cos\psi-4aL\cos^2\psi-\sin^2\psi g^2
  \mathcal{I}_2}{2B(\psi)\cos\psi}},
\ee

\be
\sin\xi=\sqrt{\frac{B(\psi)\cos\psi+4aL\cos^2\psi+\sin^2\psi g^2
  \mathcal{I}_2}{2B(\psi)\cos\psi}}.
\ee
$B(\psi)$ is given in \eqref{eq:Bpsi}.
The corresponding masses are given by 
\bea
M^2_Z &=& \frac{g^2v^2}{4L\cos^2\psi}-\frac{g^4v^4}{16L^2M^2\cos^2\psi}
\left(\frac{(\mathcal{I}^+_1)^2}{\cos^2\psi} +
(\mathcal{I}^-_1)^2\cos^2\phi\right)
\,,\\
M^2_{Z_H} &=& M^2 + 
\frac{v^2}{4L}\left(g^2\mathcal{I}_2-\frac{B(\psi)}{\cos\phi}\right)\,,\\
M_{Z'}^2 &=&  M^2 + \frac{v^2}{4L}\left(g^2\mathcal{I}_2+
\frac{B(\psi)}{\cos\phi}\right)     \,.
\eea
{Note that for $\psi=0$ the results for neutral gauge bosons reduce to
the ones for charged gauge bosons.}

\subsection{Comparison of two Alternative Approaches}\label{app:approaches}

In this Appendix we compare the approach of treating the effects of EWSB as a perturbation with the exact approach, in which EWSB is considered as a modification of the BCs on the IR brane. For a related study of this issue we refer the reader also to  \cite{Goertz:2008vr}.
We note that throughout this paper and also in our phenomenological analyses \cite{Blanke:2008zb,Blanke:2008yr,BDG} we followed the perturbative approach {which is analogous to the two-site approach presented in \cite{Contino:2006nn}.}

As a simple toy model of EWSB, we consider a $U(1)$ gauge symmetry in the RS background \eqref{eq:RS},
where a scalar field $H(x)$ resides on the {IR} brane and develops a VEV that breaks the bulk $U(1)$ symmetry. The action relevant for the gauge field is given by
\be
S_\text{gauge} = \int d^4 x \int_0^L dy\,\sqrt{G}\left[-\frac{1}{4}F_{MN}F^{MN} + \delta(y-L) (D_MH)^\dagger (D^MH)\right]\,,
\ee
where  $F_{MN}=\partial_MA_N-\partial_NA_M$, and $D_M=\partial_M+igA_M$. We choose to work in the $A_5=0$, $\partial_\mu A^\mu=0$ gauge in what follows. 

Once a potential $V(H)$ is added on the {IR} brane, the Higgs field develops a VEV
\be
\langle H(x)\rangle = \frac{v}{\sqrt{2}}\,.
\ee
Effectively, the action for the gauge field then reads
\be
S_\text{gauge} = \int d^4 x \int_0^L dy\,\sqrt{G}\left[-\frac{1}{4}F_{MN}F^{MN} + \delta(y-L)\frac{g^2v^2}{2}A_MA^M\right]\,.\label{eq:Seff}
\ee

Following the presentation in \cite{Huber:2000fh,Csaki:2002gy,Burdman:2002gr,Csaki:2003dt}, from \eqref{eq:Seff} we can now derive the bulk EOM for the gauge field and determine the shape functions and masses for the various KK modes. Finally we have to consider possible mixing between the various modes induced by the presence of the Higgs VEV.

The EOM can be derived from the variation principle for the action,
\bea
\delta S_\text{gauge} =0&=& 
\int d^4x\int_0^L dy\,\partial_R(\sqrt{G}G^{MR}G^{NS}\partial_MA_N)\delta A_S\nn\\
&&+\left.\int d^4x\,e^{-2kL}(\partial_yA_\nu + g^2v^2A_\nu)\delta A^\nu\right|_{y=L}-\left.\int d^4x\,(\partial_y A_\nu) \delta A^\nu\right|_{y=0}\,.\qquad
\eea
The bulk and boundary terms have to vanish independently of each other, so that we obtain
\bea
\text{bulk EOM:}&&\qquad \eta^{\mu\rho}\partial_\mu\partial_\rho A^\nu -\partial_y(e^{-2ky}\partial_yA^\nu)=0 \,,\label{eq:bulk-eom}\\
\text{UV brane:}&&\qquad \partial_y A_\nu\big|_{y=0}=0\,,\\
\text{IR brane:}&&\qquad \partial_y A_\nu+g^2v^2A_\nu\big|_{y=L}=0\label{eq:IR-eom}\,.
\eea

In order to understand the  effects of EWSB, we first solve \eqref{eq:bulk-eom}--\eqref{eq:IR-eom} explicitly for the zero mode by making an expansion in the small parameter $w=(m_0e^{kL}/k)^2$. We thus make the ansatz
\be\label{eq:f0-approx}
f^{(0)}(y)=1+w \,d(y)+\ord(w^2)\,.
\ee
From \eqref{eq:bulk-eom}--\eqref{eq:IR-eom} and the normalisation condition
\be
\frac{1}{L}\int_0^L dy\,f^{(0)}(y)^2=1\,,
\ee
we find 
\be
d(y) = \frac{1}{4}e^{2k(y-L)}(1-2ky)+1-\frac{1}{kL}\,,
\ee
and
\be
m_0^2 \simeq \frac{g^2}{L}v^2e^{-2kL}\equiv (g^\text{4D}v_\text{eff})^2\,.
\ee

The eigenfunctions for the massive modes (this can in principle also be used for $f^{(0)}$) are given as usual by \cite{Gherghetta:2000qt}
\be\label{eq:fn-exact}
f^{(n)}(y)=\frac{e^{ky}}{N_n}\left[J_1\left(\frac{m_n}{k}e^{ky}\right)+b_1(m_n) Y_1\left(\frac{m_n}{k}e^{ky}\right)\right]\,,
\ee
with
\be
b_1(m_n)=-\frac{J_1(\frac{m_n}{k})+\frac{m_n}{k}J'_1(\frac{m_n}{k})}{Y_1(\frac{m_n}{k})+\frac{m_n}{k}Y'_1(\frac{m_n}{k})}=-\frac{(1+r)J_1(\frac{m_n}{k}e^{kL})+\frac{m_n}{k}e^{kL}J'_1(\frac{m_n}{k}e^{kL})}{(1+r)Y_1(\frac{m_n}{k}e^{kL})+\frac{m_n}{k}e^{kL}Y'_1(\frac{m_n}{k}e^{kL})}\,,\quad r=\frac{g^2v^2}{k}\,.
\ee

Note that the EOM we have just solved, including the BCs, is a so-called linear boundary value problem of second order. For this kind of problem, it is well known that the eigenfunctions $f^{(n)}(y)$ obey the orthogonality relation
\be
\int_0^L dy\,f^{(n)}(y)f^{(m)}(y)=0\qquad\text{for }n\ne m\,.
\ee
As a direct consequence, there is {\it no mixing} between the zero and KK modes of the gauge boson. The modes we have just calculated are already the final mass eigenstates. The only effect of the spontaneous symmetry breaking within the present approach is thus the modified BC on the IR brane, resulting in shifts in the masses and a distortion of the eigenfunctions $f^{(n)}(y)$, in particular of $f^{(0)}(y)$, relative to the unbroken case.

Evaluating the mass spectrum numerically (without making any approximation or expansion in $m_0$), we find with $ke^{-kL}=1\tev,k=10^{16}\tev$,
\be
m_0=80.398\gev\,,\qquad m_1=2.55\tev\,,\qquad  m_2=5.61\tev\,,
\label{eq:m0-m4} \qquad \dots
\ee
where we have chosen $gv$ in order to satisfy $m_0=M_W$.

The question now arises whether the same results are obtained also by using our previous approach to EWSB, also followed e.\,g. by  Agashe {\it et al.} \cite{Agashe:2007ki}, where the effects of EWSB are treated as a perturbation of the unbroken case. In order to analyse this let us start from the same action $S_\text{gauge}$ as before, but solve the bulk EOM before inserting the Higgs VEV. Obviously the bulk shape functions will then be those of the unbroken case, with
\be
f^{(0)}(y)=1\,,\qquad m_0=0\,,
\ee
and
\be
f^{(n)}(y)=\frac{e^{ky}}{N_n}\left[J_1\left(\frac{m_n}{k}e^{ky}\right)+b_1(m_n) Y_1\left(\frac{m_n}{k}e^{ky}\right)\right]\,,
\ee
where
\be
b_1(m_n)=-\frac{J_1(\frac{m_n}{k})+\frac{m_n}{k}J'_1(\frac{m_n}{k})}{Y_1(\frac{m_n}{k})+\frac{m_n}{k}Y'_1(\frac{m_n}{k})}=b_1(m_ne^{kL})\,.
\ee

To determine the effects of EWSB, this solution for the gauge boson modes now has to be inserted into the interaction term with the Higgs, the latter being then replaced by its VEV. This yields
\bea
S_\text{gauge}&\ni&\int d^4x\int_0^Ldy\,\sqrt{G}\,\delta(y-L)\frac{g^2v^2}{2L}e^{2ky}\sum_{n,m}f^{(n)}(y)f^{(m)}(y)A_\mu^{(n)}(x)A^{\mu(m)}(x)\nn\\
&\simeq&\frac{g^2v^2e^{-2kL}}{2L}\int d^4x\,\Big[A_\mu^{(0)}(x)A^{\mu(0)}(x)\nn\\
&&\hspace{3.8cm}+\,2\sqrt{2kL}A_\mu^{(0)}(x)A^{\mu(1)}(x)\nn\\
&&\hspace{3.8cm}+\,{2kL}A_\mu^{(1)}(x)A^{\mu(1)}(x)+\cdots\Big]\,.
\eea

Approximating $f^{(n)}(y=L)\simeq \sqrt{2kL}$ for $n\ne0$, and denoting { $m^2=g^2v^2e^{-2kL}/L$}, we obtain the mass matrix
\be
\mathcal{M}^2=\begin{pmatrix}m^2 && m^2\sqrt{2kL}&&m^2\sqrt{2kL} && \cdots\\
m^2\sqrt{2kL} && M_1^2+m^2 2kL && m^2 2kL && \cdots \\
m^2\sqrt{2kL} && m^2 2kL && M_2^2+m^2 2kL && \cdots \\
\vdots && \vdots && \vdots && \ddots
\end{pmatrix}\,,
\ee
where $M_i$ are the KK masses in the absence of EWSB: 
\be
M_1 = 2.45\tev\,, \quad
M_2 = 5.56\tev\,,\quad \dots\,.
\ee
The off-diagonal entries now induce mixing between the zero and KK modes.

Including the modes up to $n=2$ and diagonalising $\mathcal{M}^2$, we find
\be
m_0=81.25\gev\,,\qquad m_1=2.55\tev\,,\qquad m_2=5.61\tev\,,
\ee
which differs from the result obtained from the previous approach, \eqref{eq:m0-m4} only for $m_0$, and even there the error amounts to only $1\%$.\footnote{This error should not be understood as an absolute error on $M_W$, that would be huge compared to the present experimental accuracy, but rather as a relative error which can be absorbed into the gauge KK masses by making a proper redefinition of the Higgs VEV $v$.} Note also that while the first approach yields exact results, the second approach contains the approximation of cutting the KK tower after the first few modes. In addition, since EWSB is treated as a perturbation, an expansion in $g$ is inherent.
We have also verified that the more modes are included, the smaller the errors in the mass determination become.

\begin{figure}
\center{\epsfig{file=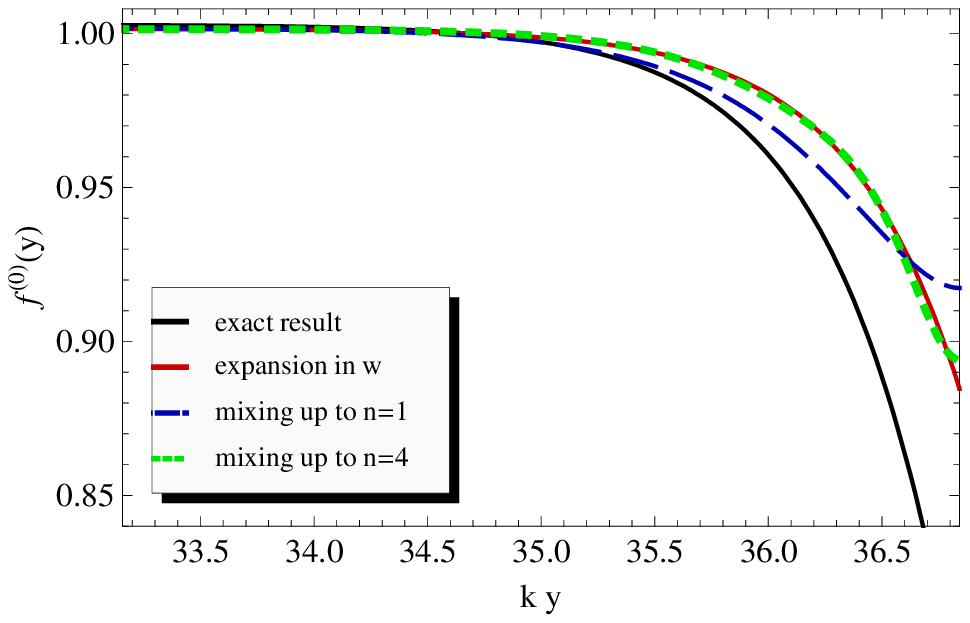,scale=.85}}
\caption{\label{fig:zeromode}\it Effective zero mode shape function $f^{(0)}(y)$ for $0.9L<y<L$ -- comparison of the exact result with different approximations. The solid black line corresponds to the exact result in \eqref{eq:f0-approx}, while the solid red line displays the expanded \eqref{eq:f0-approx}. The results obtained after diagonalisation of the mass matrix $\mathcal{M}^2$ are shown for $n\le1$ (blue, dashed line) and $n\le4$ (green, dotted line).}
\end{figure}

In addition to the mass spectrum, we have also considered the effective zero mode shape function that arises after mixing with the KK modes. By comparing numerically the results obtained by truncating the KK tower after the first few modes with the results in \eqref{eq:f0-approx} and \eqref{eq:fn-exact}, we observe that by increasing the number $n$ of modes included, the approximation quickly approaches the result of \eqref{eq:f0-approx}, see Fig.~\ref{fig:zeromode}. This confirms our expectation that inserting the effects of EWSB after performing the KK expansion effectively amounts to treating EWSB as a perturbation.

In summary, we have studied two quite distinct descriptions of EWSB effects:
\begin{enumerate}
\item {Modified EOM and wave function distortion, see e.\,g. \cite{Huber:2000fh,Csaki:2002gy,Burdman:2002gr,Csaki:2003dt,Casagrande:2008hr}.}
\bi
\item The Higgs VEV is included already for the derivation and solution of the bulk EOM.
\item This implies distortions of the wave functions and shifts in the masses of the zero and KK modes, with respect to the unbroken case.
\item Due to the orthogonality of the bulk wave functions, there is no mixing between the various modes.
\ei
\item {EWSB as a perturbation in the KK tower, see e.\,g. \cite{Agashe:2007ki,Muck:2001yv}.}
\bi
\item The bulk EOM is solved for free fields, i.\,e. the Higgs VEV is not yet taken into account and treated later as a perturbation.
\item Then the unbroken wave functions are used to calculate the effects of EWSB by replacing $H$ by its VEV.
\item
Mixing between the various modes appears, in addition to corrections to the masses.
\ei
 \end{enumerate}
We have shown that both approaches are indeed equivalent, up to the fact that the second one works only if EWSB is a small perturbation of the unbroken theory, and that exact results can in principle only be obtained by considering the whole infinite tower of KK modes. Still we have checked numerically that the results converge quickly so that it appears sufficient to take into account only the zero and the first excited modes.

A similar proof, including also the case of spontaneously broken non-abelian gauge groups, can also be found in \cite{Goertz:2008vr}.

\newsection{Effective 4D Feynman Rules}\label{app:FR}

\subsection{Preliminaries}
In this Section we list the complete set of Feynman rules in terms of gauge boson mass
eigenstates and fermion flavour eigenstates. In view of the large size of the fermion mass 
matrices in Section \ref{sec:qmm}, the rotation to fermion mass eigenstates is {best done}
numerically. The collection contains all $v^2/f^2 $
contributions for the light gauge bosons $Z$ and $W^+$, while we present only vertices 
of $\mathcal{O}(1)$ for the heavy gauge bosons $G^{(1)A}$, $A^{(1)}$, $Z_H$, $Z^{\prime}$, $W_H^{\pm}$ and 
$W^{\prime \pm}$. Note that $v^2/f^2 \sim v^2/M^2 \sim \epsilon$, where $\epsilon$ is defined in
(\ref{eq:epsilon}). There are no corrections to gluon and photon couplings.

{
\subsection{Propagators}

Our conventions for fermion and gauge boson propagators in the unitary gauge are as follows:
\bea
\begin{picture}(50,25) (70,25)
    \SetWidth{0.5}
    \SetColor{Black}
    \Text(72,32)[lb]{{\Black{$p$}}}
    \ArrowLine(50,27)(100,27)
   \Vertex(50,27){1.41}
    \Vertex(100,27){1.41}
  \end{picture}
\qquad &:&
\qquad \frac{i}{\slash p-m_f}\,,\\
\begin{picture}(50,25) (70,25)
    \SetWidth{0.5}
    \SetColor{Black}
    \Text(72,32)[lb]{{\Black{$p$}}}
    \Photon(50,27)(100,27){2}{7}
   \Vertex(50,27){1.41}
    \Vertex(100,27){1.41}
\Text(40,28)[lb]{{\Black{$\mu$}}}
\Text(104,30)[lb]{{\Black{$\nu$}}}
\Text(40,20)[lb]{{\Black{$a$}}}
\Text(104,20)[lb]{{\Black{$b$}}}
  \end{picture}
\qquad &:&
\qquad \frac{-i\delta_{ab}}{ p^2-M^2}\left[\eta^{\mu\nu}-\frac{p^\mu
    p^\nu}{M^2}\right]\,,
\eea
where $m_f$ and $M$ are the fermion and gauge boson mass, respectively.
}

\subsection{Overlap Integrals}

The overlap integrals for KK gluonic and photonic currents and for the ones for the KK modes $Z^{(1)}$ and $W^{(1)}_L$ are given by 
\begin{equation}\label{eq:intR}
\underset{nm}{\mathcal{R}_k^i}(BC)_{L,R} = \frac{1}{L} \int_0^L dy \, e^{ky}  f_{L,R}^{(n)}(y,c_k^i,BC)
 f_{L,R}^{(m)}(y,c_k^i,BC) \, g(y)\,,
\end{equation}
while for $Z_X^{(1)}$ we have
\begin{equation}
\underset{nm}{\mathcal{P}_k^i}(BC)_{L,R}  = \frac{1}{L} \int_0^L dy \, e^{ky}  f_{L,R}^{(n)}(y,c_k^i,BC)
 f_{L,R}^{(m)}(y,c_k^i,BC) \,\tilde g(y)
\label{overlapR3}
\end{equation}
with $\tilde g(y) \neq g(y)$ as the shape functions depend weakly on BCs.
For charged currents mediated by $W_R^{(1)}$ we also have 
\begin{equation}\label{eq:intS}
\underset{nm}{\mathcal{S}_k^i}(BC)(\widetilde{BC})_{L,R}  = \frac{1}{L} \int_0^L dy \, e^{ky}  f_{L,R}^{(n)}(y,c_k^i,BC)
 f_{L,R}^{(m)}(y,c_k^i,\widetilde{BC})\, \tilde g(y)\,.
\end{equation}

\subsection{Fermion Couplings to Gluon and KK Gluon}

All couplings to the zero mode gluons are SM-like. That is the relevant term
in the QCD Lagrangian is
\begin{equation}
-\frac{g_s}{\sqrt{L}} \bar{\psi} \gamma^{\mu}\, t^A \psi G^{(0)A}_{\mu} \, ,
\end{equation}
where we suppressed the colour indices.

In Tables \ref{tab:zero-KKgluon}--\ref{tab:KKfermion-KKgluon} we list all non-vanishing couplings to the KK gluon. {The  last} column in these tables denotes the entry  in the coupling matrices defined in Section
\ref{sec:currents}.

\begin{table}%[htbp]
\renewcommand{\arraystretch}{1.5}
\begin{center}
\begin{tabular}{|c|c|c|} 
\hline
\multicolumn{3}{|c|}{\bf\boldmath Zero mode couplings to the KK gluon}\\ 
\hline\hline
\multicolumn{3}{|l|}{$Q=2/3$ quarks} \\\hline
 
 $  \bar q_L^{u_i(0)} q_L^{u_i(0)} G^{(1)A} $ &
$- i \frac{g_s}{\sqrt{L}} \gamma^{\mu}\, t^A 
\underset{00}{\mathcal{R}_1^i}(++)_L $ & (0,0) \\\hline
 $  \bar  u_R^{i(0)} u_R^{i(0)} G^{(1)A} $ &
$- i \frac{g_s}{\sqrt{L}} \gamma^{\mu}\, t^A 
\underset{00}{\mathcal{R}_2^i}(++)_R$ &  (0,0)\\\hline
\multicolumn{3}{|l|}{$Q=-1/3$ quarks} \\\hline %zero mode
 $  \bar  q_L^{d_i(0)} q_L^{d_i(0)} G^{(1)A} $ &
$- i \frac{g_s}{\sqrt{L}} \gamma^{\mu}\, t^A \underset{00}{\mathcal{R}_1^i}(++)_L $ &  (0,0)  
\\\hline
 $  \bar D_R^{i(0)} D_R^{i(0)} G^{(1)A} $ &
$- i \frac{g_s}{\sqrt{L}} \gamma^{\mu}\, t^A 
\underset{00}{\mathcal{R}_3^i}(++)_R$  & (0,0)
\\\hline
\end{tabular} 
\end{center}
\renewcommand{\arraystretch}{1.0}
\caption{\it Vertices involving left-handed and
 right-handed zero modes and the KK gluon. These zero modes
correspond  to the SM quark fields when the rotation to fermion mass eigenstates is performed.\label{tab:zero-KKgluon}}
\end{table}

\begin{table}[htbp]
\renewcommand{\arraystretch}{1.5}
\begin{center}
\begin{tabular}{|c|c|c|} 
 \hline
\multicolumn{3}{|c|}{\bf\boldmath Off-diagonal couplings to the KK gluon}\\ 
\hline\hline
\multicolumn{3}{|l|}{$Q=2/3$ quarks} \\\hline
$\bar q_L^{u_i(0)} q_L^{u_i} G^{(1)A} $ &
$ - i \frac{g_s}{\sqrt{L}} \gamma^{\mu}\, t^A  \underset{01}{\mathcal{R}_1^i}(++)_L$ & (0,1) \\\hline
 $\bar q_L^{u_i} q_L^{u_i(0)} G^{(1)A} $ &
$- i \frac{g_s}{\sqrt{L}} \gamma^{\mu}\, t^A  \underset{10}{\mathcal{R}_1^i}(++)_L$ & (1,0) \\\hline
 $ \bar  u_R^{i(0)} u_R^{i} G^{(1)A} $&
$ -i \frac{g_s}{\sqrt{L}} \gamma^{\mu}\, t^A  \underset{01}{\mathcal{R}_2^i}(++)_R$ & (0,5) \\\hline
 $  \bar  u_R^{i} u_R^{i(0)} G^{(1)A}$ &
$ -i \frac{g_s}{\sqrt{L}} \gamma^{\mu}\, t^A  \underset{10}{\mathcal{R}_2^i}(++)_R $  & (5,0)  \\\hline
\multicolumn{3}{|l|}{$Q=-1/3$ quarks} \\\hline
 $\bar  q_L^{d_i(0)} q_L^{d_i} G^{(1)A}$ &
$ -i \frac{g_s}{\sqrt{L}} \gamma^{\mu}\, t^A  \underset{01}{\mathcal{R}_1^i}(++)_L$& (0,1) \\\hline
 $  \bar  q_L^{d_i} q_L^{d_i(0)} G^{(1)A} $&
$- i \frac{g_s}{\sqrt{L}} \gamma^{\mu}\, t^A  \underset{10}{\mathcal{R}_1^i}(++)_L$ & (1,0)\\\hline
 $  \bar D_R^{i(0)} D_R^{i} G^{(1)A} $ &
$ - i \frac{g_s}{\sqrt{L}} \gamma^{\mu}\, t^A   \underset{01}{\mathcal{R}_3^i}(++)_R $  & (0,3)  \\\hline
 $ \bar D_R^{i} D_R^{i(0)} G^{(1)A} $ &
$ -i \frac{g_s}{\sqrt{L}} \gamma^{\mu}\, t^A  \underset{10}{\mathcal{R}_3^i}(++)_R $   & (3,0)
\\\hline
\end{tabular} 
\end{center}
\renewcommand{\arraystretch}{1.0}
\caption{\it Vertices involving the KK gluon and a single zero mode. 
In terms of the matrices given in Section \ref{se:KKphot-glu}, we are talking about the 
off-diagonal elements.}
\end{table}

\begin{table}[htbp]
\renewcommand{\arraystretch}{1.5}
\begin{center}
\begin{tabular}{|c|c|c|} 
\hline
\multicolumn{3}{|c|}{\bf\boldmath Heavy fermion couplings to the KK gluon}\\ 
\hline\hline
\multicolumn{3}{|l|}{$Q=5/3$ quarks} \\\hline
 $ \bar \chi^{u_i}_L  \chi^{u_i}_L G^{(1)A}$ &
$ - i \frac{g_s}{\sqrt{L}} \gamma^{\mu}\, t^A  \underset{11}{\mathcal{R}_1^i}(-+)_L $ & (1,1) \\\hline 
$ \bar \psi^{\prime i}_L \psi^{\prime i}_L G^{(1)A}$ &
 $-  i \frac{g_s}{\sqrt{L}} \gamma^{\mu}\, t^A  \underset{11}{\mathcal{R}_3^i}(+-)_L  $ & (2,2)  \\\hline 
$ \bar \psi^{\prime\prime i}_L  \psi^{\prime\prime i}_L G^{(1)A}$ &  
$ -i \frac{g_s}{\sqrt{L}} \gamma^{\mu}\, t^A \underset{11}{\mathcal{R}_3^i}(+-)_L  $ & (3,3)  \\\hline
\multicolumn{3}{|l|}{$Q=2/3$ quarks} \\\hline
 $  \bar q_L^{u_i} q_L^{u_i} G^{(1)A}$ &
$ - i \frac{g_s}{\sqrt{L}} \gamma^{\mu}\, t^A  \underset{11}{\mathcal{R}_1^i}(++)_L $&  (1,1)  \\\hline
$ \bar U_L^{\prime i} U_L^{\prime i} G^{(1)A}$ &
$ -i \frac{g_s}{\sqrt{L}} \gamma^{\mu}\, t^A  \underset{11}{\mathcal{R}_3^i}(+-)_L$  & (2,2) \\\hline
$  \bar U_L^{\prime \prime i} U_L^{\prime\prime i} G^{(1)A}$ &
$ - i \frac{g_s}{\sqrt{L}} \gamma^{\mu}\, t^A   \underset{11}{\mathcal{R}_3^i}(+-)_L$ &  (3,3)  \\\hline
$  \bar \chi^{d_i}_L \chi^{d_i}_L G^{(1)A}$ &
$ -i \frac{g_s}{\sqrt{L}} \gamma^{\mu}\, t^A  \underset{11}{\mathcal{R}_1^i}(-+)_L$   &(4,4) \\\hline
$  \bar u_L^i  u_L^i G^{(1)A}$ &
$-  i \frac{g_s}{\sqrt{L}} \gamma^{\mu}\, t^A   \underset{11}{\mathcal{R}_2^i}(--)_L$ & (5,5)   \\\hline
\multicolumn{3}{|l|}{$Q=-1/3$ quarks} \\\hline
$  \bar  q_L^{d_i} q_L^{d_i} G^{(1)A}$ &
$  -i \frac{g_s}{\sqrt{L}} \gamma^{\mu}\, t^A  \underset{11}{\mathcal{R}_1^i}(++)_L$  & (1,1)  \\\hline
$  \bar D_L^{\prime i} D_L^{\prime i} G^{(1)A}  $ &
$ - i \frac{g_s}{\sqrt{L}} \gamma^{\mu}\, t^A  \underset{11}{\mathcal{R}_3^i}(+-)_L$ &  (2,2)  \\\hline
$  \bar D_L^{i} D_L^{i} G^{(1)A}$ &
$ - i \frac{g_s}{\sqrt{L}} \gamma^{\mu}\, t^A  \underset{11}{\mathcal{R}_3^i}(--)_L$ &  (3,3)
\\\hline
\end{tabular} 
\end{center}
\renewcommand{\arraystretch}{1.0}
\caption{\it Couplings of the KK gluon 
to heavy left-handed fermions. Note that for 
each entry in this table there exists a coupling to the corresponding right-handed 
fermion fields. The translation from left-handed to right-handed vertices is given in Table
\ref{tab:scheme1}.\label{tab:KKfermion-KKgluon}}
\end{table}

Diagonal couplings of right-handed heavy fermion fields can be easily obtained from the 
left-handed ones by replacing the boundary conditions according to the scheme given in 
 Table \ref{tab:scheme1}. In addition the index $L$ has to be replaced by $R$ in the overlap integrals in \eqref{eq:intR}--\eqref{eq:intS}.
These replacements are also valid for the heavy neutral gauge bosons $Z$, $Z_H$
 and $Z^{\prime}$ discussed below.

\begin{table}[htbp]
\renewcommand{\arraystretch}{1.5}
\begin{center}
\begin{tabular}{|lcc|}
\hline 
$L$ & $ \to $& $R$ \\
\hline 
$+$& $ \to $& $-$\\
$-$& $ \to $& $+$\\
\hline 
\end{tabular} 
\end{center}
\renewcommand{\arraystretch}{1.0}
\caption{\it Substitution scheme from  heavy left-handed to  heavy right-handed fermions. \label{tab:scheme1}}
\end{table}

%\afterpage\clearpage

\subsection{Fermion Couplings to Photon and KK Photon}

The couplings to photon and KK photon can be read off from the results for gluon and KK gluon 
with a simple modification: $t^A$ has to be removed and the coupling $g_s$ must be replaced by 
the coupling $g_Q$ as the photon and KK photon couple through electric charge. In particular we have
\begin{eqnarray}
g_{5/3 }&= &g_X Q_X \cos\phi \cos\psi + g \sin\psi = \frac{5}{3} \,e \,,\label{eq:g53}\\
g_{2/3 }&=&g_X Q_X \cos\phi \cos\psi = \frac{2}{3} \,e \,,\\
g_{-1/3 }&=& g_X Q_X \cos\phi \cos\psi - g \sin\psi =-\frac{1}{3} \,e \,,\label{eq:g13}
\end{eqnarray}
where we defined $e=g \sin\psi= g_X \cos\phi  \cos\psi$ with $e$ being the 5D
coupling:
\begin{equation}
e^{4D}=\frac{e}{\sqrt{L}}\,.
\end{equation}

%\afterpage\clearpage

\subsection{Fermion Couplings to Neutral Gauge Bosons}

We introduce the small parameter
\begin{equation}
\epsilon = \frac{g^2 v^2}{4 L M^2}\, ,
\label{eq:epsilon}
\end{equation}
where $M$ is defined in \eqref{eq:Ma},  and the couplings 
\begin{equation}\label{eq:gZ}
g_Z(\Psi) = \frac{g}{\sqrt{L}\cos\psi} \left(T^3_L - (\sin\psi)^2\, Q \right)\,,
\end{equation}
\bea
\kappa_1 &=& g_{Z_X}(q^{u_i}) = g_{Z_X}(q^{d_i}) 
=  \frac{1}{\sqrt{L}} \left(-g_X Q_X \sin\phi  - \frac{1}{2} g \cos\phi  \right)  \,,\\
\kappa_2 &=& g_{Z_X}(\chi^{u_i})= g_{Z_X}(\chi^{d_i}) 
=\frac{1}{\sqrt{L}} \left(-g_X Q_X \sin\phi  + \frac{1}{2} g \cos\phi  \right)  \,,\\
\kappa_3 &=& g_{Z_X}(u^i) =g_{Z_X}(\psi^{\prime i}) =  g_{Z_X}(U^{\prime i})
 = g_{Z_X}(D^{\prime i})  =  g_{Z_X}(U^{\prime\prime i}) \\
&=& \frac{1}{\sqrt{L}} \left(-g_X Q_X \sin\phi  \right) \,,\\
\kappa_4 &=& g_{Z_X}(\psi^{\prime \prime i})
= \frac{1}{\sqrt{L}} \left(-g_X Q_X \sin\phi  + g \cos\phi  \right)  \,,\\
\kappa_5 &=& g_{Z_X}( D^{ i})
= \frac{1}{\sqrt{L}} \left(-g_X Q_X \sin\phi- g \cos\phi  \right) \,.\label{eq:kappa5}
\eea
Furthermore, the expressions for $\kappa_i$ ($i=1,\dots,5$) can be combined in the following formula:
\begin{equation}
\kappa= \frac{1}{\sqrt{L}} \left(T_R^3 -(T_R^3+ Q_X) \sin^2\phi \right) \frac{g}{\cos\phi}  \,,
\end{equation}
{where $T^3_R$ is the $SU(2)_R$-isospin of the fermion related to a given $\kappa_i$, and $Q_X=2/3$ for quarks, while $Q_X=0$ for leptons, as discussed in Section \ref{sec:leptons}.}

\begin{table}[htbp]
\renewcommand{\arraystretch}{1.5}
\begin{center}
\begin{tabular}{|c|c|c|} 
 \hline
\multicolumn{3}{|c|}{\bf\boldmath Zero mode couplings to the $Z$ boson}\\ 
\hline\hline
\multicolumn{3}{|l|}{$Q=2/3$ quarks} \\\hline
$   \bar q_L^{u_i(0)} q_L^{u_i(0)} Z $ &
$ -i \gamma^{\mu} \Big[
 g_Z(q^{u_i}) -  \epsilon  g_Z(q^{u_i}) \frac{1}{\cos^2\psi}\mathcal{I}_1^+ 
\underset{00}{\mathcal{R}_1^i}(++)_L  +  \epsilon \frac{\cos\phi}{\cos\psi} \mathcal{I}_1^- \kappa_1 
\underset{00}{\mathcal{P}_1^i}(++)_L \Big]$ & (0,0)\\\hline
 $   \bar  u_R^{i(0)} u_R^{i(0)} Z $ &
$-i \gamma^{\mu} \Big[  g_Z(u^i) - \epsilon  g_Z(u^i)
 \frac{1}{\cos^2\psi}\mathcal{I}_1^+ 
\underset{00}{\mathcal{R}_2^i}(++)_R + \epsilon \frac{\cos\phi}{\cos\psi} \mathcal{I}_1^- \kappa_3 
\underset{00}{\mathcal{P}_2^i}(++)_R  \Big]$ & (0,0) \\\hline
\multicolumn{3}{|l|}{$Q=-1/3$ quarks} \\\hline
 $   \bar  q_L^{d_i(0)} q_L^{d_i(0)}  Z$ &
$-i \gamma^{\mu} \Big[  g_Z(q^{d_i}) - \epsilon  g_Z(q^{d_i})
 \frac{1}{\cos^2\psi}\mathcal{I}_1^+ 
\underset{00}{\mathcal{R}_1^i}(++)_L\
+ \epsilon \frac{\cos\phi}{\cos\psi} \mathcal{I}_1^- \kappa_1 
\underset{00}{\mathcal{P}_1^i}(++)_L  \Big]$  & (0,0) \\\hline
 $   \bar D_R^{i(0)} D_R^{i(0)} Z $ &
$-i \gamma^{\mu} \Big[  g_Z(D^{i}) - \epsilon  g_Z(D^{i})
 \frac{1}{\cos^2\psi}\mathcal{I}_1^+ 
\underset{00}{\mathcal{R}_3^i}(++)_R
+ \epsilon \frac{\cos\phi}{\cos\psi} \mathcal{I}_1^- \kappa_5
\underset{00}{\mathcal{P}_3^i}(++)_R \Big]$ & (0,0)
\\\hline
\end{tabular} 
\end{center}
\renewcommand{\arraystretch}{1.0}
\caption{\it Couplings involving zero modes 
and the $Z$ boson. These zero modes correspond  to the SM quark field when the rotation to fermion mass eigenstates is performed.}
\end{table}

\begin{table}[htbp]
\renewcommand{\arraystretch}{1.5}
\begin{center}
\begin{tabular}{|c|c|c|} 
\hline
\multicolumn{3}{|c|}{\bf\boldmath Off-diagonal couplings to the $Z$ boson}\\ 
\hline\hline
\multicolumn{3}{|l|}{$Q=2/3$ quarks} \\\hline
 $ \bar q_L^{u_i(0)} q_L^{u_i}  Z $ &
$-i \gamma^{\mu} \Big[  - \epsilon  g_Z(q^{u_i}) \frac{1}{\cos^2\psi}\mathcal{I}_1^+ 
\underset{01}{\mathcal{R}_1^i}(++)_L\
  + \epsilon \frac{\cos\phi}{\cos\psi} \mathcal{I}_1^- \kappa_1 
\underset{01}{\mathcal{P}_1^i}(++)_L \Big]$  & (0,1) \\\hline
 $\bar q_L^{u_i} q_L^{u_i(0)} Z $ &
$-i \gamma^{\mu} \Big[ - \epsilon  g_Z(q^{u_i}) \frac{1}{\cos^2\psi}\mathcal{I}_1^+ 
\underset{10}{\mathcal{R}_1^i}(++)_L\
 + \epsilon \frac{\cos\phi}{\cos\psi} \mathcal{I}_1^- \kappa_1 
\underset{10}{\mathcal{P}_1^i}(++)_L \Big]$  & (1,0) \\\hline
 $  \bar  u_R^{i(0)} u_R^{i} Z $ &
$-i \gamma^{\mu} \Big[  - \epsilon  g_Z(u^i)
 \frac{1}{\cos^2\psi}\mathcal{I}_1^+ 
\underset{01}{\mathcal{R}_2^i}(++)_R
+ \epsilon \frac{\cos\phi}{\cos\psi} \mathcal{I}_1^- \kappa_3 
\underset{01}{\mathcal{P}_2^i}(++)_R \Big]$  & (0,5) \\\hline
 $  \bar  u_R^{i} u_R^{i(0)} Z$ &
$-i \gamma^{\mu} \Big[  - \epsilon  g_Z(u^i)
 \frac{1}{\cos^2\psi}\mathcal{I}_1^+ 
\underset{10}{\mathcal{R}_2^i}(++)_R 
 + \epsilon \frac{\cos\phi}{\cos\psi} \mathcal{I}_1^- \kappa_3 
\underset{10}{\mathcal{P}_2^i}(++)_R\Big] $  &  (5,0) \\\hline
\multicolumn{3}{|l|}{$Q=-1/3$ quarks} \\\hline
$ \bar  q_L^{d_i(0)} q_L^{d_i} Z $ &
$-i \gamma^{\mu} \Big[  - \epsilon  g_Z(q^{d_i})
 \frac{1}{\cos^2\psi}\mathcal{I}_1^+ 
\underset{01}{\mathcal{R}_1^i}(++)_L
+ \epsilon \frac{\cos\phi}{\cos\psi} \mathcal{I}_1^- \kappa_1 
\underset{01}{\mathcal{P}_1^i}(++)_L\Big] $ & (0,1) \\\hline
 $ \bar  q_L^{d_i} q_L^{d_i(0)} Z $ &
$-i \gamma^{\mu} \Big[   - \epsilon  g_Z(q^{d_i})
 \frac{1}{\cos^2\psi}\mathcal{I}_1^+ 
\underset{10}{\mathcal{R}_1^i}(++)_L + \epsilon \frac{\cos\phi}{\cos\psi} \mathcal{I}_1^- \kappa_1 
\underset{10}{\mathcal{P}_1^i}(++)_L \Big]$ & (1,0) \\\hline
% offdiagonal RH
 $ \bar D_R^{i(0)} D_R^{i} Z $ &
$-i \gamma^{\mu} \Big[  - \epsilon  g_Z(D^{i})
 \frac{1}{\cos^2\psi}\mathcal{I}_1^+ 
\underset{01}{\mathcal{R}_3^i}(++)_R 
 + \epsilon \frac{\cos\phi}{\cos\psi} \mathcal{I}_1^- \kappa_5
\underset{01}{\mathcal{P}_3^i}(++)_R\Big] $  &  (0,3) \\\hline
% offdiagonal RH
 $  \bar D_R^{i} D_R^{i(0)} Z$ &
$-i \gamma^{\mu} \Big[  - \epsilon  g_Z(D^{i})
 \frac{1}{\cos^2\psi}\mathcal{I}_1^+ 
\underset{10}{\mathcal{R}_3^i}(++)_R
+ \epsilon \frac{\cos\phi}{\cos\psi} \mathcal{I}_1^- \kappa_5
\underset{10}{\mathcal{P}_3^i}(++)_R\Big] $   & (3,0)
\\\hline
\end{tabular} 
\end{center}
\renewcommand{\arraystretch}{1.0}
\caption{\it Couplings involving the $Z$ boson and a single zero mode.}
\end{table}

\begin{table}[htbp]
\renewcommand{\arraystretch}{1.5}
\begin{center}
\begin{tabular}{|c|c|c|} 
 \hline
\multicolumn{3}{|c|}{\bf\boldmath Heavy fermion couplings to the $Z$ boson}\\ 
\hline\hline
\multicolumn{3}{|l|}{$Q=5/3$ quarks} \\\hline
 $ \bar \chi^{u_i}_L  \chi^{u_i}_L  Z$ &
$-i \gamma^{\mu} \Big[  g_Z(\chi^{u_i}) - \epsilon  g_Z(\chi^{u_i}) \frac{1}{\cos^2\psi}\mathcal{I}_1^+ 
\underset{11}{\mathcal{R}_1^i}(-+)_L
+ \epsilon \frac{\cos\phi}{\cos\psi} \mathcal{I}_1^- \kappa_2 
\underset{11}{\mathcal{P}_1^i}(-+)_L \Big]$ & (1,1) \\\hline 
 $ \bar \psi^{\prime i}_L \psi^{\prime i}_L  Z$ &
 $-i \gamma^{\mu} \Big[ g_Z( \psi^{\prime i})  
- \epsilon  g_Z(\psi^{\prime i}) \frac{1}{\cos^2\psi}\mathcal{I}_1^+
\underset{11}{\mathcal{R}_3^i}(+-)_L 
+ \epsilon \frac{\cos\phi}{\cos\psi} \mathcal{I}_1^- \kappa_3 
\underset{11}{\mathcal{P}_3^i}(+-)_L \Big] $ & (2,2)  \\\hline 
 $ \bar \psi^{\prime\prime i}_L  \psi^{\prime\prime i}_L Z  $ &  
$-i \gamma^{\mu} \Big[ g_Z( \psi^{\prime\prime i})  -
 \epsilon  g_Z(\psi^{\prime\prime i}) \frac{1}{\cos^2\psi}
 \mathcal{I}_1^+ \underset{11}{\mathcal{R}_3^i}(+-)_L  
+ \epsilon \frac{\cos\phi}{\cos\psi} \mathcal{I}_1^- \kappa_4
 \underset{11}{\mathcal{P}_3^i}(+-)_L \Big] $  & (3,3) \\\hline
\multicolumn{3}{|l|}{$Q=2/3$ quarks} \\\hline
 $ \bar q_L^{u_i} q_L^{u_i}  Z$ &
$-i \gamma^{\mu} \Big[ g_Z(q^{u_i}) - \epsilon  g_Z(q^{u_i})
 \frac{1}{\cos^2\psi}\mathcal{I}_1^+ \underset{11}{\mathcal{R}_1^i}(++)_L
 + \epsilon \frac{\cos\phi}{\cos\psi} \mathcal{I}_1^- \kappa_1 
\underset{11}{\mathcal{P}_1^i}(++)_L \Big]$  & (1,1) \\\hline
 $ \bar U_L^{\prime i} U_L^{\prime i}  Z$ &
$-i \gamma^{\mu} \Big[ g_Z(U^{\prime i}) - \epsilon  g_Z(U^{\prime i}) 
\frac{1}{\cos^2\psi}\mathcal{I}_1^+ \underset{11}{\mathcal{R}_3^i}(+-)_L
 + \epsilon \frac{\cos\phi}{\cos\psi} \mathcal{I}_1^- \kappa_3 
\underset{11}{\mathcal{P}_3^i}(+-)_L \Big]$  & (2,2) \\\hline
 $  \bar U_L^{\prime \prime i} U_L^{\prime\prime i} Z$ &
$-i \gamma^{\mu} \Big[ g_Z(U^{\prime\prime i}) - \epsilon  g_Z(U^{\prime\prime i})
 \frac{1}{\cos^2\psi}\mathcal{I}_1^+ \underset{11}{\mathcal{R}_3^i}(+-)_L
 + \epsilon \frac{\cos\phi}{\cos\psi} \mathcal{I}_1^- \kappa_3 
\underset{11}{\mathcal{P}_3^i}(+-)_L \Big]$  & (3,3)  \\\hline
 $  \bar \chi^{d_i}_L \chi^{d_i}_LZ $ &
$-i \gamma^{\mu} \Big[ g_Z(\chi^{d_i}) - \epsilon  g_Z(\chi^{d_i})
 \frac{1}{\cos^2\psi}\mathcal{I}_1^+ \underset{11}{\mathcal{R}_1^i}(-+)_L
 + \epsilon \frac{\cos\phi}{\cos\psi} \mathcal{I}_1^- \kappa_2 
\underset{11}{\mathcal{P}_1^i}(-+)_L \Big]$  & (4,4) \\\hline
 $ \bar u_L^i  u_L^i  Z$ &
$-i \gamma^{\mu} \Big[ g_Z(u^i) - \epsilon  g_Z(u^i)
 \frac{1}{\cos^2\psi}\mathcal{I}_1^+ \underset{11}{\mathcal{R}_2^i}(--)
_L + \epsilon \frac{\cos\phi}{\cos\psi} \mathcal{I}_1^- \kappa_3 
\underset{11}{\mathcal{P}_2^i}(--)_L \Big]$   & (5,5) \\\hline
\multicolumn{3}{|l|}{$Q=-1/3$ quarks} \\\hline
 $ \bar  q_L^{d_i} q_L^{d_i} Z $ &
$-i \gamma^{\mu} \Big[ g_Z(q^{d_i}) - \epsilon  g_Z(q^{d_i})
 \frac{1}{\cos^2\psi}\mathcal{I}_1^+ \underset{11}{\mathcal{R}_1^i}(++)_L
 + \epsilon \frac{\cos\phi}{\cos\psi} \mathcal{I}_1^- \kappa_1 
\underset{11}{\mathcal{P}_1^i}(++)_L\Big] $  & (1,1)  \\\hline
 $ \bar D_L^{\prime i} D_L^{\prime i} Z $ &
$-i \gamma^{\mu} \Big[ g_Z(D^{\prime i}) - \epsilon  g_Z(D^{\prime i})
 \frac{1}{\cos^2\psi}\mathcal{I}_1^+ \underset{11}{\mathcal{R}_3^i}(+-)_L
 + \epsilon \frac{\cos\phi}{\cos\psi} \mathcal{I}_1^- \kappa_3
\underset{11}{\mathcal{P}_3^i}(+-)_L\Big] $  & (2,2)  \\\hline
 $  \bar D_L^{i} D_L^{i} Z$ &
$ -i \gamma^{\mu} \Big[ g_Z(D^{i}) - \epsilon  g_Z(D^{i})
 \frac{1}{\cos^2\psi}\mathcal{I}_1^+ \underset{11}{\mathcal{R}_3^i}(--)_L
 + \epsilon \frac{\cos\phi}{\cos\psi} \mathcal{I}_1^- \kappa_5
\underset{11}{\mathcal{P}_3^i}(--)_L\Big] $  & (3,3)
\\\hline
\end{tabular} 
\end{center}
\renewcommand{\arraystretch}{1.0}
\caption{\it Couplings involving the $Z$ boson and the heavy 
left-handed fermions. In analogy to the couplings of the KK gluon 
to heavy fermions, one has to complete these rules with the couplings of the $Z$ boson 
to the heavy right-handed fermions according to the scheme given in Table \ref{tab:scheme1}.}
\end{table}

\begin{table}[htbp]
\renewcommand{\arraystretch}{1.5}
\begin{center}
\begin{tabular}{|c|c|c|} 
 \hline
\multicolumn{3}{|c|}{\bf\boldmath Zero mode couplings to the $Z_H$  boson}\\ 
\hline\hline
\multicolumn{3}{|l|}{$Q=2/3$ quarks} \\\hline
 %Nullmode 
 $ \bar q_L^{u_i(0)} q_L^{u_i(0)}  Z_H$ &
$ -i \gamma^{\mu} \Big[     g_Z(q^{u_i}) \cos\xi \,
\underset{00}{\mathcal{R}_1^i}(++)_L
+ \sin\xi \,\kappa_1 
\underset{00}{\mathcal{P}_1^i}(++)_L\Big] $ & (0,0) \\\hline
%zero mode RH
 $  \bar  u_R^{i(0)} u_R^{i(0)} Z_H$ &
$ - i \gamma^{\mu} \Big[    g_Z(u^i)
 \cos\xi \,
\underset{00}{\mathcal{R}_2^i}(++)_R
+ \sin\xi \,\kappa_3 
\underset{00}{\mathcal{P}_2^i}(++)_R \Big]$ & (0,0)  \\\hline
\multicolumn{3}{|l|}{$Q=-1/3$ quarks} \\\hline
 %zero mode
 $ \bar  q_L^{d_i(0)} q_L^{d_i(0)} Z_H $ &
$  -i \gamma^{\mu} \Big[    g_Z(q^{d_i})
 \cos\xi \,
\underset{00}{\mathcal{R}_1^i}(++)_L
+ \sin\xi \,\kappa_1 
\underset{00}{\mathcal{P}_1^i}(++)_L \Big]$  & (0,0)   \\\hline
% zero mode RH
 $ \bar D_R^{i(0)} D_R^{i(0)}  Z_H$ &
$ - i \gamma^{\mu} \Big[   g_Z(D^{i})
 \cos\xi \,
\underset{00}{\mathcal{R}_3^i}(++)_R
+   \sin\xi \,\kappa_5
\underset{00}{\mathcal{P}_3^i}(++)_R \Big]$  & (0,0) 
\\\hline
\end{tabular} 
\end{center}
\renewcommand{\arraystretch}{1.0}
\caption{\it Couplings involving zero modes 
and the $Z_H$ boson. These zero modes
correspond after rotation to fermion mass eigenstates to the SM quark fields.}
\end{table}

\begin{table}[htbp]
\renewcommand{\arraystretch}{1.5}
\begin{center}
\begin{tabular}{|c|c|c|} 
 \hline
\multicolumn{3}{|c|}{\bf\boldmath Off-diagonal couplings to the $Z_H$ boson}\\ 
\hline\hline
%offdiagonal
\multicolumn{3}{|l|}{$Q=2/3$ quarks} \\\hline
 $ \bar q_L^{u_i(0)} q_L^{u_i}  Z_H $ &
$- i \gamma^{\mu} \Big[    g_Z(q^{u_i}) \cos\xi \,
\underset{01}{\mathcal{R}_1^i}(++)_L
 + \sin\xi \,\kappa_1 
\underset{01}{\mathcal{P}_1^i}(++)_L \Big]$  & (0,1) \\\hline
%offdiagonal
 $ \bar q_L^{u_i} q_L^{u_i(0)}  Z_H $ &
$ -i \gamma^{\mu} \Big[    g_Z(q^{u_i}) \cos\xi \,
\underset{10}{\mathcal{R}_1^i}(++)_L
 + \sin\xi \,\kappa_1 
\underset{10}{\mathcal{P}_1^i}(++)_L\Big] $  & (1,0) \\\hline
%offdiagonal RH
 $ \bar  u_R^{i(0)} u_R^{i}  Z_H$ &
$ -i \gamma^{\mu} \Big[   g_Z(u^i)
 \cos\xi \,
\underset{01}{\mathcal{R}_2^i}(++)_R
 + \sin\xi \,\kappa_3 
\underset{01}{\mathcal{P}_2^i}(++)_R \Big]$  & (0,5) \\\hline
%offdiagonal RH
 $ \bar  u_R^{i} u_R^{i(0)}  Z_H$ &
$ -i \gamma^{\mu} \Big[    g_Z(u^i)
 \cos\xi \,
\underset{10}{\mathcal{R}_2^i}(++)_R
 + \sin\xi \,\kappa_3 
\underset{10}{\mathcal{P}_2^i}(++)_R\Big] $    & (5,0)
 \\\hline
%off-diagonal
\multicolumn{3}{|l|}{$Q=-1/3$ quarks} \\\hline
 $ \bar  q_L^{d_i(0)} q_L^{d_i}  Z_H$ &
$ -i \gamma^{\mu} \Big[   g_Z(q^{d_i})
 \cos\xi \,
\underset{01}{\mathcal{R}_1^i}(++)_L
 + \sin\xi \,\kappa_1 
\underset{01}{\mathcal{P}_1^i}(++)_L\Big] $  & (0,1) \\\hline
%off-diagonal
 $ \bar  q_L^{d_i} q_L^{d_i(0)}  Z_H$ &
$- i \gamma^{\mu} \Big[    g_Z(q^{d_i})
 \cos\xi \,
\underset{10}{\mathcal{R}_1^i}(++)_L
 + \sin\xi \,\kappa_1 
\underset{10}{\mathcal{P}_1^i}(++)_L\Big] $  & (1,0)\\\hline
% offdiagonal RH
 $  \bar D_R^{i(0)} D_R^{i} Z_H$ &
$ - i \gamma^{\mu} \Big[  g_Z(D^{i})
 \cos\xi \,
\underset{01}{\mathcal{R}_3^i}(++)_R
 + \sin\xi \,\kappa_5
\underset{01}{\mathcal{P}_3^i}(++)_R\Big] $   & (0,3)  \\\hline
% offdiagonal RH
 $ \bar D_R^{i} D_R^{i(0)}  Z_H$ &
$ -i \gamma^{\mu} \Big[    g_Z(D^{i})
 \cos\xi \,
\underset{10}{\mathcal{R}_3^i}(++)_R
 + \sin\xi \,\kappa_5
\underset{10}{\mathcal{P}_3^i}(++)_R\Big] $    & (3,0)
\\\hline
\end{tabular} 
\end{center}
\renewcommand{\arraystretch}{1.0}
\caption{\it Couplings involving the $Z_H$ boson and a
 single zero mode.}
\end{table}

%\clearpage

\begin{table}[htbp]
\renewcommand{\arraystretch}{1.5}
\begin{center}
\begin{tabular}{|c|c|c|} 
\hline
\multicolumn{3}{|c|}{\bf\boldmath Heavy fermion couplings to the $Z_H$ boson}\\ 
\hline\hline
\multicolumn{3}{|l|}{$Q=5/3$ quarks} \\\hline
$  \bar \chi^{u_i}_L  \chi^{u_i}_L Z_H$ &
$ - i \gamma^{\mu} \Big[   g_Z(\chi^{u_i}) \cos\xi \,
\underset{11}{\mathcal{R}_1^i}(-+)_L
+ \sin\xi \,\kappa_2 
\underset{11}{\mathcal{P}_1^i}(-+)_L\Big] $ & (1,1) \\\hline 
 $  \bar \psi^{\prime i}_L \psi^{\prime i}_L Z_H$ &
 $ - i \gamma^{\mu} \Big[  g_Z(\psi^{\prime i}) \cos\xi \,
\underset{11}{\mathcal{R}_3^i}(+-)_L  
+ \sin\xi \,\kappa_3 
\underset{11}{\mathcal{P}_3^i}(+-)_L \Big] $ & (2,2) \\\hline 
 $\bar \psi^{\prime\prime i}_L  \psi^{\prime\prime i}_L Z_H  $ &  
$ -i \gamma^{\mu} \Big[   g_Z(\psi^{\prime\prime i}) \cos\xi \,
 \underset{11}{\mathcal{R}_3^i}(+-)_L  
+ \sin\xi \,\kappa_4
 \underset{11}{\mathcal{P}_3^i}(+-)_L\Big]  $ & (3,3) \\\hline
\multicolumn{3}{|l|}{$Q=2/3$ quarks} \\\hline
 $ \bar q_L^{u_i} q_L^{u_i} Z_H $ &
$- i \gamma^{\mu} \Big[   g_Z(q^{u_i}) \cos\xi \,
\underset{11}{\mathcal{R}_1^i}(++)_L
 + \sin\xi \,\kappa_1 
\underset{11}{\mathcal{P}_1^i}(++)_L\Big] $  & (1,1) \\\hline
 $ \bar U_L^{\prime i} U_L^{\prime i} Z_H  $ &
$- i \gamma^{\mu} \Big[  g_Z(U^{\prime i}) \cos\xi \,
\underset{11}{\mathcal{R}_3^i}(+-)_L
 + \sin\xi \,\kappa_3 
\underset{11}{\mathcal{P}_3^i}(+-)_L \Big]$ & (2,2) \\\hline
 $ \bar U_L^{\prime \prime i} U_L^{\prime\prime i} Z_H$ &
$ - i \gamma^{\mu} \Big[   g_Z(U^{\prime\prime i})
 \cos\xi \,
\underset{11}{\mathcal{R}_3^i}(+-)_L
 + \sin\xi \,\kappa_3 
\underset{11}{\mathcal{P}_3^i}(+-)_L \Big]$ & (3,3)  \\\hline
 $ \bar \chi^{d_i}_L \chi^{d_i}_L  Z_H$ &
$- i \gamma^{\mu} \Big[   g_Z(\chi^{d_i})
 \cos\xi \,
\underset{11}{\mathcal{R}_1^i}(-+)_L
 + \sin\xi \,\kappa_2 
\underset{11}{\mathcal{P}_1^i}(-+)_L \Big]$ & (4,4) \\\hline
 $ \bar u_L^i  u_L^i Z_H $ &
$ - i \gamma^{\mu} \Big[    g_Z(u^i)
 \cos\xi \,
\underset{11}{\mathcal{R}_2^i}(--)_L
 + \sin\xi \,\kappa_3 
\underset{11}{\mathcal{P}_2^i}(--)_L\Big] $ & (5,5)  \\\hline
\multicolumn{3}{|l|}{$Q=-1/3$ quarks} \\\hline
 $ \bar  q_L^{d_i} q_L^{d_i}  Z_H$ &
$ - i \gamma^{\mu} \Big[   g_Z(q^{d_i})
 \cos\xi \,
\underset{11}{\mathcal{R}_1^i}(++)_L
 + \sin\xi \,\kappa_1 
\underset{11}{\mathcal{P}_1^i}(++)_L \Big]$  & (1,1) \\\hline
 $\bar D_L^{\prime i} D_L^{\prime i}  Z_H $ &
$- i \gamma^{\mu} \Big[  g_Z(D^{\prime i})
 \cos\xi \,
\underset{11}{\mathcal{R}_3^i}(+-)_L
 + \sin\xi \,\kappa_3
\underset{11}{\mathcal{P}_3^i}(+-)_L \Big]$ & (2,2)  \\\hline
 $ \bar D_L^{i} D_L^{i}  Z_H$ &
$ - i \gamma^{\mu} \Big[  g_Z(D^{i})
 \cos\xi \,
\underset{11}{\mathcal{R}_3^i}(--)_L
 +  \sin\xi \,\kappa_5
\underset{11}{\mathcal{P}_3^i}(--)_L \Big]$  & (3,3)
\\\hline
\end{tabular} 
\end{center}
\renewcommand{\arraystretch}{1.0}
\caption{\it Couplings involving the $Z_H$ boson 
and the heavy left-handed fermions. The rules for the couplings of the $Z_H$ boson 
to the heavy right-handed fermions can be found by means of the scheme given in 
Table \ref{tab:scheme1}.}
\end{table}

In order to obtain the $Z^{\prime}$ couplings one can use the results of the $Z_H$ couplings
with the replacements summarised in Table \ref{tab:scheme2}.
\begin{table}[h!]
\renewcommand{\arraystretch}{1.3}
\begin{center}
\begin{tabular}{|lcc|}
\hline 
 $Z_H$ & $\to $&  $Z^{\prime}$\\
\hline 
$\cos\xi$& $ \to $& $-\sin\xi$\\
$\sin\xi$& $ \to $& $\cos\xi$\\
\hline 
\end{tabular} 
\end{center}
\renewcommand{\arraystretch}{1.0}
\caption{\it Replacement rules for obtaining the $Z'$ couplings from the $Z_H$ couplings. \label{tab:scheme2}}
\end{table}

%\clearpage

\subsection{Fermion Couplings to Charged Gauge Bosons}

In Tables \ref{tab:W+SM}--\ref{tab:chargedKK} we give all  fermion couplings to $W^+$ and $W_H^+$.
Similarly to the case of heavy neutral gauge bosons, the $W^{\prime +}$ couplings are 
easy to get from $W_H^+$ couplings by making the replacements summarised in Table \ref{tab:scheme3}.

\begin{table}%[h!]
\renewcommand{\arraystretch}{1.3}
\begin{center}
\begin{tabular}{|lcc|}
\hline 
 $W_H$ & $\to $&  $W^{\prime}$\\
\hline 
$\cos\chi$& $ \to $& $-\sin\chi$\\
$\sin\chi$& $ \to $& $\cos\chi$\\
\hline 
\end{tabular}
\end{center} 
\renewcommand{\arraystretch}{1.0}
\caption{\it Replacement rules for obtaining the $W'$ couplings from the $W_H$ couplings.\label{tab:scheme3}}
\end{table}

Tables \ref{tab:W+SM}--\ref{tab:chargedKK} give automatically the couplings of $W^-$,  $W_H^-$ and $W^{\prime -}$ to
fermionic flavour eigenstates. But as e.\,g. 
$ \bar q_L^{u_i(0)} q_L^{d_i(0)} W^+$ is now replaced by 
$ \bar q_L^{d_i(0)} q_L^{u_i(0)} W^-$, after rotation to fermionic mass eigenstates through 
complex matrices  $\mathcal{U}_{L,R}$  and $\mathcal{D}_{L,R}$, the 
couplings of $W^+$,  $W_H^+$ and $W^{\prime +}$ and  $W^-$,  $W_H^-$ and $W^{\prime -}$ will
differ from each other by complex conjugation of the relevant mixing matrix. For instance 
$V_{td}$ in the vertex $\bar t d W^+ $ will be changed to $V_{td}^{\ast}$ in $\bar d t W^-$.

\begin{table}%[htbp]
\renewcommand{\arraystretch}{1.5}
\begin{center}
\begin{tabular}{|c|c|c|} 
\hline
\multicolumn{3}{|c|}{\bf\boldmath Zero mode coupling to the $W^+$ boson}\\ 
\hline\hline
\multicolumn{3}{|l|}{$Q=-1/3 \to Q=2/3$ transitions} \\\hline
%zero modes
 $ \bar q_L^{u_i(0)} q_L^{d_i(0)} W^+$ &
$ -i \frac{g}{\sqrt{2L}} \gamma^{\mu} \Big[     
1- \epsilon\, \mathcal{I}_1^+ \underset{00}{\mathcal{R}_1^i}(++)_L \Big] $  & (0,0)  
\\\hline
\end{tabular} 
\end{center}
\renewcommand{\arraystretch}{1.0}
\caption{\it In case of the $W^+$ we have only a single SM-like coupling 
including two zero mode fermion fields.\label{tab:W+SM}}
\end{table}

\begin{table}[htbp]
\renewcommand{\arraystretch}{1.5}
\begin{center}
\begin{tabular}{|c|c|c|} 
 \hline
\multicolumn{3}{|c|}{\bf\boldmath  $W^+$ boson couplings involving a single zero mode }\\ 
\hline\hline
\multicolumn{3}{|l|}{$Q=2/3 \to Q=5/3$ transitions} \\\hline
 %one zero mode
 $ \bar \chi^{u_i}_L  q_L^{u_i(0)} W^+$ &
$ -i \frac{g}{\sqrt{2L}} \gamma^{\mu} 
\epsilon\, \mathcal{I}_1^- \underset{10}{\mathcal{S}_1^i}(-+)(++)_L $  & (1,0)  
\\\hline
\multicolumn{3}{|l|}{$Q=-1/3 \to Q=2/3$ transitions} \\\hline
%one zero mode
 $ \bar q_L^{u_i(0)} q_L^{d_i} W^+$ &
$ i \frac{g}{\sqrt{2L}} \gamma^{\mu}   
\epsilon\, \mathcal{I}_1^+ \underset{01}{\mathcal{R}_1^i}(++)_L  $     & (0,1) 
\\\hline
%one zero mode
 $ \bar q_L^{u_i} q_L^{d_i(0)} W^+$ &
$ i \frac{g}{\sqrt{2L}} \gamma^{\mu}      
\epsilon\, \mathcal{I}_1^+ \underset{10}{\mathcal{R}_1^i}(++)_L $   & (1,0)   
\\\hline
%one zero mode
 $ \bar   \chi^{d_i}_L  q_L^{d_i(0)} W^+$ &
$ -i \frac{g}{\sqrt{2L}} \gamma^{\mu}      
\epsilon\, \mathcal{I}_1^- \underset{10}{\mathcal{S}_1^i}(-+)(++)_L  $     & (4,0) 
\\\hline
%one zero mode
 $ \bar   U_R^{\prime\prime i}  D_R^{i(0)} W^+$ &
$ -i \frac{g}{\sqrt{2L}} \gamma^{\mu}   
\epsilon\, \mathcal{I}_1^- \underset{10}{\mathcal{S}_3^i}(-+)(++)_R  $     & (3,0) 
\\\hline
\end{tabular} 
\end{center}
\renewcommand{\arraystretch}{1.0}
\caption{\it Couplings involving the
 $W^+$ boson and a single zero mode.}
\end{table}

\begin{table}[htbp]
\renewcommand{\arraystretch}{1.5}
\begin{center}
\begin{tabular}{|c|c|c|} 
\hline
\multicolumn{3}{|c|}{\bf\boldmath Heavy fermion couplings to the $W^+$ boson}\\ 
\hline\hline
\multicolumn{3}{|l|}{$Q=2/3 \to Q=5/3$ transitions} \\\hline
 $ \bar \chi^{u_i}_L  \chi^{d_i}_L  W^+$ &
$ -i \frac{g}{\sqrt{2L}} \gamma^{\mu} \Big[     
1-\epsilon\, \mathcal{I}_1^+ \underset{11}{\mathcal{R}_1^i}(-+)_L \Big] $  & (1,4)  
\\\hline
 $ \bar \psi^{\prime i}_L  U_L^{\prime i}  W^+$ &
$ -i \frac{g}{\sqrt{L}} \gamma^{\mu} \Big[     
-1+ \epsilon\, \mathcal{I}_1^+ \underset{11}{\mathcal{R}_3^i}(+-)_L \Big] $   & (2,2)   
\\\hline
 $ \bar \chi^{u_i}_L  q_L^{u_i}  W^+$ &
$ -i \frac{g}{\sqrt{2L}} \gamma^{\mu}    
\epsilon\, \mathcal{I}_1^- \underset{11}{\mathcal{S}_1^i}(-+)(++)_L  $  & (1,1)      
\\\hline
 $ \bar \psi^{\prime\prime i}_L  U_L^{\prime \prime i}  W^+$ &
$ i \frac{g}{\sqrt{L}} \gamma^{\mu}   
\epsilon \mathcal{I}_1^- \underset{11}{\mathcal{S}_3^i}(+-)(+-)_L  $    & (3,3)  
\\\hline
\multicolumn{3}{|l|}{$Q=-1/3 \to Q=2/3$ transitions} \\\hline
 $ \bar q_L^{u_i} q_L^{d_i} W^+$ &
$ -i \frac{g}{\sqrt{2L}} \gamma^{\mu} \Big[     
1 - \epsilon\, \mathcal{I}_1^+ \underset{11}{\mathcal{R}_1^i}(++)_L \Big] $  & (1,1)    
\\\hline
 $ \bar U_L^{\prime i}  D_L^{\prime i} W^+$ &
$ -i \frac{g}{\sqrt{L}} \gamma^{\mu} \Big[     
1 - \epsilon \mathcal{I}_1^+ \underset{11}{\mathcal{R}_3^i}(+-)_L \Big] $ & (2,2)     
\\\hline
 $ \bar   \chi^{d_i}_L  q_L^{d_i} W^+$ &
$ -i \frac{g}{\sqrt{2L}} \gamma^{\mu}    
\epsilon\, \mathcal{I}_1^- \underset{11}{\mathcal{S}_1^i}(-+)(++)_L  $    & (4,1)  
\\\hline
 $ \bar U_L^{\prime\prime i} D_L^{ i}W^+$ &
$ -i \frac{g}{\sqrt{L}} \gamma^{\mu}     
\epsilon \mathcal{I}_1^- \underset{11}{\mathcal{S}_3^i}(+-)(--)_L $   & (3,3)   
\\\hline
\end{tabular} 
\end{center}
\renewcommand{\arraystretch}{1.0}
\caption{\it Couplings including the $W^+$ boson
 and the heavy left-handed fermions. For each coupling to left-handed heavy 
fermion fields there exist one with right-handed fermions.
The corresponding couplings can be read off from this table by 
making the substitution according to the scheme given in Table \ref{tab:scheme1}.}
\end{table}

\begin{table}[htbp]
\renewcommand{\arraystretch}{1.5}
\begin{center}
\begin{tabular}{|c|c|c|} 
 \hline
\multicolumn{3}{|c|}{\bf\boldmath Zero mode coupling to the $W_H^+$ boson}\\ 
\hline\hline
\multicolumn{3}{|l|}{$Q=-1/3 \to Q=2/3$ transitions} \\\hline
 $ \bar q_L^{u_i(0)} q_L^{d_i(0)} W_H^+$ &
$ -i \frac{g}{\sqrt{2L}} \gamma^{\mu}      
\cos\chi\, \underset{00}{\mathcal{R}_1^i}(++)_L  $    & (0,0)   
\\\hline
\end{tabular} 
\end{center}
\renewcommand{\arraystretch}{1.0}
\caption{\it In case of the $W_H^+$ we have only a single SM-like coupling 
involving two zero mode fermion fields.}
\end{table}

\begin{table}[htbp]
\renewcommand{\arraystretch}{1.5}
\begin{center}
\begin{tabular}{|c|c|c|} 
 \hline
\multicolumn{3}{|c|}{\bf \boldmath  $W_H^+$ boson couplings involving a single zero mode }\\ 
\hline\hline
\multicolumn{3}{|l|}{$Q=2/3 \to Q=5/3$ transitions} \\\hline
 $ \bar \chi^{u_i}_L  q_L^{u_i(0)} W_H^+$ &
$ -i \frac{g}{\sqrt{2L}} \gamma^{\mu} 
\sin\chi\, \underset{10}{\mathcal{S}_1^i}(-+)(++)_L $   & (1,0)  
\\\hline
\multicolumn{3}{|l|}{$Q=-1/3 \to Q=2/3$ transitions} \\\hline
 $ \bar q_L^{u_i(0)} q_L^{d_i} W_H^+$ &
$ - i \frac{g}{\sqrt{2L}} \gamma^{\mu}   
\cos\chi\, \underset{01}{\mathcal{R}_1^i}(++)_L  $  & (0,1)    
\\\hline
%one zero mode
 $ \bar q_L^{u_i} q_L^{d_i(0)} W_H^+$ &
$ - i \frac{g}{\sqrt{2L}} \gamma^{\mu}      
\cos\chi\, \underset{10}{\mathcal{R}_1^i}(++)_L $    & (1,0)  
\\\hline
%one zero mode
 $ \bar   \chi^{d_i}_L  q_L^{d_i(0)} W_H^+$ &
$ -i \frac{g}{\sqrt{2L}} \gamma^{\mu}      
\sin\chi\, \underset{10}{\mathcal{S}_1^i}(-+)(++)_L $    & (4,0)   
\\\hline
%one zero mode
 $ \bar   U_R^{\prime\prime i}  D_R^{i(0)} W_H^+$ &
$ -i \frac{g}{\sqrt{2L}} \gamma^{\mu}   
\sin\chi\, \underset{10}{\mathcal{S}_3^i}(-+)(++)_R  $   & (3,0)   
\\\hline
\end{tabular} 
\end{center}
\renewcommand{\arraystretch}{1.0}
\caption{\it Here we show all couplings involving the
 $W_H^+$ boson and a single zero mode.}
\end{table}

\begin{table}[htbp]
\renewcommand{\arraystretch}{1.5}
\begin{center}
\begin{tabular}{|c|c|c|} 
 \hline
\multicolumn{3}{|c|}{\bf\boldmath Heavy fermion couplings to the $W_H^+$ boson}\\ 
\hline\hline
\multicolumn{3}{|l|}{$Q=2/3 \to Q=5/3$ transitions} \\\hline
 $ \bar \chi^{u_i}_L  \chi^{d_i}_L  W_H^+$ &
$ -i \frac{g}{\sqrt{2L}} \gamma^{\mu}     
\cos\chi\, \underset{11}{\mathcal{R}_1^i}(-+)_L  $    & (1,4)  
\\\hline
 $ \bar \psi^{\prime i}_L  U_L^{\prime i}  W_H^+$ &
$ i \frac{g}{\sqrt{L}} \gamma^{\mu}      
\cos\chi\,  \underset{11}{\mathcal{R}_3^i}(+-)_L  $    & (2,2)  
\\\hline
 $ \bar \chi^{u_i}_L  q_L^{u_i}  W_H^+$ &
$ -i \frac{g}{\sqrt{2L}} \gamma^{\mu}    
\sin\chi\, \underset{11}{\mathcal{S}_1^i}(-+)(++)_L  $    & (1,1)    
\\\hline
 $ \bar \psi^{\prime\prime i}_L  U_L^{\prime \prime i}  W_H^+$ &
$ i \frac{g}{\sqrt{L}} \gamma^{\mu}   
\sin\chi\, \underset{11}{\mathcal{S}_3^i}(+-)(+-)_L  $    & (3,3)  
\\\hline
\multicolumn{3}{|l|}{$Q=-1/3 \to Q=2/3$ transitions} \\\hline
 $ \bar q_L^{u_i} q_L^{d_i} W_H^+$ &
$ -i \frac{g}{\sqrt{2L}} \gamma^{\mu}     
 \cos\chi\, \underset{11}{\mathcal{R}_1^i}(++)_L $    & (1,1)  
\\\hline
 $ \bar U_L^{\prime i}  D_L^{\prime i} W_H^+$ &
$ -i \frac{g}{\sqrt{L}} \gamma^{\mu}     
 \cos\chi\, \underset{11}{\mathcal{R}_3^i}(+-)_L  $   & (2,2)   
\\\hline
 $ \bar   \chi^{d_i}_L  q_L^{d_i} W_H^+$ &
$ -i \frac{g}{\sqrt{2L}} \gamma^{\mu}    
\sin\chi\, \underset{11}{\mathcal{S}_1^i}(-+)(++)_L  $    & (4,1)  
\\\hline
 $ \bar U_L^{\prime\prime i} D_L^{ i}W_H^+$ &
$ -i \frac{g}{\sqrt{L}} \gamma^{\mu}     
\sin\chi\, \underset{11}{\mathcal{S}_3^i}(+-)(--)_L $    & (3,3)  
\\\hline
\end{tabular} 
\end{center}
\renewcommand{\arraystretch}{1.0}
\caption{\it Couplings involving the $W_H^+$ boson 
and the heavy left-handed 
fermions. For each coupling to left-handed heavy fermion fields there exist one with 
right-handed fermions that can be obtained using Table \ref{tab:scheme1}.\label{tab:chargedKK}}
\end{table}

\clearpage

%\clearpage

{
\subsection{Triple Gauge Boson Couplings}

In this section we list the triple gauge boson couplings, where we give the
SM-like couplings up to the order of $\ord(\epsilon)$, while the couplings involving a
heavy gauge boson are given at $\mathcal{O}(1)$.
Therefore we define the following overlap integrals:
\be
\mathcal{T}_3^{+++} = \frac{1}{L} \int_0^L dy \,g(y)^3\,,\qquad
\mathcal{T}_3^{--+} = \frac{1}{L} \int_0^L dy \,\tilde g(y)^2 g(y)\,\qquad
\mathcal{T}_3^{---} = \frac{1}{L} \int_0^L dy \,\tilde g(y)^3\,.
\ee
The corresponding overlap integrals with one or two shape functions of the first
KK mode simplify because of the orthonormality condition.

The following Feynman rules are given in gauge boson mass eigenstates. The Dirac structure of all vertices is the same,
\begin{center}
\begin{minipage}{1cm}
\qquad
\end{minipage}
\begin{minipage}{1cm}
\begin{picture}(-150,100) (100,0)
    \SetWidth{0.5}
    \SetColor{Black}
    \Photon(10,10)(50,35){2}{7}
     \LongArrow(22,10)(41,21)
\LongArrow(78,10)(59,21)
\LongArrow(56,70)(56,47)
    \Photon(90,10)(50,35){2}{7}
     \Photon(50,82)(50,35){2}{7}
   \Vertex(50,35){1.41}
     \Text(94,2)[lb]{{\Black{$V_\nu^-$}}}
     \Text(-9,2)[lb]{{\Black{$V_\mu^+$}}}
     \Text(46,87)[lb]{{\Black{$V_\rho^0$}}}
     \Text(21,24)[lb]{{\Black{$k$}}}
     \Text(77,23)[lb]{{\Black{$p$}}}
     \Text(39.5,55)[lb]{{\Black{$q$}}}
    \end{picture}
\end{minipage}
\begin{minipage}{7.5cm}
$\mathcal{C}\left[\eta_{\mu\nu}(k-p)_\rho +\eta_{\nu\rho}(p-q)_\mu+\eta_{\rho\mu}(q-k)_\nu\right]\,,$
\end{minipage}
\end{center}
where $V_\mu^+ =W_\mu^+,W_{H\mu}^+,W_\mu^{\prime+}$, $V_\nu^-=W_\nu^-, W_{H\nu}^-,W_\nu^{\prime-}$, $V_\rho^0=A_\rho^{(0)},A_\rho^{(1)}, Z_\rho,Z_{H\rho},Z_\rho'$, and $k,p,q$ are their incoming momenta. Therefore in Tables \ref{tab:WWZ}--\ref{tab:WWA1} we collect only the coefficients $\mathcal{C}$ of the respective couplings.

In Table \ref{tab:WWZ} we give a subset of the vertices involving the $Z$ boson, from which the remaining $Z$ vertices can be obtained performing the replacements in Table \ref{tab:scheme3}.

\begin{table}[htbp]
\renewcommand{\arraystretch}{1.5}
\centering
\begin{tabular}{|c|c|} 
\hline
\multicolumn{2}{|c|}{\bf\boldmath Couplings to the $Z$ boson}\\ 
\hline\hline
 $ W^+W^-Z $ &
$ i \frac{g}{\sqrt{L}} \cos\psi +\mathcal{O}(\epsilon^2)$ \\\hline
%zero mode RH
 $  W^+_HW^-Z $ &
$\mathcal{O}(\epsilon)$ \\\hline
 $  W^+_HW^-_HZ $ &
$i \frac{g}{\sqrt{L}} \left(\cos\psi \cos^2\chi -\sin\phi \sin\psi \sin^2\chi \right) $  
\\\hline
 $ W^{\prime +}W^-_HZ  $ &
 $- i \frac{g}{\sqrt{L}} \sin\chi \cos\chi \left(\cos\psi +\sin\phi \sin\psi  \right) $ 
\\\hline
\end{tabular} 
\renewcommand{\arraystretch}{1.0}
\caption{\it Triple gauge boson couplings to the $Z$ boson. The remaining
  vertices $ W^{\prime +}W^-Z $ and $ W^{\prime +}W^{\prime -}Z$ can be simply
derived by making use of the replacements of table \ref{tab:scheme3}. Furthermore the
coupling of $  W^+W^-_HZ $ is equal to $  W^+_HW^-Z$, the same ist valid for $
W^{+}W^{\prime -}Z $ and $ W^{\prime +}W^-Z $ as well as $ W^+_H W^{\prime -}Z
$ and $ W^{\prime +}W^-_HZ$.\label{tab:WWZ}}
\end{table}

The corresponding $Z_H$ vertices are given in Table \ref{tab:WWZH}.
In order to obtain the triple gauge boson couplings involving $Z^{\prime}$ one can
use the results of the couplings to $Z_H$
with the replacements of Table \ref{tab:scheme2}.

\begin{table}[htbp]
\renewcommand{\arraystretch}{1.5}
\centering
\begin{tabular}{|c|c|} 
\hline
\multicolumn{2}{|c|}{\bf\boldmath Couplings to the $Z_H$ boson}\\ 
\hline\hline
 $ W^+W^-Z_H $ &
$\mathcal{O}(\epsilon)$ \\\hline
%zero mode RH
 $  W^+_HW^-Z_H $ &
$i \frac{g}{\sqrt{L}} \cos\psi \cos\chi\cos\xi $ \\\hline
 $  W^+_HW^-_HZ_H $ &
$i \frac{g}{\sqrt{L}} \big(\cos\psi \cos^2\chi \cos\xi\,
  \mathcal{T}_3^{+++} + \cos\phi \sin^2\chi \sin\xi\, \mathcal{T}_3^{---}$ \\
& \hfill $
-\sin\phi \sin\psi \sin^2\chi \cos\xi\, \mathcal{T}_3^{--+} \big) $  
\\\hline
 $ W^{\prime +}W^-_HZ_H  $ &
$i \frac{g}{\sqrt{L}} \sin\chi \cos\chi \big(-\cos\psi  \cos\xi\,
  \mathcal{T}_3^{+++} + \cos\phi  \sin\xi\, \mathcal{T}_3^{---}$ \\
& \hfill $ -\sin\phi \sin\psi \cos\xi\, \mathcal{T}_3^{--+} \big)$ 
\\\hline
\end{tabular} 
\renewcommand{\arraystretch}{1.0}
\caption{\it Triple gauge boson couplings to the $Z_H$ boson. The remaining
  vertices can again be 
derived by making use of the replacements of Tables \ref{tab:scheme2} and \ref{tab:scheme3}.\label{tab:WWZH}}
\end{table}

Finally we give in Tables \ref{tab:WWA0} and \ref{tab:WWA1} the vertices involving the photon and its first KK mode.

\begin{table}[htbp]
\renewcommand{\arraystretch}{1.5}
\centering
\begin{tabular}{|c|c|} 
\hline
\multicolumn{2}{|c|}{\bf Couplings to the photon}\\ 
\hline\hline
 $ W^+W^-A^{(0)} $ &
$i \frac{g}{\sqrt{L}} \sin\psi $ \\\hline
%zero mode RH
 $  W^+_HW^-A^{(0)} $ &
0 \\\hline
 $  W^+_HW^-_HA^{(0)}$ &
$i \frac{g}{\sqrt{L}} \sin\psi  $  
\\\hline
 $ W^{\prime +}W^-_HA^{(0)}$ &
0 
\\\hline
\end{tabular} 
\renewcommand{\arraystretch}{1.0}
\caption{\it Triple gauge boson couplings to the photon zero mode. The remaining
  vertices can again be 
derived by making use of the replacements of Table \ref{tab:scheme3}.\label{tab:WWA0}}
\end{table}

\begin{table}[htbp]
\renewcommand{\arraystretch}{1.5}
\centering
\begin{tabular}{|c|c|} 
\hline
\multicolumn{2}{|c|}{\bf Couplings to the KK photon}\\ 
\hline\hline
 $ W^+W^-A^{(1)} $ &
$\mathcal{O}(\epsilon)$  \\\hline
%zero mode RH
 $  W^+_HW^-A^{(1)} $ &
$i \frac{g}{\sqrt{L}} \sin\psi \cos\chi $ \\\hline
 $ W^+_HW^-_HA^{(1)}$ &
$i \frac{g}{\sqrt{L}}\sin\psi \left(\cos^2\chi\, \mathcal{T}_3^{+++}
+\sin^2\chi\, \mathcal{T}_3^{--+} \right) $  
\\\hline
 $ W^{\prime +}W^-_HA^{(1)}$ &
$i \frac{g}{\sqrt{L}} \sin\chi \cos\chi  \sin\psi\left(-\mathcal{T}_3^{+++}
+\mathcal{T}_3^{--+}  \right)$ 
\\\hline
\end{tabular} 
\renewcommand{\arraystretch}{1.0}
\caption{\it Triple gauge boson couplings to the first KK photon mode. The remaining
  vertices can again be 
derived by making use of the replacements of Table \ref{tab:scheme3}.\label{tab:WWA1}}
\end{table}

}

\clearpage

\end{appendix}

%\bibliography{bibliography}
%\bibliographystyle{JHEP}

\providecommand{\href}[2]{#2}\begingroup\raggedright\endgroup

\end{document}